\documentclass[twoside,11pt]{article}

\usepackage{jmlr2e}

\usepackage{lmodern}
\usepackage[T1]{fontenc}
\usepackage[utf8]{inputenc}
\usepackage[english]{babel}
\usepackage{booktabs}
\usepackage{mathtools}
\usepackage{amsmath}
\usepackage{amssymb}
\usepackage{graphicx}
\usepackage{setspace}
\usepackage{xcolor,soul}

\providecommand{\tabularnewline}{\\}

\usepackage{psfrag}
\usepackage{epsf}
\usepackage{enumerate}
\usepackage{breakcites}
\usepackage{breqn}
\usepackage{placeins}

\usepackage{epstopdf}
\usepackage[caption=false]{subfig}

\ShortHeadings{Combining Geometric and Topological Information for Boundary Estimation}{Luo and Strait}
\firstpageno{1}
 
\begin{document}

\title{Combining Geometric and Topological Information for Boundary Estimation}

\author{\name Hengrui Luo \email luo.619@osu.edu \\
       \addr Department of Statistics\\
       The Ohio State University\\
       Columbus, OH 43210, USA
       \AND
       \name Justin D. Strait \email justin.strait@uga.edu \\
       \addr Department of Statistics\\
       University of Georgia\\
       Athens, GA 30602, USA}


\maketitle

\begin{abstract}
A fundamental problem in computer vision is boundary estimation, where the goal is to delineate the boundary of objects in an image.  In this paper, we propose a method which jointly incorporates geometric and topological information within an image to simultaneously estimate boundaries for objects within images with more complex topologies. We use a topological clustering-based method to assist initialization of the Bayesian  active contour model. This combines pixel clustering, boundary smoothness, and potential prior shape information to produce an estimated object boundary. Active contour methods are known to be extremely sensitive to algorithm initialization, relying on the user to provide a reasonable starting curve to the algorithm. In the presence of images featuring objects with complex topological structures, such as objects with holes or multiple objects, the user must initialize separate curves for each boundary of interest.  Our proposed topologically-guided method can provide an interpretable, smart initialization in these settings, freeing up the user from potential pitfalls associated with objects of complex topological structure. We provide a detailed simulation study comparing our initialization to boundary estimates obtained from standard segmentation algorithms. The method is demonstrated on artificial image datasets from computer vision, as well as real-world applications to skin lesion and neural cellular images, for which multiple topological features can be identified.
\end{abstract}

\newpage
\section{Introduction}

\subsection{The Boundary Estimation Problem}

Image segmentation is a field with a rich literature, comprised of numerous methods addressing various sub-tasks which generally aim to split a 2-dimensional image into meaningful segments. One such sub-task is \emph{boundary estimation}, which refers to the problem of delineating the contour(s) of objects in an image. This important problem has been extensively
studied in the engineering, computer science, and statistics literature (see e.g., 
\citet{huang1995quantitative,zhang1996survey,AMFM2011,bryner2013elastic}).
Classical statistical shape analysis methods \citep{Kendall} and
recently developed topology-based methods \citep{Carlsson2009} can both
be applied to this problem. However, many existing methods use either geometric or topological information exclusively. Intuitively, the topology and shape of objects within images are both important pieces of information which can be used in conjunction to guide boundary estimation. However, it is not immediately clear how statistical shape analysis methods \citep{Kendall,joshisrivBAC,bryner2013elastic} and topology-based methods \citep{Paris&Durand2007,letscher2007image,Carlsson2009} can be combined together to accomplish this task. 

In this paper, we propose one such model for fusing topological information within a boundary estimation model from the shape analysis literature.  We attempt to combine topological and shape information for boundary estimation, based on the following considerations:
\begin{enumerate}
    \item For some image datasets, one type of information may dominate and an analysis that ignores the other type of information would be sufficient. For example, active contour methods work well in estimating boundaries of the noiseless MPEG-7 computer vision dataset, as minimal noise is observed \citep{MPEG}.
    \item For other image datasets, both kinds of information can be used together to improve boundary estimates. For example, topological methods \citep{Paris&Durand2007} may perform better in terms of visual validation, boundary smoothness and performance measures when joined with shape analysis.
    \item It is not immediately clear at the moment how to identify datasets for which shape analysis methods or topology-based methods will suffice on their own. To determine which approach is better, or whether an interpretable combination of the two is needed, requires a careful comparison.
\end{enumerate}
Our proposed method, which we deem to be ``topologically-aware,'' provides an approach for boundary estimation of complex topological objects within images. Instead of providing an expert segmentation boundary as initialization, the user may simply tune parameters of topological methods to yield a reasonable initialization. This can help reduce user error within the existing estimation method, as well as provide additional guidance to the practitioner in more general settings. To help motivate the proposed method, we proceed by outlining two classes of boundary estimation procedures.

\subsection{Contours and Shape Analysis}
\label{subsec:StatShape}
One approach to boundary estimation of images is to treat the problem as a  \emph{curve estimation} problem. 
In this setting, the goal is to estimate object boundaries within images as curve(s) in $\mathbb{R}^{2}$, which separates the targeted object(s) of interest from the image background. 

Within curve estimation, one popular approach is called active contours \citep{Zhu&Yuille1996}. For a 2-dimensional grayscale image, the goal of the active contour model is to find a 2-dimensional curve $u$ in the (larger) region $\Omega\subset\mathbb{R}^{2}$
that encompasses the clusters consisting of region $C$ by minimizing an energy functional defined by image and auxiliary curve-dependent terms  \citep{mumford1989optimal,joshisrivBAC,bryner2013elastic}. The former uses pixel value information both inside and outside the current curve estimate to ``push'' regions of the curve towards important image features. Typically, the other term tries to balance the image energy with a term which encourages smoothness of the boundary. 

An extension of active contours, called Bayesian active contours, was proposed by \citet{joshisrivBAC} and adapted by \citet{bryner2013elastic}. This supervised statistical shape analysis method uses an additional set of training images related to the test image (i.e., the image for which one desires boundary estimates) to improve boundary curve estimation by re-defining the energy functional to be minimized.
In particular, Bayesian active contours introduces a prior term, which aims to update curves towards a common shape identified within the training images. Placing higher weight on this term can have significant influence on the final boundary estimate. This formulation is useful if the image features objects of interest which are difficult to identify, perhaps due to large amounts of pixel noise. As our proposed method involves use of Bayesian active contours, we briefly discuss existing literature on statistical shape analysis, fundamental to the formulation of the prior term.

\emph{Shape} can be defined as a geometric feature of an object which is invariant to certain classes of transformations deemed shape-preserving: rigid-motion and scaling. In other words, applying any combination
of translation, rotation, and scale transformations may change the
image of the object's contour in $\mathbb{R}^{2}$, but preserves
its shape. Shape analysis research focuses on the representation of
these objects in a way which respects these invariances, along with
the development of statistical procedures for shape data. Early work
in the field focused on the representation of shape by a set of finite,
labeled points known as landmarks. If landmark labelings are known, statistical
modeling proceeds using traditional multivariate techniques, with
mathematical adjustments to account for the non-linear structure of the data.
\citet{Kendall} pioneered the landmark representation, with other
key work by \citet{bookstein1986} and \citet{DrydenBook}.

However, in the boundary estimation setting, the underlying contour is a closed curve in $\mathbb{R}^{2}$, parameterized by $\mathcal{D}=\mathbb{S}^{1}$
(the unit circle), rather than a known, finite landmark set. Correspondence between important shape features of curves is driven by how the curves are parameterized, rather than labeling of the landmarks. In the study of shapes of curves, invariance to parameterization is often desired, and is difficult to account for mathematically. Early work addressed this by standardizing
curves to be arc-length parameterized \citep{Zahn,younes-elastic-distance},
which has been shown to be sub-optimal for many problems \citep{JoshiSRVF},
as it imposes a strict, potentially unnatural correspondence of features
across a collection of curves. Recent work in \emph{elastic shape
analysis} resolves these issues by exploiting parameterization invariance
to allow for flexible re-parameterizations which optimally match points
between curves, commonly referred to as \textit{registration} \citep{SrivESA}.
This allows for a more natural correspondence of features and improved
statistical modeling of curves and shapes. The shape prior term in the energy functional for Bayesian active contours relies on this elastic view of shape.

Returning to Bayesian active contours, the desired energy functional is minimized via gradient-descent. This requires specification of an initial curve, which is then updated sequentially until a stable estimate is obtained. The method can be quite sensitive to this initialization, particularly since gradient-descent is only able to find local extrema. Thus, active contour methods require initializations to be ``reasonable.'' In the setting of images of objects with complex topological structures, this can be difficult, particularly as multiple curves must be pre-specified initially. We seek a different method for initialization, which learns the topological structure of the images. In a broader sense, this follows the general idea of providing pre-conditioners for an algorithm to improve its convergence and performance.  To explain this, we next discuss an alternate view to the boundary estimation problem, which provides insight into the application of topological data analysis to boundary estimation.

\subsection{Clustering and Topological Data Analysis}

\label{subsec:TDA}

A different perspective to boundary estimation can be obtained from treating image segmentation as an unsupervised \emph{point clustering problem}. \citet{AMFM2011} present a unified approach to the 2-dimensional image segmentation problem and an exhaustive
comparison between their proposed global-local methods along with existing
classical methods. Following their insights, image segmentation can be formulated as an unsupervised clustering problem of feature points. 

For a 2-dimensional grayscale image, the dataset comes in the form of pixel positions $(x_{i},y_{i})$ representing the 2-dimensional coordinates, and the pixel value $f_{i}$
representing grayscale values between 0 and 1 at this particular pixel
position (without loss of generality, one can re-scale pixel values to lie in this interval). The term feature space is used to describe the product space
of the position space and pixel value space, whose coordinates
are $((x_{i},y_{i}),f_{i})$. 
The most straightforward approach to cluster pixel points is according
to their pixel value $f_{i}$. This naive clustering method works
well when the image is binary and without any noise, i.e., $f_{i}\in\{0,1\}$.
Finer clustering criteria on a grayscale image usually lead to better performance for
a broader class of images. The final result is a clustering of pixel
points, equivalent to grouping of the points by this criterion. As
pointed out by \citet{Paris&Durand2007}, this hierarchical
clustering structure is intrinsically related to multi-scale analysis,
which can be studied by looking at level sets of pixel density functions. One can connect this to boundary estimation through the identification of curves of interest by boundaries formed from the clustering of pixel points. We propose using this latter hierarchical clustering structure to produce an initial boundary estimate, which can then be refined via the Bayesian active contours method discussed earlier.

The aforementioned discussion of constructing a hierarchical clustering structure based on pixel values has links to the rapidly growing field of topological data analysis (TDA). Many interdisciplinary applications make use of topological summaries as a new characterization of
the data, which provides additional insight to data analysis. TDA techniques
have also been an essential part of applied research focusing
on image processing, and in particular, the segmentation problem \citep{Paris&Durand2007,letscher2007image,Gao_et_al2013,Wu2017, clough2020topological}.
The topological summary is algebraically endowed with a multi-scale hierarchical structure
that can be used in various clustering tasks, including image segmentation.

In the procedure given by \citet{Paris&Durand2007}, a simplicial
complex can be constructed as a topological space that approximates
the underlying topology of the level sets of pixel density \citep{Luo_etal2019}. Nested complexes called \emph{filtrations}
can be introduced to describe the topology of the dataset at different
scales. In a filtration, we control the scale by adjusting a scale
parameter. A special kind of complex (i.e., Morse-Smale complex)
can be constructed from the level sets of a density function $f$,
whose scale parameter is chosen to be the level parameter of level
sets. 

Based on the grayscale image, we can estimate its grayscale density
$f$ using a (kernel) density estimator $\hat{f}$. Complexes
based on super-level sets of estimated grayscale density function
$\hat{f}$ of images are constructed as described above. In \citet{Paris&Durand2007},
the scale parameter $\tau$ is chosen to be related to level sets of
the density estimator, known as a boundary persistence parameter. By Morse theory, the topology of the level sets $\hat{S}_{\lambda}\coloneqq \hat{f}^{-1}{(\lambda,\infty)}$ changes when and only when a critical point of $\hat{f}$ enters (or exits) the level set. When $\lambda$ increases to cross a non-modal critical point (e.g., a saddle point), a new connected component is generated in $\hat{S_\lambda}$ (a topological feature is born). When $\lambda$ increases to cross a modal critical point (e.g., a maximum) from $\hat{S}_\lambda$, two connected components merge in $\hat{S}_\lambda$ (a topological feature dies). 

Therefore, each topological feature in the nested sequence $\hat{S}_{\lambda}$ of complexes with a ``birth-death'' pair $(a, b)$ corresponds to two critical points of $\hat{f}$. The persistence of a topological feature becomes the difference $\mid \hat{f}(b)-\hat{f}(a)\mid$ in terms of density estimator values and shows the ``contrast'' between two modes. The modal behavior of $\hat{f}$ is depicted by the topological changes in the level sets of $\hat{f}$, and the ``sharpness'' between modes is characterized in terms of boundary persistence. 

Then, this persistence-based
hierarchy indexed by $\tau$ describes the multi-scale topological structure of the feature space.
Clustering can be performed using this hierarchical structure, and
then we select those modal regions which persist long enough in terms
of the boundary persistence. A region whose boundary shows sharp contrast
to its neighboring regions will be identified by this method. We will
choose this as our primary approach for extracting the topological
information from the image, which is also the topology of the pixel density
estimator $\hat{f}$. 

We also point out that there exist other ``low-level'' competing topological methods for segmentation based on certain types of representative
cycles \citep{Deay_etal2010,Gao_et_al2013}. These methods are
more closely related to the philosophy of treating contours as curves,
instead of the clustering philosophy. 
There is a major difference between ``low-level'' \citep{Paris&Durand2007,Gao_et_al2013} and ``high-level'' topological methods \citep{carriere2020perslay,hu2019topology, clough2020topological}, which use the TDA summaries (e.g. persistence diagrams) for designing advanced statistical models (e.g., neural networks) for segmentation tasks is the interpretability. ``Low-level'' methods rely on structural information provided by the TDA pipeline, which uses the topological information directly from the image as pixel densities and representative cycles, while ``high-level'' methods do not necessarily tie back to the features in the image. Although these ``high-level'' methods perform exceptionally in some scenarios, we do not attempt to include them in the current paper, due to the loss of interpretability of ``low-level'' methods. To be more precise, we focus on the approaches that use the topology of the image directly. 

Finally, we note that topological methods have the inherent advantage of stability. With regards to image segmentation, this means that slight perturbations of images (in terms of pixel values and/or underlying boundaries) do not result in extreme changes in the estimated clusters obtained from topological methods. In addition, topological methods for image segmentation estimate the underlying topologies of objects lying within, rather than requiring the user to ``pre-specify'' topologies (as is the case when initializing active contour methods for boundary estimation). Thus, we believe that use of this philosophy can provide a robust initialization for active contour methods, which can then be refined through minimization of its corresponding energy functional.

\subsection{Organization}

The primary contribution of this paper is to present a novel topological initialization of the Bayesian active contour model for boundary estimation within images. The rest of the paper is organized as follows. In Section \ref{sec:Model-Specification-and},
we specify the Bayesian active contours (BAC) and topology-based (TOP)
boundary estimation methods, and their underlying model assumptions. Then,
we discuss the merits of our proposed method, which we call TOP+BAC. 
In Section \ref{sec:Data-Analysis}, we examine the proposed TOP+BAC using artificial images of objects with varying complexity in topology, which are perturbed by various types of noise. We provide a quantitative comparison to some existing methods (including TOP and BAC separately), along with qualitative guidance under different noise settings. Finally, we look at skin lesion images \citep{codella2018} and neural cellular (section) images \citep{jurrus2013semi}  to apply our method to estimate boundaries within images which potentially contain multiple connected components, which is very common in medical and biological images. We
conclude and discuss future work in Section \ref{sec:Conclusion}. 

\section{Model Specification and Implementation}

\label{sec:Model-Specification-and} In this section, we outline three
different models for boundary estimation of a 2-dimensional image. The method of Bayesian active contours (BAC), as described by \citet{joshisrivBAC} and \citet{bryner2013elastic},
is discussed in Section \ref{subsec:BAC}. Section \ref{subsec:TOP}
describes the topology-based method (TOP) of \citet{Paris&Durand2007}.  Finally, Section
\ref{subsec:TOPBAC} outlines our proposed method, TOP+BAC, which
combines TOP with BAC in a very specific way to produce a ``topology-aware''
boundary estimation. The initialization of the active contour method is guided by the coarser topological result. 

\subsection{Bayesian Active Contours (BAC)}

\label{subsec:BAC}

\subsubsection{Energy Functional Specification}

\label{subsubsec:Energy} In this section, we describe the Bayesian
active contours (BAC) approach to boundary estimation. Active contours seek
a contour curve $u$ delineating the target object from the background
by minimizing an energy functional. The model proposed by \citet{joshisrivBAC}
and \citet{bryner2013elastic} uses the following functional: 
\begin{align}
F(u) & \coloneqq\lambda_{1}E_{\text{image}}(u)+\lambda_{2}E_{\text{smooth}}(u)+\lambda_{3}E_{\text{prior}}(u),\label{eq:minimizing energy functional}
\end{align}
where $\lambda_{1},\lambda_{2},\lambda_{3}>0$ are user-specified
weights. Contrary to the unsupervised TOP method of Section \ref{subsec:TOP},
the supervised BAC model assumes that we have a set of training images
with known object boundaries, and a set of test images on which we desire boundary estimates. 

The image energy term, $E_{\text{image}}$, is given by the following
expression, 
\begin{align*}
E_{\text{image}}(u) & =-\int_{\text{int }u}\log(p_{\text{int}}(f(x,y)))\ dx\ dy-\int_{\text{ext }u}\log(p_{\text{ext}}(f(x,y)))\ dx\ dy,
\end{align*}
where $f(x,y)$ denotes the pixel value at location $(x,y)$,
and $\text{int }u,\ \text{ext }u$ denotes the interior and exterior
regions of the image as delineated by the closed curve $u$, respectively. The quantities $p_{\text{int}},\ p_{\text{ext}}$
are kernel density estimates of pixel values for the interior and
exterior, respectively. In \citet{joshisrivBAC}
and \citet{bryner2013elastic}, these densities are estimated from
training images, for which true contours of objects are known. 
Intuitively, this term measures the contrast in the image between the interior and exterior regions. 

The smoothing energy term, $E_{\text{smooth}}$, for contour $u$
is given by, 
\begin{align*}
E_{\text{smooth}}(u) & =\int_{u}|\nabla u(x,y)|^{2}\ dx\ dy,
\end{align*}
where $|\nabla u(x,y)|$ is the Jacobian of the curve at pixel coordinates
$(x,y)$. This quantity is approximated numerically, given the discrete
nature of an image. In general, the more wiggly $u$ is, the larger
$E_{\text{smooth}}$ will be. Thus, the prescribed weight $\lambda_{2}$ can be used to penalize boundary estimates which are not smooth.

The final term, $E_{\text{prior}}$, is the primary contribution of
\citet{joshisrivBAC} and \citet{bryner2013elastic}. This quantifies
the difference between the contour $u$ and the mean shape of training
curves, and the choice of $\lambda_{3}$ determines how much weight is associated with this difference. Briefly, assume we have $M$ boundary curves extracted from training images, $\beta_1,\beta_2,\ldots,\beta_M$, where $\beta_{i}:\mathcal{D}\rightarrow\mathbb{R}^{2}$ and $\mathcal{D}=\mathbb{S}^{1}$. We map each curve as an element into elastic shape space \citep{SrivESA}. A brief description of this mapping can be found in Appendix \ref{appendixESA}. Typically, one then projects shape space elements (a non-linear space) into a linear tangent space at the intrinsic mean shape of $\beta_1,\beta_2,\ldots,\beta_M$. This mean shape is identified to the origin of the tangent space. We compute the sample covariance matrix of finite-dimensional representations of shapes on this tangent space, where such linear operations are readily available. This covariance is decomposed into eigenvectors (contained in $U_J$) and eigenvalues (the diagonal of $\Sigma_J$). The columns of $U_J$ describe shape variation, ordered by proportion of total variation explained.

Suppose $u$ is the input contour into the energy functional. Referring to the image of $u$ in the aforementioned tangent space as $w$, the prior energy term is, 
\begin{align*}
E_{\text{prior}}(u) & =\frac{1}{2}w^{T}(U_{J}\Sigma_{J}^{-1}U_{J}^{\top})w+\frac{1}{2\delta^{2}}|w-U_{J}U_{J}^{\top}w|^2,
\end{align*}
where $\delta$ is selected to be a small enough quantity that is less than the smallest eigenvalue in $\Sigma_{J}$. The first term measures a covariance-weighted distance of the current contour from the sample mean shape of the training contours (with respect to the first $J$ modes of shape variation), and the second term accounts for the remaining difference after accounting for the $J$ modes of shape variation. This prior energy specification is equivalent to the negative log-likelihood for a truncated wrapped Gaussian model placed on the space of finite-dimensional approximations of curves. Further details about this term's construction are found in Appendix \ref{appendixESA}. Use of a shape prior
can help in boundary estimation within images where the object is very noisy or not fully observable,
by pushing the contour towards the desired shape. This is a consequence of the term's lack of reliance on test image pixel information.

\subsubsection{Gradient-Descent}

\label{subsubsec:GD} To minimize the energy functional
in Equation \ref{eq:minimizing energy functional}, we use a gradient-descent
algorithm. Given current contour $u^{(i)}$, we update by moving along
the negative gradient of $F$ as: 
\begin{align*}
u^{(i+1)}(t) & =u^{(i)}(t)-\lambda_{1}\nabla E_{\text{image}}(u^{(i)}(t))-\lambda_{2}\nabla E_{\text{smooth}}(u^{(i)}(t))-\lambda_{3}\nabla E_{\text{prior}}(u^{(i)}(t)).
\end{align*}
Updates proceed sequentially until the energy term converges (i.e.,
the rate of change of the total energy is below some pre-specified
convergence tolerance), or a maximum number of iterations is reached. This approach requires specification of an initial contour $u^{(0)}$, which can be chosen in many different ways. Unfortunately, as gradient-descent
is only guaranteed to find local extrema, which depends on the choice of initial point, final estimates can
be extremely sensitive to this initialization choice, as well as the parameters $\lambda_1,\lambda_2,\lambda_3$.

The gradient terms are given by: 
\begin{align*}
\nabla E_{\text{image}}(u^{(i)}(t)) & =-\log\Bigg(\frac{p_{\text{int}}(f(u^{(i)}(t)))}{p_{\text{ext}}(f(u^{(i)}(t)))}\Bigg)n(t)\\
\nabla E_{\text{smooth}}(u^{(i)}(t)) & =\kappa_{u^{(i)}}(t)n(t)\\
\nabla E_{\text{prior}}(u^{(i)})(t) & =\frac{1}{\epsilon}\big(u^{(i)}(t)-\beta_{\text{new}}(t)\big).
\end{align*}
where $n(t)$ is the outward unit normal vector to $u^{(i)}(t)$,
$\kappa_{u^{(i)}}(t)$ is the curvature function for $u^{(i)}(t)$,
$\beta_{\text{new}}$ is a function which depends on the sample
mean shape of training contours $\{\beta_{1},\ldots,\beta_{M}\}$, and $\epsilon>0$ controls the step size (typically chosen to be small, e.g., $\epsilon=0.3$) of the update from $u^{(i)}$ towards the mean shape.
At a particular point along the contour $u^{(i)}(t)$, the image update
moves this point inward along the normal direction if $p_{\text{int}}(f(u^{(i)}(t)))<p_{\text{ext}}(f(u^{(i)}(t)))$,
meaning that this current image location is more likely to be outside
the contour than inside, with respect to the estimated pixel densities
$p_{\text{int}}$ and $p_{\text{ext}}$. If the opposite is true,
then the point on the contour is updated outward along the normal
direction, ``expanding'' the contour. The smoothness update uses
curvature information to locally smooth the curve. Without the action
of other energy updates, this term will push any contour towards a
circle. Finally, the prior update term aims to evolve the current
contour $u^{(i)}$ a small amount in the direction of the sample mean
shape of the training data. Hereafter, we will indicate the parameters
of BAC as $(\lambda_{1},\lambda_{2},\lambda_{3})$. The explicit form
for $\beta_{\text{new}}$ in the shape prior is found in Appendix \ref{appendixESA}. We also refer readers to \citet{JoshiSRVF,SrivESA,KurtekJASA} and \citet{FSDA} for  details.

\subsection{Topological Segmentation (TOP)}

\label{subsec:TOP} In this section, we describe a topological segmentation
method based on a mean-shift model \citep{Paris&Durand2007}. This
is an unsupervised clustering method which operates on sets of pixels.
As pointed out in Section \ref{subsec:TDA}, the most straightforward approach to cluster feature points is according
to their pixel values $f_{i}$. 
However, one can also cluster these points according to which pixel value density mode they are ``closer'' to by the following procedure: 
\begin{itemize}
    \item First, we estimate the pixel density and extract the modes of the estimator using a mean-shift algorithm. \citet{Paris&Durand2007} designed such an algorithm that sorts out the modes $m_1,m_2,m_3\cdots$ of a kernel density estimator $\hat{f}$ (such that $\hat{f}(m_1)\geq\hat{f}(m_2)\geq\hat{f}(m_3)\geq\ldots$) and a batch of critical points $s_{12}\text{ (between $m_1$ and $m_2$)}$, $s_{13}\text{ (between $m_1$ and $m_3$)}$, $s_{23}\text{ (between $m_2$ and $m_3$)}\ldots$ between modes.
    \item Next, we segment using the modes and create clusters surrounding these modes. Then, we form a hierarchical structure between these clusters represented by modes using the notion of boundary persistence. The \emph{boundary persistence} $p_{12}$ between modes $m_1,m_2$ and the critical point $s_{12}$ between these two modes, is defined to be $p_{12}\coloneqq \hat{f}(m_{2})-\hat{f}(s_{12})$, if $\hat{f}(m_{1})>\hat{f}(m_{2})$. We merge clusters corresponding to those modes with persistence below a threshold $\tau$.
\end{itemize}
Suppose that we know all locations of maxima from the pixel density.
Then, one can compute the \emph{boundary persistence} $p_{b}(m_{1},m_{2})\coloneqq\hat{f}(m_{2})-\hat{f}(s_{12})$
between two local maxima $m_{1},m_{2}$, satisfying $\hat{f}(m_{1})>\hat{f}(m_{2})$
with a saddle point $s_{12}$ between these two points. This value
$p_{b}$ conveys relations between modal sets around $m_{1},m_{2}$
of the estimated density $\hat{f}$. Following this intuition, sets
of pixels corresponding to modes with boundary persistence greater
than a pre-determined threshold $T$ can be collected to form a hierarchical
structure of clusters, ordered by their persistence. 
Thus, we can proceed by clustering based on
boundary persistence, rather than calculating the whole filtration
of super-level sets of density estimator $\hat{f}$.

A natural question which arises with boundary persistence is how to obtain
locations of pixel density maxima. We can extract these
pixel density modes by using the \emph{mean-shift algorithm}.  
Intuitively, we first take the ``average
center'' $m(x)$ of those observations that are lying within a $\lambda$-ball
of the $x\in\mathcal{X}$, and then compute the deviation of $x$
to this center. We can generalize these notions by replacing the flat
kernel $K_{x}(y)=\boldsymbol{1}\{\|x-y\|\leq\lambda\}$ centered at
$x$ with the 2-dimensional Gaussian kernel $K_{\sigma_1,\sigma_2}$ parameterized by $\sigma_1,\sigma_2$, which we use exclusively in the rest
of paper.

The mean-shift algorithm
will iterate a point $x\in\mathcal{X}$ 
 and generate a sequence of points $x_{1}=x, \ x_{2}=m(x), \ x_{3}=m(m(x)),\ldots$.
It can be shown that the iterative algorithm stops when $m(\mathcal{X})=\mathcal{X}$.
The mean-shift sequence $x_{1},x_{2},x_{3},\ldots$ corresponds to
a steepest descent on the density $D(x)=\sum_{y\in\mathcal{X}}\tilde{K}(x-y)w(y)$,
where $\tilde{K}$ is the so-called shadow kernel associated with
the kernel $K$ \citep{cheng1995mean}. As specified above, after
performing the mean-shift algorithm to extract pixel density modes,
we can then compute the boundary persistence. Merging modes with boundary persistence less than certain threshold $t$, a hierarchical structure can be obtained by varying this threshold within a user-specified range $t \in [0,T]$. 
We will indicate the parameters of TOP as $(\sigma_{1},\sigma_{2},T)$ hereafter. Appendix \ref{appendixTOP} shows the output of this method, and contains further discussion about how clustering varies with the value of $T$. A benefit of this method, as with other topological methods, is that it produces clusters which are stable under pixel-valued perturbations.

\subsection{Bayesian Active Contours with Topological Initialization (TOP+BAC)}

\label{subsec:TOPBAC} As noted in Section \ref{subsec:BAC}, Bayesian
active contours require a gradient-descent algorithm with an initially-specified
contour $u^{(0)}$ to produce a final boundary estimate. Since gradient-descent searches for local extrema,
there is no guarantee that the final contour represents the target
of interest. In addition, many images feature objects with more complex
topological structure (e.g., objects with multiple connected components
and/or holes within).

We propose a new approach, Bayesian active contours
with topological initialization (TOP+BAC), as a combination of the
methods from Sections \ref{subsec:TOP} and \ref{subsec:BAC}. TOP+BAC uses the segmentation result obtained from TOP as the initial contour in the BAC model. This has the following benefits: 
\begin{enumerate}
\item TOP can provide an automatic way of initializing a single contour,
in the hope that the initialization-sensitive, gradient-descent algorithm converges
efficiently to the correct target. 
\item The results from TOP segmentation can be used to initialize multiple active contours simultaneously, for object boundaries which cannot be represented by a single closed curve due to its complex topology.
\item BAC can incorporate information a priori from the training dataset (with ground truth) into its prior term.
\item The smoothness term within the BAC energy functional allows one to control the smoothness of the final boundary estimate, which is generally desirable for interpretability. 
\end{enumerate}
TOP+BAC can be thought of as first using TOP to provide a coarse boundary estimate which is able to identify topological features of interest, followed by refinement of these results using BAC in order to produce finer estimates. This allows for increased interpretability and consistency in terms of topological properties of boundary estimate.

We note that one could certainly initialize multiple contours under BAC, and update these simultaneously. However, we argue through experimental results that TOP provides a more robust, data-driven approach to initialization, as it learns the underlying topological structure automatically. Using the proposed two-step method, we can extend most current active contour methods, which depends on an optimization method, to multiple-object images; as TOP provides a natural partition of an image into several regions, where only one object lies in each region. We can also segment a single topologically complex object by performing separate active contour algorithms on components identified by the topological method. 

In addition, by treating each topological feature as a separate curve, one can then incorporate prior information in BAC and control smoothness separately for each curve, which cannot be achieved by using TOP alone. If an interpretable boundary estimate which respects the underlying topological structure is of utmost importance, TOP+BAC will enhance individual results from
TOP or BAC alone.

In general, the idea of using TOP to initialize the Bayesian active contour model fits into the broader practice of forming a pre-conditioner for an optimization algorithm. Therefore, we typically expect faster convergence and better performance, particularly compared to initializations which are not guided. The primary difficulty in achieving such a topologically-guided procedure is to identify how these two sharply different philosophies can be combined. Unlike the ``high-level'' topologically-guided methods \citep{clough2020topological,carriere2020perslay,hu2019topology}, which borrow strength from more advanced models, our combined method uses  topological information at a ``low-level'', which preserves its interpretability.

\section{Data Analysis}
\label{sec:Data-Analysis}

In this section, we 
first argue for the use of a topologically-guided initialization by demonstrating the proposed method (TOP+BAC) on image datasets where both noise and ground truth are known and controlled. Then, we apply the method to two real-world applications where images contain features with complex topological structures. One is a skin lesion dataset \citep{codella2018}, which contains images of skin rashes comprised of multiple connected components. The other is a neuron imaging dataset, which contains typical neural section images from biological experiemnts \citep{jurrus2013semi}. With the latter, we demonstrate how one can use TOP+BAC to approach boundary estimation in the presence of numerous noisy connected components with very little separation between. 

\subsection{Simulated Images}
\label{subsec:SimImages}
We first consider an artificial dataset consisting of ten user-created binary donut images. These images were generated from a known large, outer contour and a small, inner contour, thus allowing for the ground truth curves to be known. We then apply Gaussian blurring using MATLAB function \texttt{imgaussfilt} (with standard deviation of 15) to make the image non-binary. We know the noise nature and ground truth of this image dataset.

\begin{figure}
\caption{Interior (blue) and exterior (red) estimated pixel densities constructed using Gaussian kernel density estimators for the donut dataset (test image shown in Figure \ref{fig:BlurDonutTOP}) when $j=1,2$ (left and right, respectively). A bandwidth of 0.05 was used for the kernel density estimates in both plots.}

\begin{center}
\begin{tabular}{cc}
\toprule
$j=1$ & $j=2$ \\
\hline
\hline
\includegraphics[width=2in]{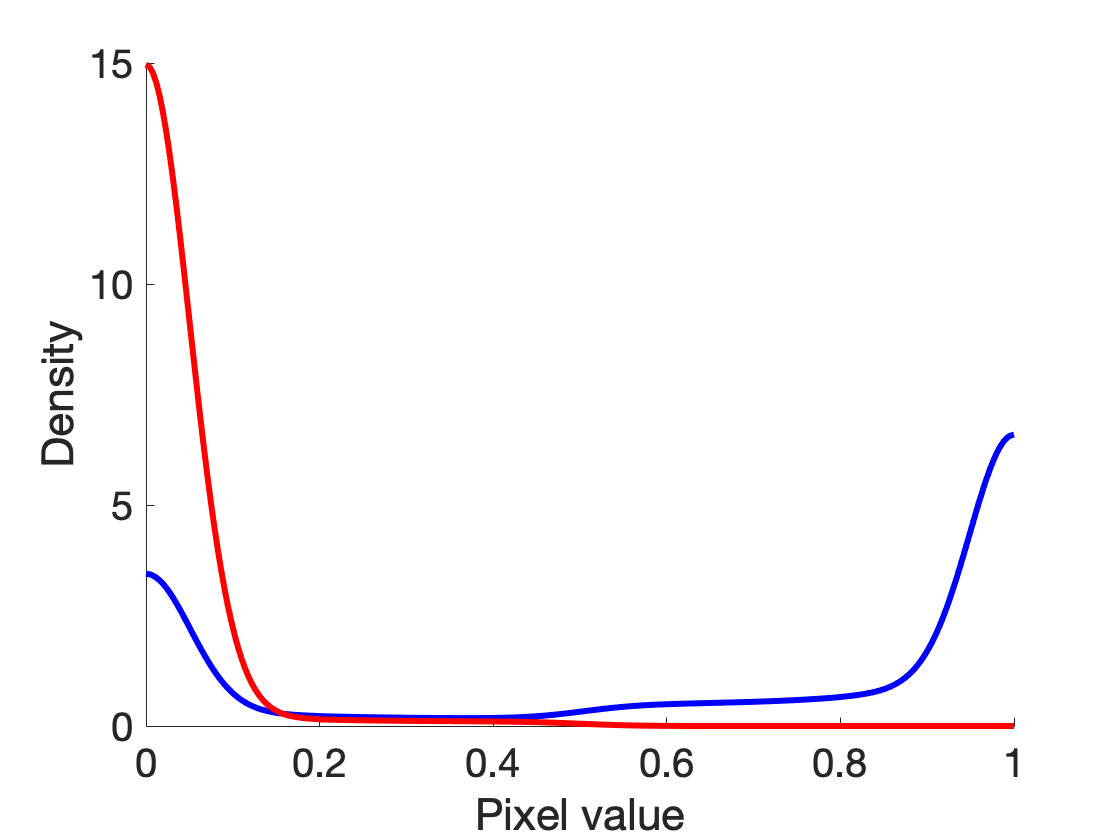} & \includegraphics[width=2in]{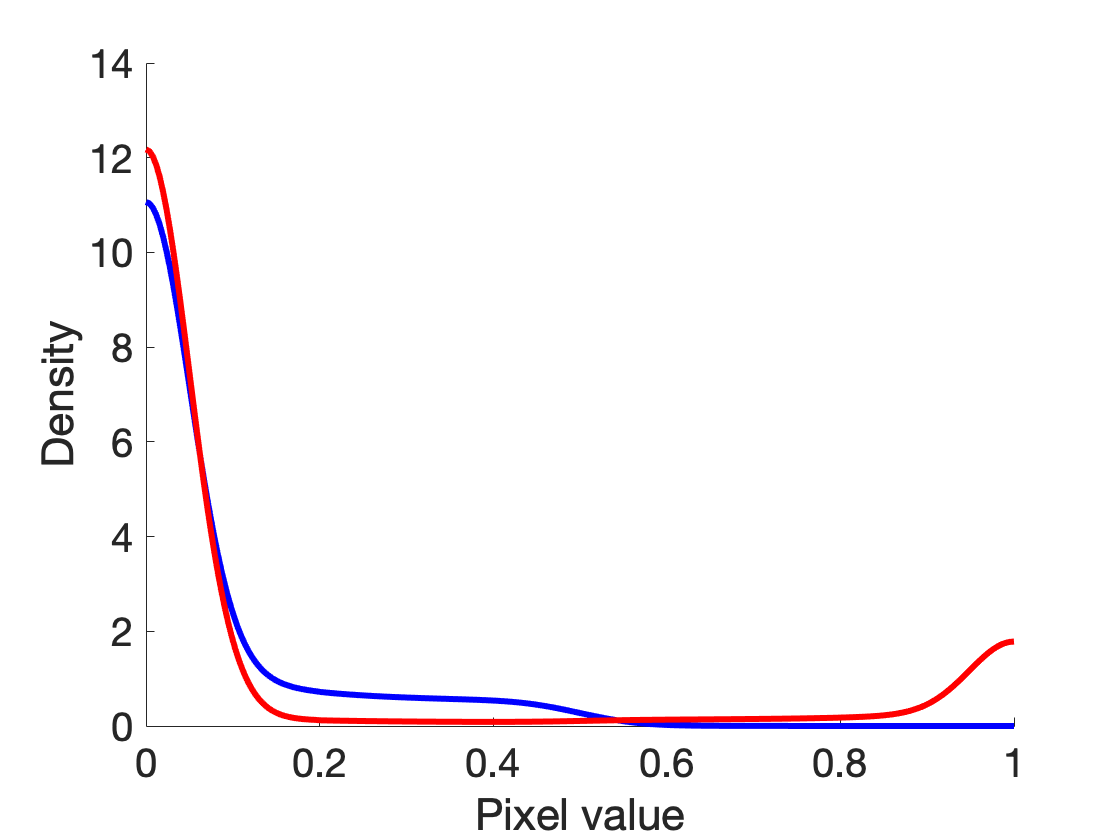} \\
\bottomrule
\end{tabular}
\end{center}
\label{fig:DonutDensity}
\end{figure}

\begin{figure}
\caption{Top: Boundaries estimated from TOP for the blurred donut image under various choices of parameters $(\sigma_1,\sigma_2,T)$ (labeled as a triple in the top row). Bottom: Final segmented contours (in blue) obtained using TOP+BAC under TOP settings corresponding to the top row. The parameters of BAC are $\lambda_{1}=0.15,\ \lambda_{2}=0.3,\ \lambda_{3}=0$, with convergence reached after 91, 133, 170, and 195 iterations, respectively, in total using a convergence tolerance of $10^{-7}$, and bandwidth 0.05 for pixel kernel density estimates.}

\begin{centering}
\begin{tabular}{cccc}
\toprule 

TOP1 $(1,5,3)$ & TOP2 $(1,5,5)$ & TOP3 $(1,5,7)$ & TOP4 $(3,5,5)$ \\
\hline
\hline
\includegraphics[width=1.3in]{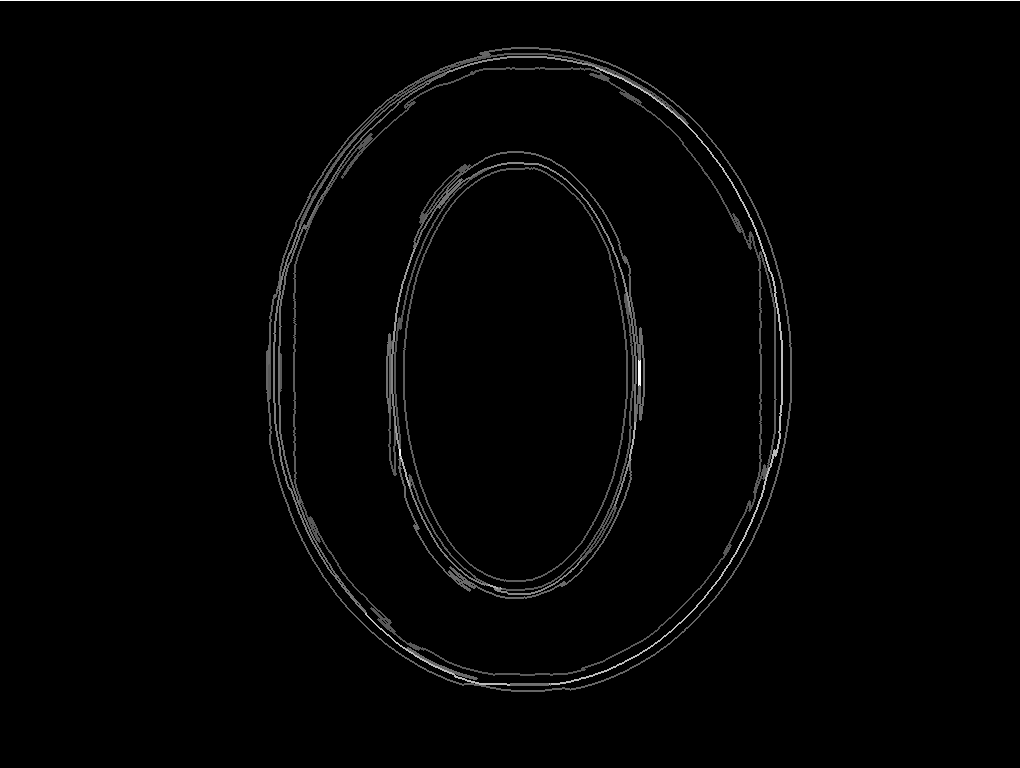} & \includegraphics[width=1.3in]{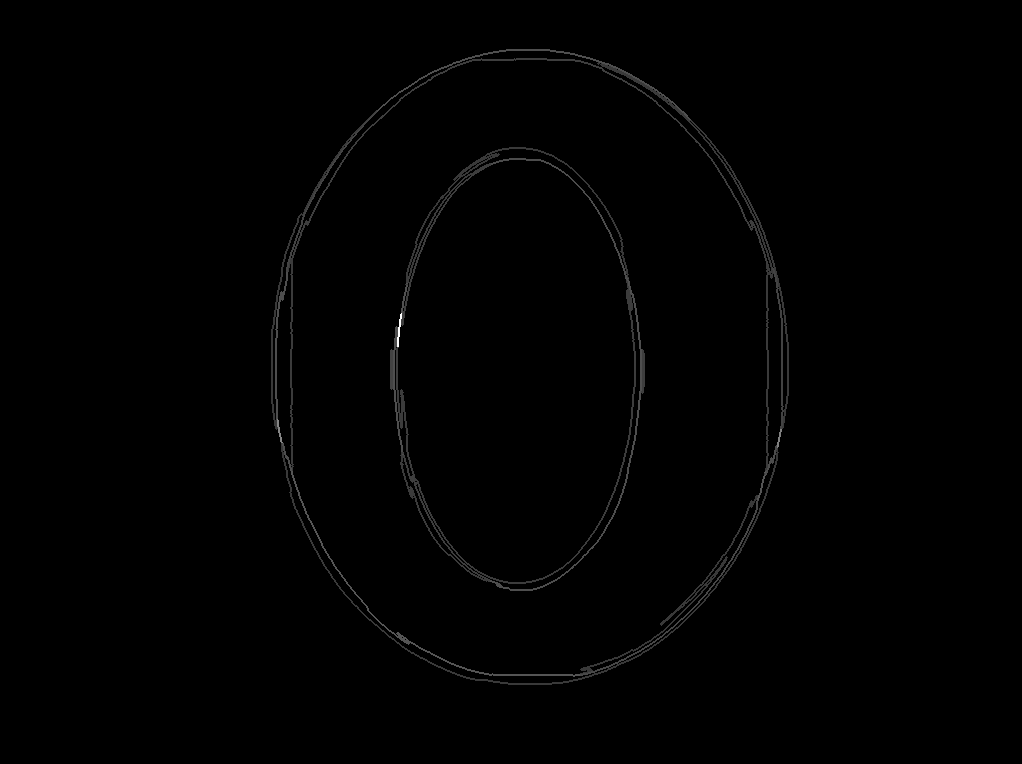} & \includegraphics[width=1.3in]{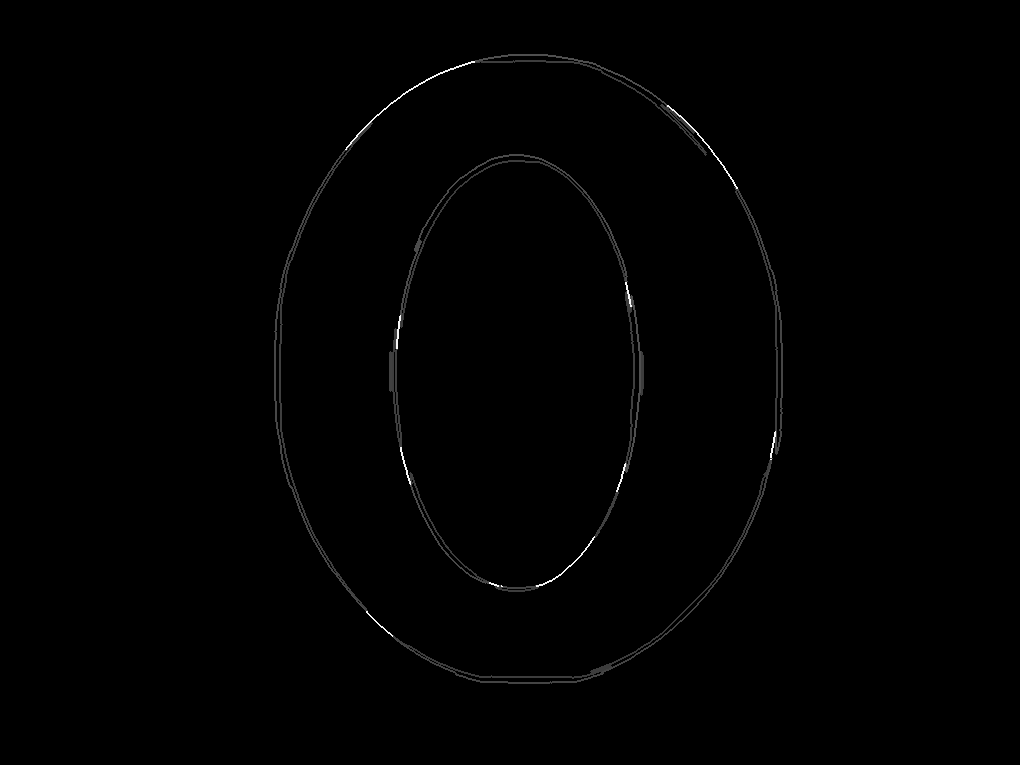} & \includegraphics[width=1.3in]{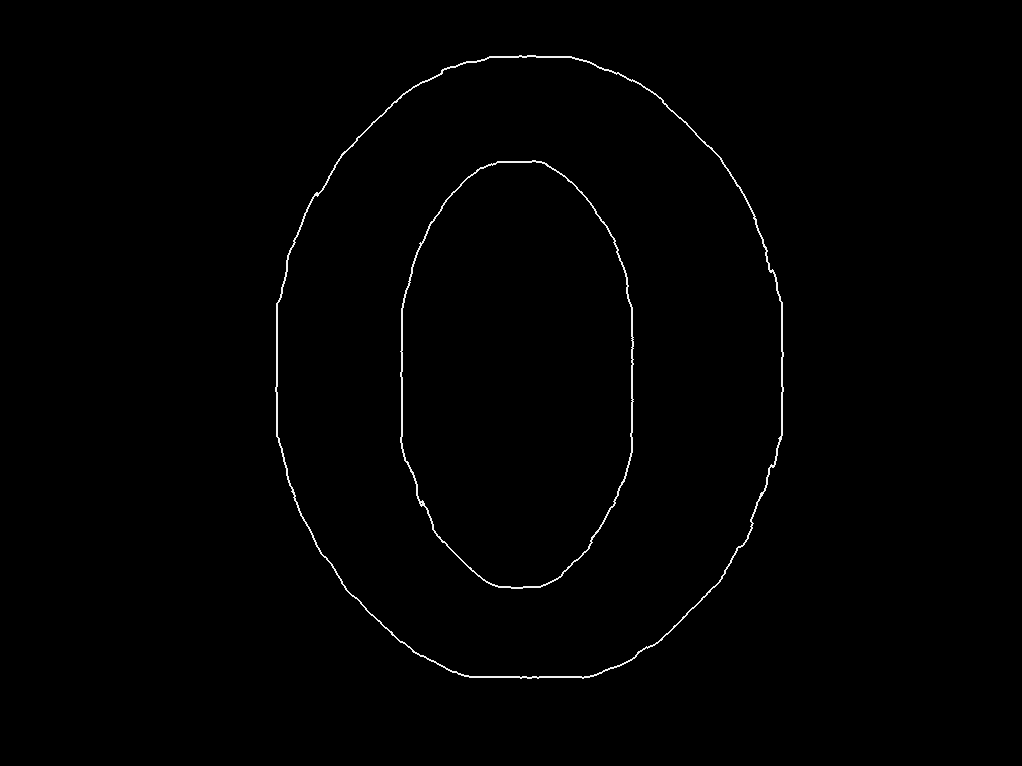}\\
\hline
\includegraphics[width=1.3in]{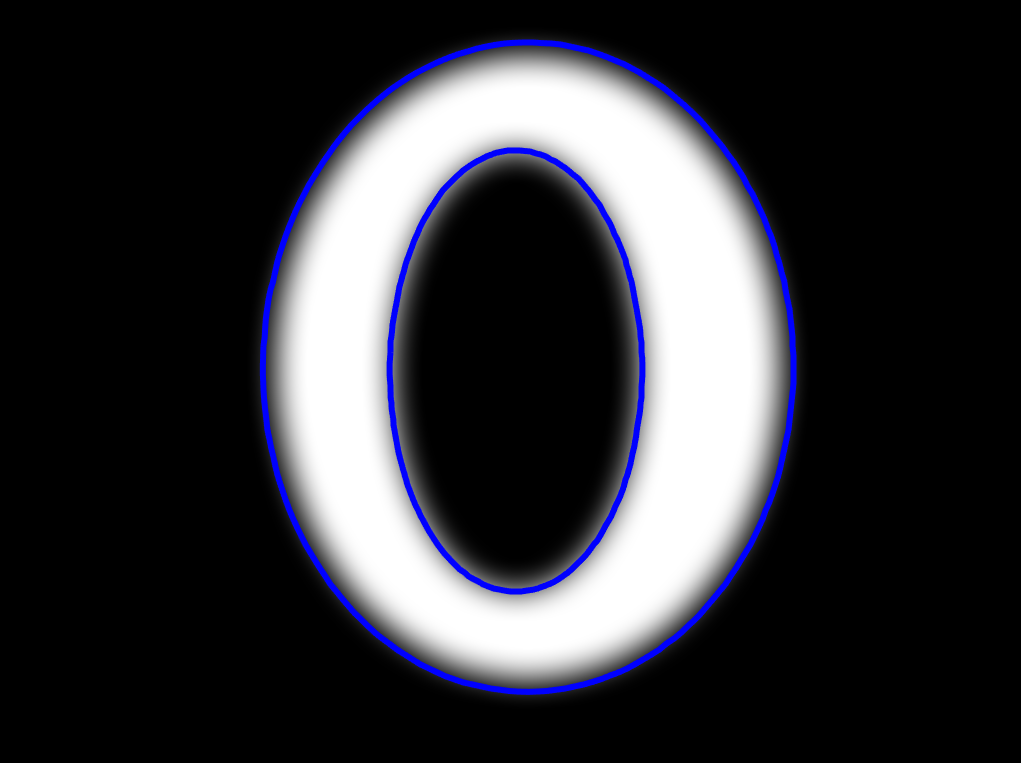} & \includegraphics[width=1.3in]{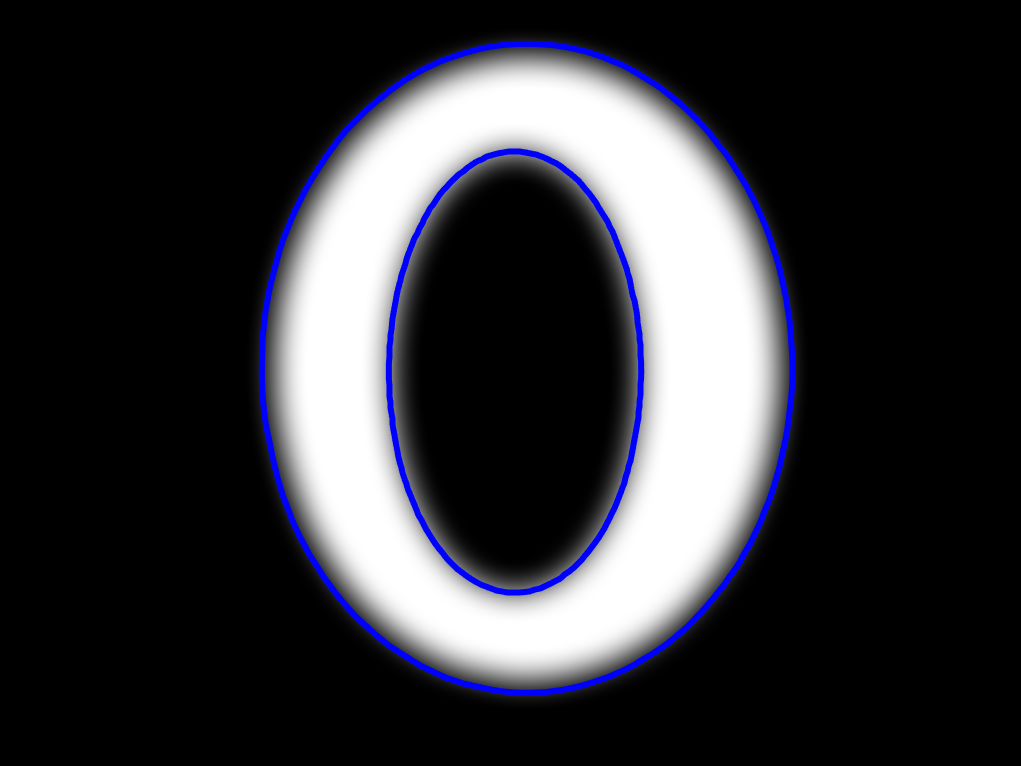} & \includegraphics[width=1.3in]{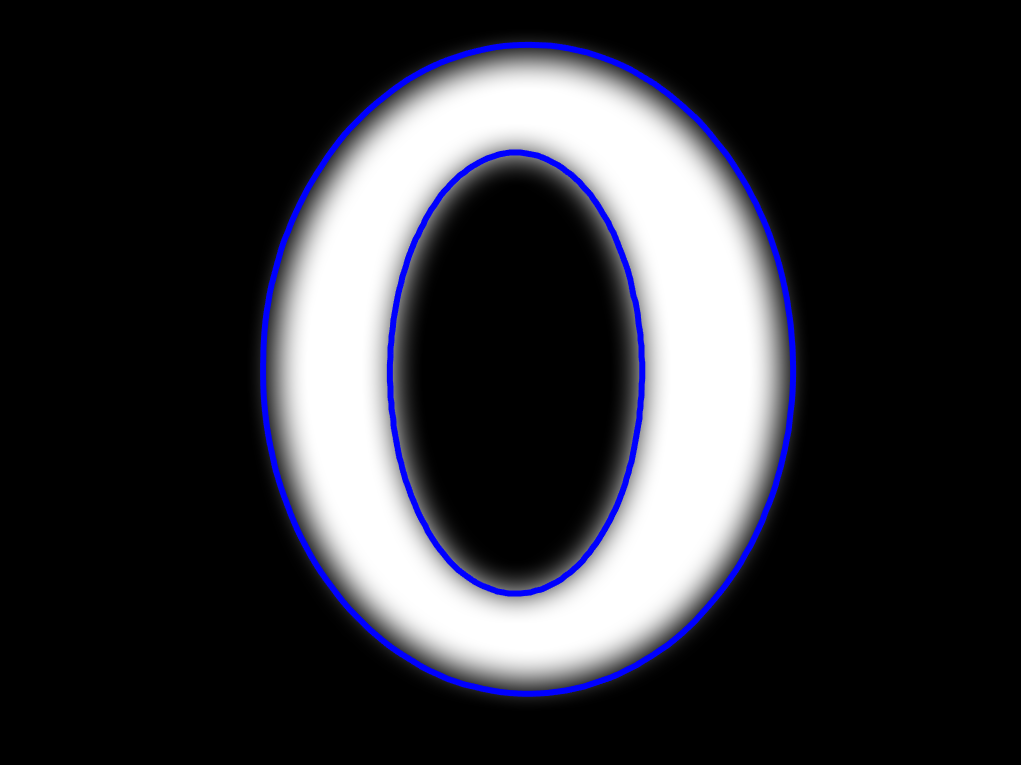} & \includegraphics[width=1.3in]{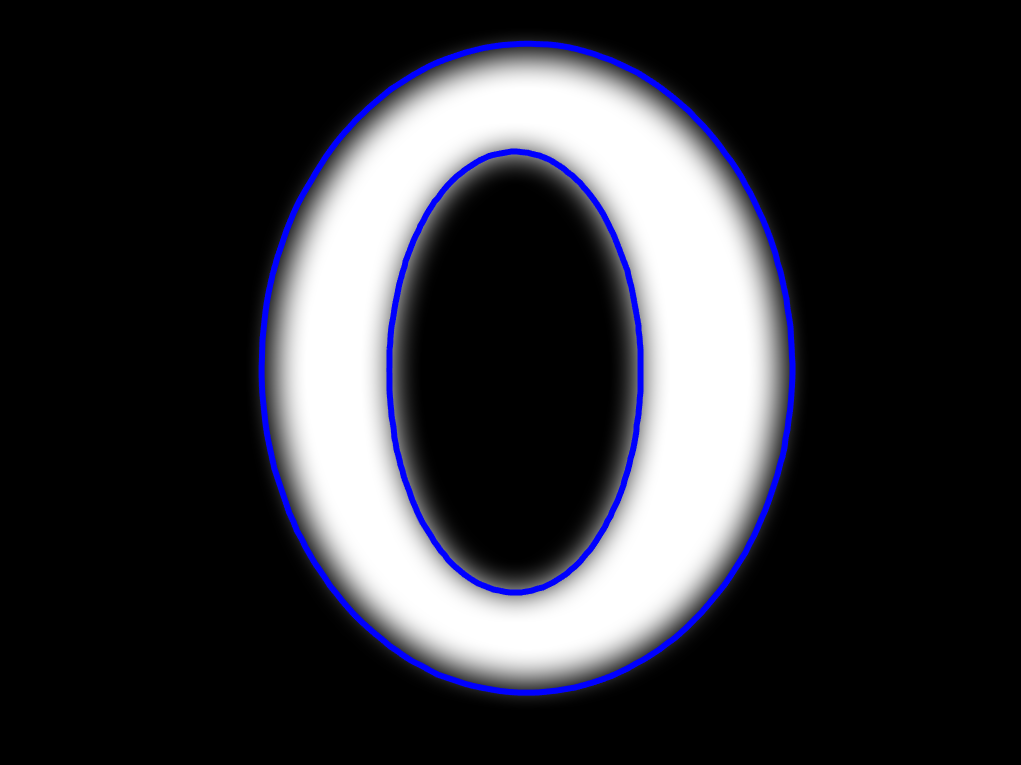}\\

\bottomrule
\end{tabular}
\par\end{centering}
\label{fig:BlurDonutTOP}
\end{figure}

The dataset is split into one test image and nine training images. 
We estimate interior ($j=1$) and exterior ($j=2$) pixel densities $p_{\text{int}}^{(j)}$ and $p_{\text{ext}}^{(j)}$ separately in the energy functional of BAC. The interior and exterior density estimates for $j=1,2$ are found in Figure \ref{fig:DonutDensity}. Note that we use a fixed bandwidth of 0.05 for the kernel density estimators for both features, as this performs generally well across   simulations we considered. In general, this parameter needs to be tuned depending on the images. This is a choice which is up to the user, and we describe ramifications of this choice in Section \ref{subsec:Lesion} and Appendix \ref{appendixEx}.
To start, we perform TOP+BAC in Figure \ref{fig:BlurDonutTOP} using four different TOP parameter settings, coded as TOP1 ($\sigma_1=1$, $\sigma_2=5$ and $T=3$), TOP2 ($\sigma_1=1$, $\sigma_2=5$ and $T=5$), TOP3 ($\sigma_1=1$, $\sigma_2=5$ and $T=7$), TOP4 ($\sigma_1=3$, $\sigma_2=5$ and $T=5$) below. The top row shows the resulting boundary maps obtained by performing TOP to identify candidates for initial contours to initialize  BAC. For setting TOP1, there are numerous boundaries obtained due to the noise introduced by blurring at the donut's boundaries; however, the boundaries generally tend to capture the correct contours (from TOP1 to TOP3). 
The setting TOP4, where $\sigma_1$ is changed to 3 and $T$ is changed to 5, yields the most clear boundary map, with only two contours (corresponding to the outer and inner rings) identified. 

In the bottom row of Figure \ref{fig:BlurDonutTOP}, we initialize two contours (largest and smallest in terms of enclosed area) from the TOP segmentation in the top row to initialize the BAC portion of the TOP+BAC algorithm. Despite the extraneous contours in these TOP initialization result, all four settings qualitatively target the inner and outer boundaries correctly. 
To compare the performance of different methods quantitatively, we evaluate the final estimated contours via Hausdorff distance, Hamming distance, Jaccard distance, and performance measure via discovery rate, as well as the elastic shape distance (see details in Appendix \ref{appendixPM}, computed for individual contours). The first four columns of Table \ref{tab:DonutPM} confirm similarity in performance of TOP+BAC to obtain the final contour estimates in Figure \ref{fig:BlurDonutTOP}.

\begin{figure}
\caption{Final segmented contours (in blue) obtained using TOP+BAC as the prior term $\lambda_3$ in BAC is varied. In all three images, TOP with parameters $\sigma_1=3$, $\sigma_2=5$, and $T=5$ are used to extract two initial contours; then, for BAC, $\lambda_1=0.05$ and $\lambda_{2}=0.15$ is fixed. Convergence is reached after 195, 38, and 7 iterations, respectively, in total using a convergence tolerance of $10^{-7}$, and bandwidth 0.05 for pixel kernel density estimates.}

\begin{centering}
\begin{tabular}{ccc}
\toprule 
P1: $\lambda_3=0$ & P2: $\lambda_3=0.05$ & P3: $\lambda_3=0.15$ \\
\hline
\hline
\includegraphics[width=1.8in]{Figure/BlurDonutTOPBACPrior0.png} & \includegraphics[width=1.8in]{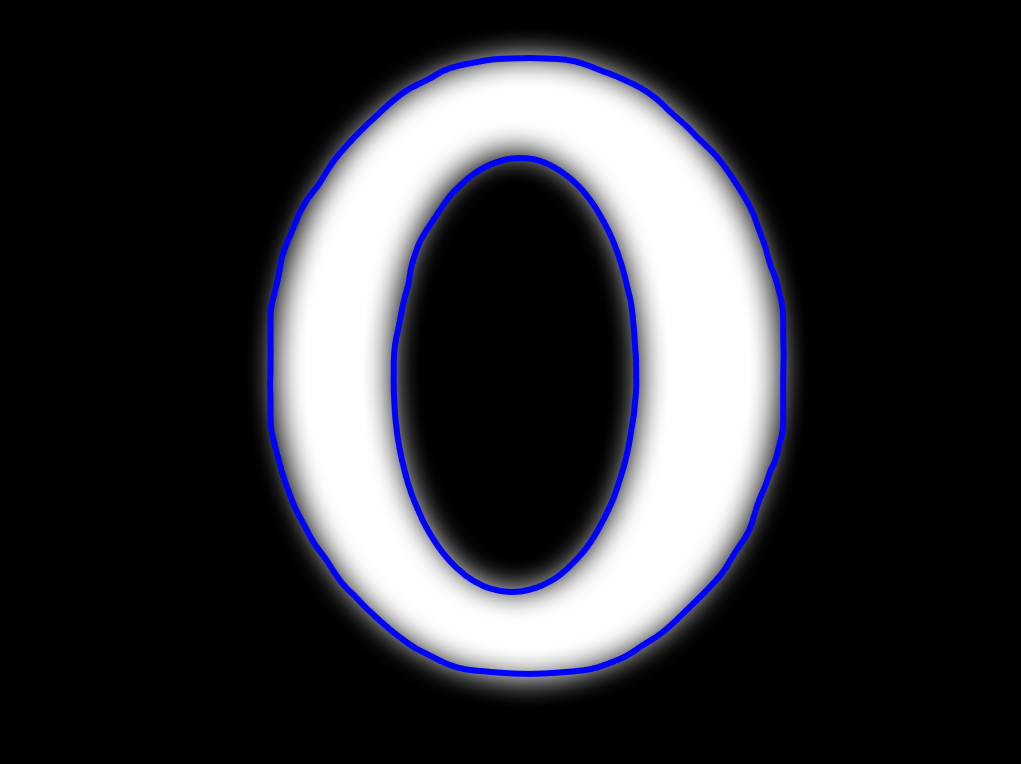} & \includegraphics[width=1.8in]{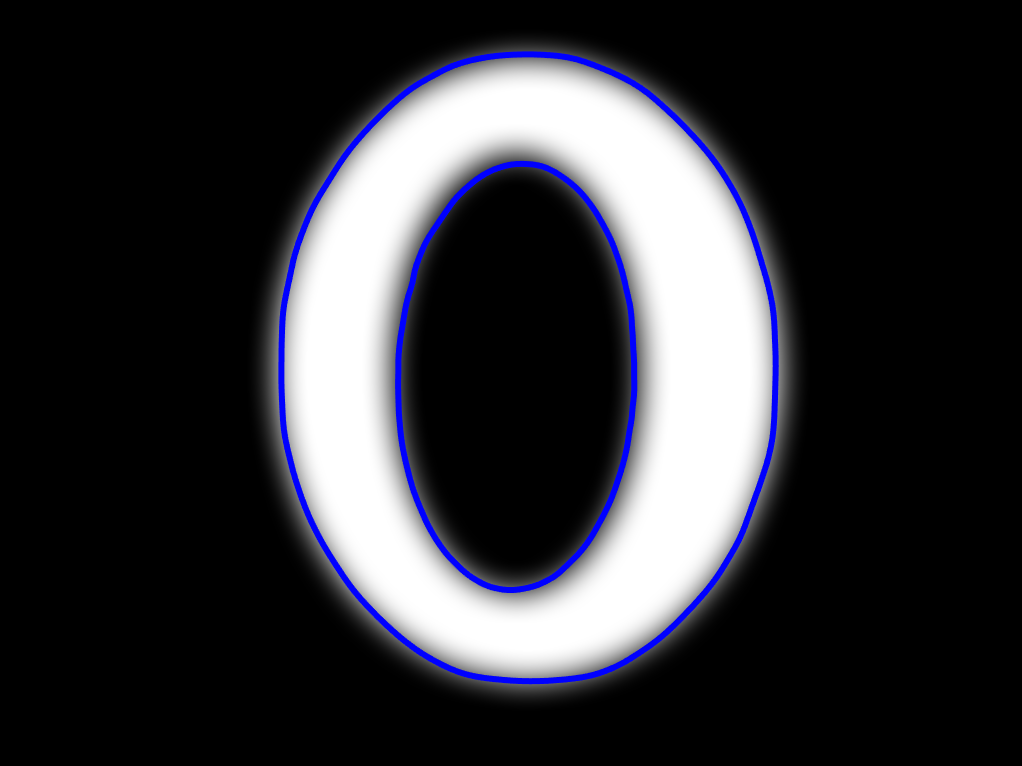}\\
\bottomrule
\end{tabular}
\par\end{centering}
\label{fig:BlurDonutPrior}
\end{figure}

Regarding the choice of parameters in BAC, setting $\lambda_1$ between 0.10 and 0.50 tends to produce reasonable results for most images considered in this paper, although as $\lambda_1$ increases, one typically has to counteract with a large smoothing parameter $\lambda_2$ to ensure updates are not too large. By setting $\lambda_3=0$ in BAC models, we do not incorporate prior information, as the targeted object is relatively clear, and not heavily obstructed by extraneous objects or noise -- we recommend setting $\lambda_3=0$ when the testing image is of high quality. 
When the test image is of low quality, selecting $\lambda_3>0$ uses information from the known shape of contours. For the other donut training images, which exhibit various contour perturbations along both the inner and outer boundaries, this forces BAC to balance pixel and smoothness updates while simultaneously pushing the contour towards the estimated training mean shape. Figure \ref{fig:BlurDonutPrior} shows the impact as $\lambda_3$ is varied, for fixed $\lambda_1=0.15$ and $\lambda_2=0.3$ and TOP parameters $\sigma_1=3$, $\sigma_2=3$, and $T=5$. Note that as $\lambda_3$ is increased, the estimated contours of both the outer and inner rings are pushed further into the white region of the donut, as the mean shape for these boundaries in the low-quality training images exhibit more eccentricity than this particular testing image suggests. Compared to the true boundaries, the estimates under $\lambda_3>0$ will often perform slightly worse for high-quality testing images.


\begin{table}
\caption{Performance measures for TOP+BAC methods applied to the blurred donut image presented in Figures \ref{fig:BlurDonutTOP} and \ref{fig:BlurDonutPrior} for various choices of TOP parameters as well as choice of prior update term $\lambda_3$ in BAC. 
Labels of the TOP and prior update parameter choices correspond to those in Figures \ref{fig:BlurDonutTOP} and \ref{fig:BlurDonutPrior}.}

\begin{centering}
\begin{tabular}{c|cccc|ccc}
\toprule 
Measure & TOP1 & TOP2 & TOP3 & TOP4 & P1 & P2 & P3 \\
\hline
\hline
$d_{H}$ (Hausdorff) & 5.8310 & 5.9161 & 5.9161 & 5.8310 & 5.8310 & 5.2915 & 4.3589 \\
$p_{H}$ (Hamming)& 0.0326 & 0.0324 & 0.0322 & 0.0324 & 0.0324 & 0.0141 & 0.0140 \\
$d_{J}$ (Jaccard)& 0.1154 & 0.1145 & 0.1140 & 0.1147 & 0.1147 & 0.0500 & 0.0546 \\
PM (discovery rate)& 0.1392 & 0.1384 & 0.1378 & 0.1385 & 0.1385 & 0.0651 & 0.0651 \\
Outer ESD & 0.0291 & 0.0292 & 0.0280 & 0.0287 & 0.0310 & 0.0647 & 0.0436 \\
Inner ESD & 0.0649 & 0.0634 & 0.0632 & 0.0622 & 0.0602 & 0.0653 & 0.0645 \\
\bottomrule
\end{tabular}
\par\end{centering}
\label{tab:DonutPM}
\end{table}

\begin{figure}
\caption{Example of a donut image where $k$-means clustering does not yield an adequate initialization for BAC. On the left is the original image (with the six true contours in red), and the right three panels show grayscale images of pixel cluster assignments when $k=2,3,4$ (each cluster corresponding to white, black, or a shade of gray).}

\begin{centering}
\begin{tabular}{cccc}
\toprule 
Image & $k=2$ & $k=3$ & $k=4$\\
\hline
\hline
\includegraphics[width=1.3in]{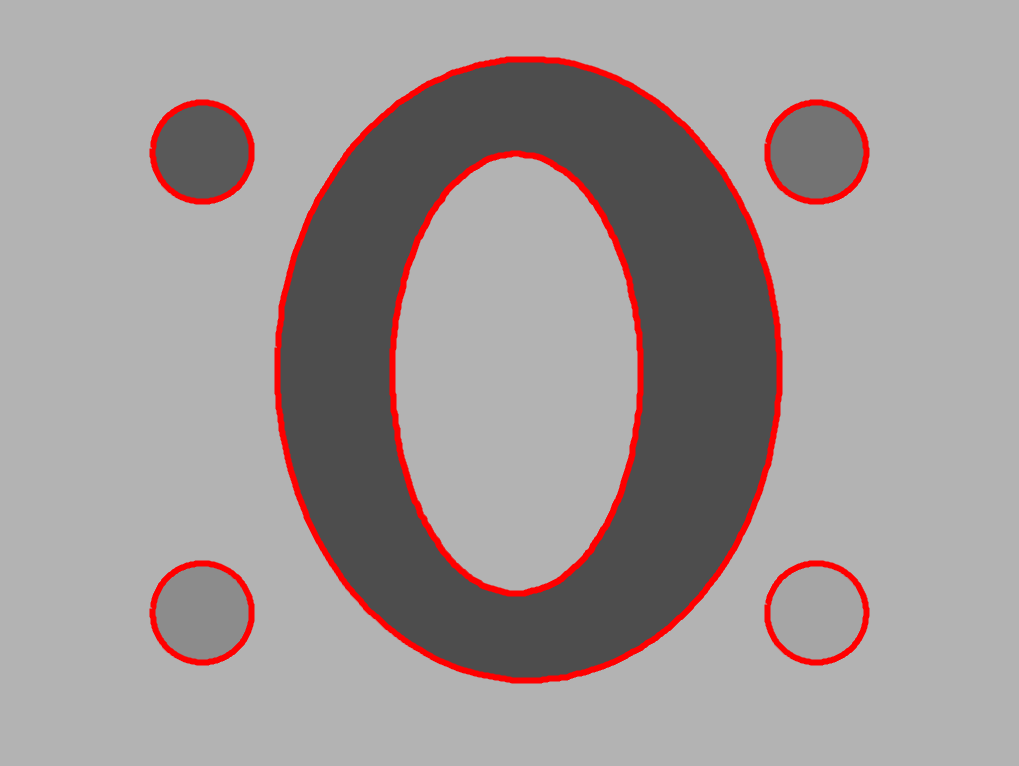} & \includegraphics[width=1.3in]{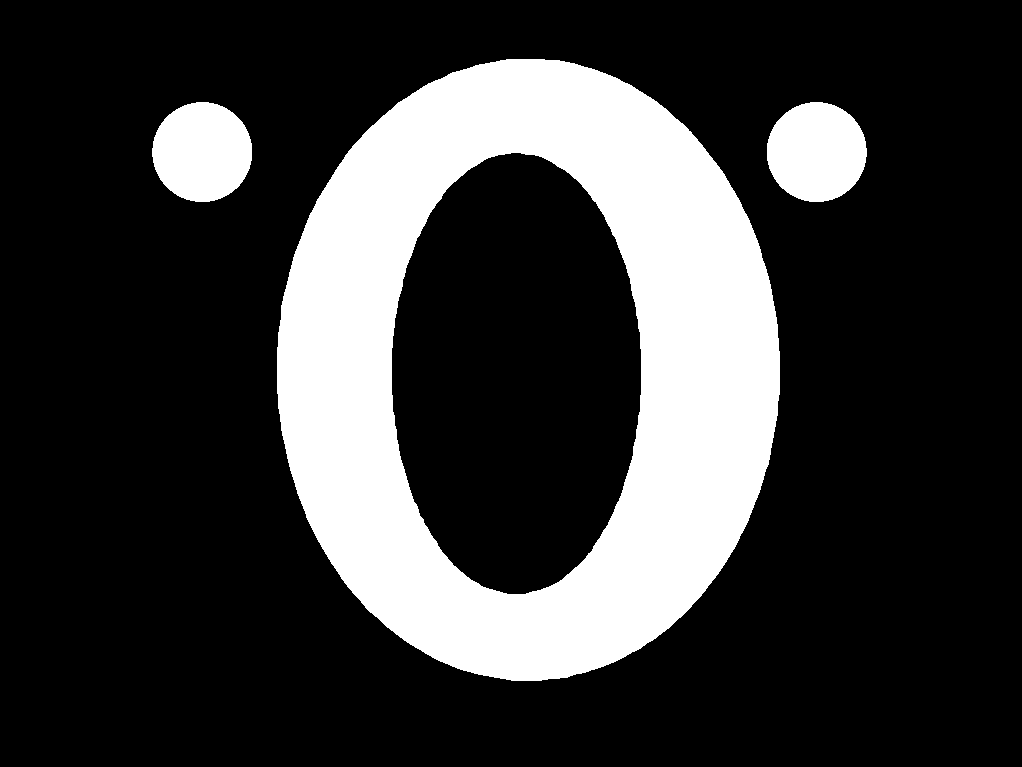} &
\includegraphics[width=1.3in]{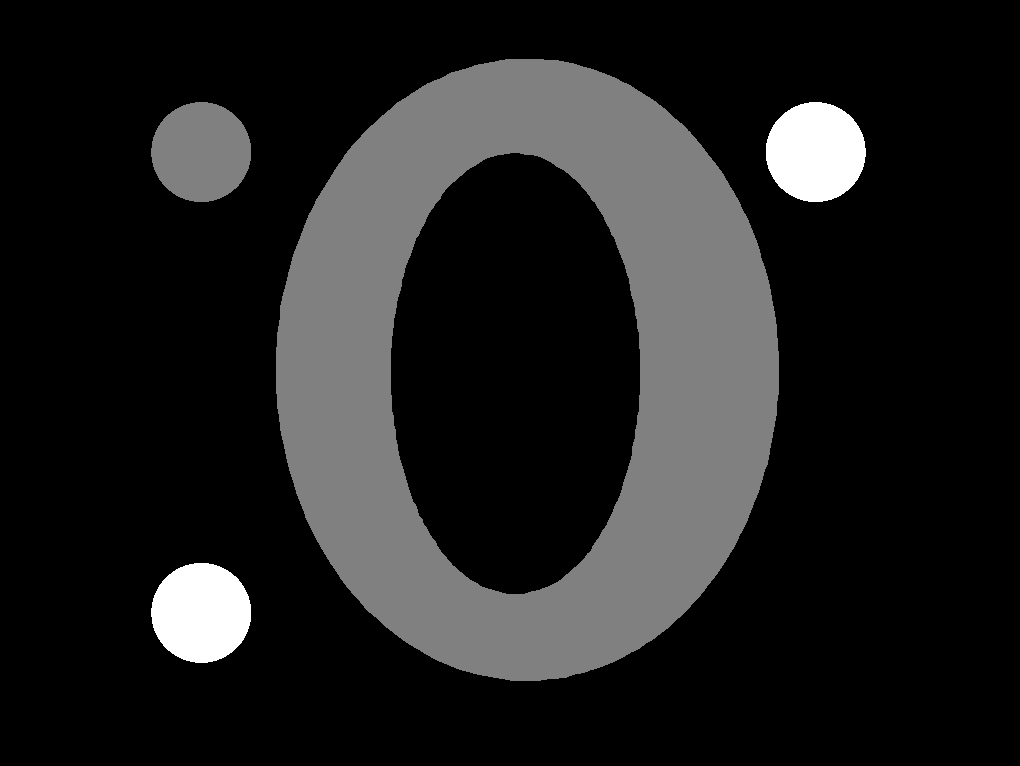} &
\includegraphics[width=1.3in]{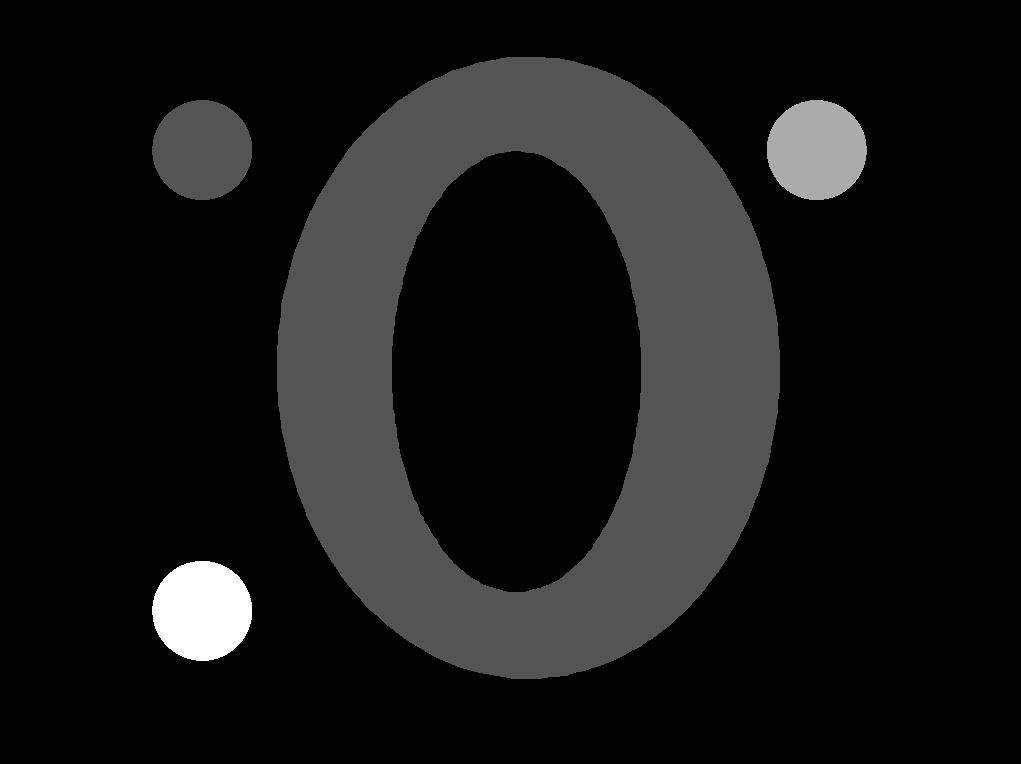}\\
\bottomrule 
\end{tabular}
\par\end{centering}
\label{fig:DonutCounterexample} 
\end{figure}

An alternative approach for obtaining contours to initialize BAC is through $k$-means, another pixel clustering-based method for boundary estimation. This method simultaneously estimates $k$ centroids (in pixel space) while clustering each image pixel to the closest pixel centroid; the resulting cluster boundaries can be viewed as contour estimates for the image. It is well known that $k$-means is sensitive to starting point, as well as choice of the number of clusters $k$. For a simple binary image, selecting $k=2$ will perfectly cluster pixels. However, for grayscale images with widely-varying pixel values representing objects of interest, choice of $k$ is highly non-trivial. Consider an alternative donut image in Figure \ref{fig:DonutCounterexample}, where other objects of different grayscale  have been introduced, each of which has a fixed pixel value lying in between the pixel values of the donut and object exterior. Note that for $k=2$, the cluster boundaries would only identify 4 contours to initialize for BAC; whereas $k=3$ and $k=4$ both would yield 5 contours. However, these values of $k$ assign the lower right dot into the same cluster as the background, making it impossible for $k$-means to identify an initial contour for this feature. For more complex images with multiple objects perturbed by noise, this choice of parameter becomes exceedingly difficult compared to selecting parameters for TOP, BAC, or TOP+BAC. 

We provide further details on smaller issues related to TOP and BAC in the Appendix. 
TOP behaves poorly as an initialization for BAC when pixel noise is extremely high -- an example is given in Appendix \ref{appendixTOP}. Convergence of TOP+BAC is discussed in Appendix \ref{appendixConv}. Finally, applications of TOP+BAC to toplogically trivial examples derived from the MPEG-7 computer vision dataset \citep{MPEG} are presented in Appendix \ref{appendixEx}.

\subsection{Skin Lesion Images}
\label{subsec:Lesion}
For images with multiple objects and hence nontrivial topological connected components, TOP+BAC can be used to identify initial contours sequentially, corresponding to the largest features (in terms of enclosed area).  Then, one can perform separate Bayesian active contour algorithms on each boundary to refine the estimates. We illustrate this on the ISIC-1000 skin lesion dataset from \citet{codella2018}. 
The lesion images are taken from patients presented for skin cancer screening. These patients were reported to have a wide variety of dermatoscopy types, from many anatomical sites. We choose a subset of these images with ground truth boundaries provided by a human expert.
Although we know the ground truth, the noise generation process of this image dataset is unknown to us.

\begin{figure}[t]
\caption{Results of a benign nevus lesion image from the ISIC-1000 dataset. (Left) Boundary map obtained from TOP, used to initialize two active contours. (Middle) Estimated final contours based on the topological initialization (TOP+BAC), using parameters $\lambda_{1}=0.3,\ \lambda_{2}=0.3,\ \lambda_{3}=0$ for BAC; parameters $\sigma_{1}=1,\sigma_{2}=5,T=5$ for TOP. (Right) Gaussian kernel density estimators for interior (blue) and exterior (red) using a bandwidth of 0.05.}

\begin{centering}
\begin{tabular}{ccc}
\includegraphics[width=1.8in]{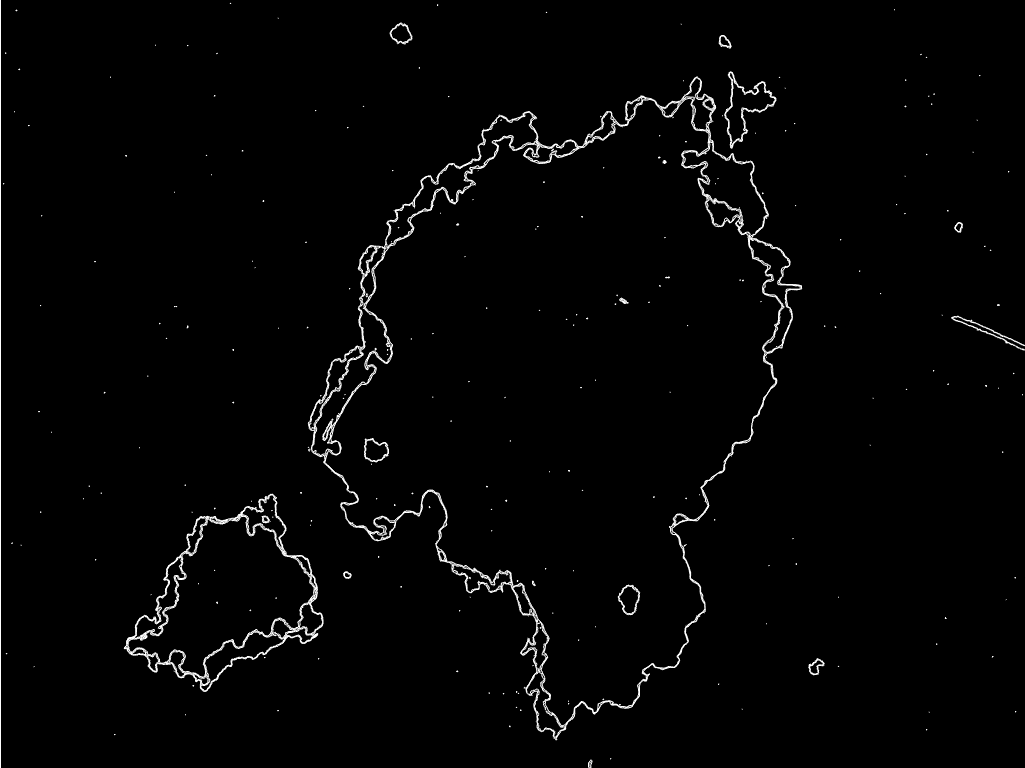} & \includegraphics[width=1.8in]{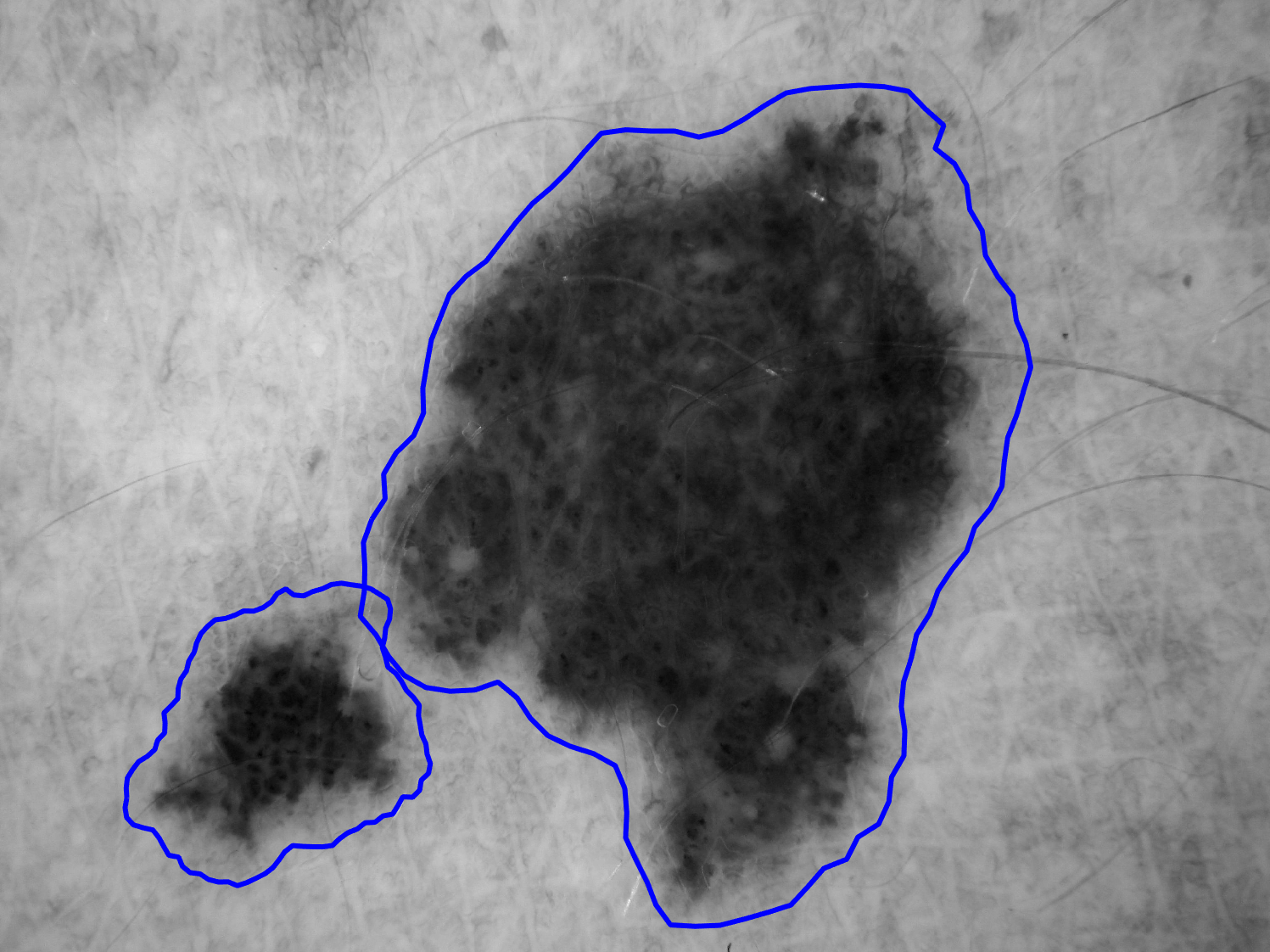} & \includegraphics[width=1.8in]{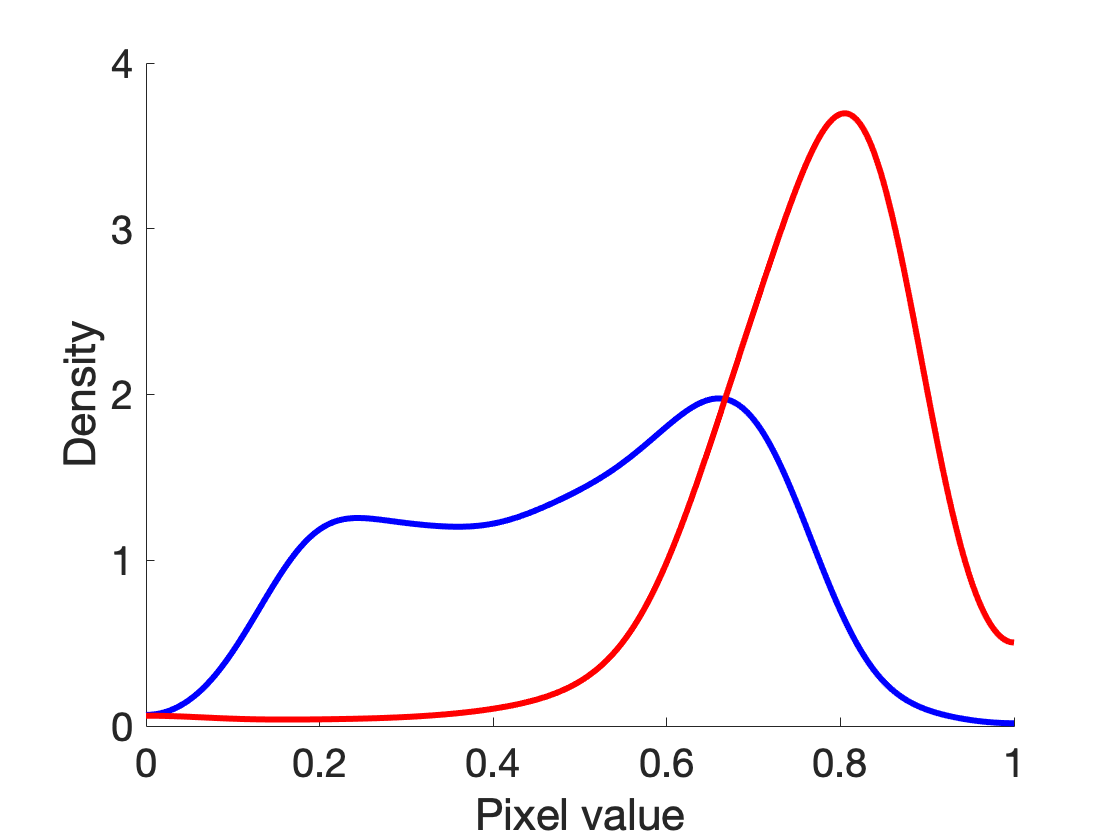}\\
\end{tabular}
\par\end{centering}
\label{fig:SkinLesion1} 
\end{figure}

It is important to note that all of the ``ground truth'' boundaries by human experts were identified as one connected component with no holes. However, there are many lesion images that consist of multiple well-defined connected components, thus making TOP+BAC beneficial for this task. Practically, it is of medical importance to identify whether a lesion consists of multiple smaller regions or one large region \citep{takahashi2002differences}. We examine two images featuring benign nevus lesion diagnoses with TOP+BAC. The lesion in the first image, shown in Figure \ref{fig:SkinLesion1}, appears to be best represented by two connected components. In order to apply TOP+BAC, we require a training set of images to estimate interior and exterior pixel densities. To do this, we select nine other benign nevus skin lesion images with an identified ground truth boundary. As noted, these expert segmentations are marked as one connected component with no holes, leading to a mismatch with our belief about the topological structure of the test image. To remedy this, we estimate just one set of pixel densities $p_{\text{int}}$ and $p_{\text{ext}}$, as opposed to separate densities for each connected component.

Figure \ref{fig:SkinLesion1} shows the result of TOP+BAC applied to a benign nevus lesion. The left panel shows boundaries obtained via the TOP algorithm. While there are some extraneous boundaries, the largest two coincide with the two separate connected components. 
Then, we filter to obtain the two outer boundaries as those which enclose the largest area. If we initialize with these contours, and run two separate active contours using weights $\lambda_{1}=0.3,\ \lambda_{2}=0.3,\ \lambda_{3}=0$, we obtain the final boundary estimates in the middle panel of the figure after 300 iterations. For this data, we justify setting $\lambda_{3}=0$ for the following reasons. First, the ground truth is variable in quality, as some of the hand-drawn contours are much coarser or noisier than others. In addition, the prior term requires shape registration, which can be challenging in the presence of noisy boundaries. Finally, after exploratory data analysis of the various lesion types present in this dataset, it does not appear that lesions share common shape features.

It is interesting to note that while two connected components are easily identifiable visually from the image, the two estimated boundaries from TOP+BAC in terms of physical distance could be very close to each other. Changing the bandwidth of the density estimator for interior and exterior pixel values from the one displayed in the right panel of the figure (with bandwidth 0.05) can result in contours which more tightly capture the darkest features within the lesions (see Appendix \ref{appendixEx}). Unfortunately, we are not able to quantitatively compare the performance of TOP+BAC to the ground truth, as the ground truth incorrectly represents the topology of the skin lesions in this image. It does not make sense to evaluate the resulting boundary estimates from TOP+BAC, which may consist of multiple contours, to the single contour provided by the ground truth.

\begin{figure}[t]
\caption{Results of TOP+BAC for another benign nevus lesion image from the ISIC-1000 dataset. (Left) Boundary map obtained from TOP, used to initialize two active contours. (Middle) Estimated final contours based on the topological initialization (TOP+BAC), using parameters $\lambda_{1}=0.3,\ \lambda_{2}=0.3,\ \lambda_{3}=0$ for BAC; parameters $\sigma_{1}=1,\sigma_{2}=5,T=5$ for TOP. (Right) Gaussian kernel density estimators for interior (blue) and exterior (red) using a bandwidth of 0.05.}

\begin{centering}
\begin{tabular}{ccc}
\includegraphics[width=1.8in]{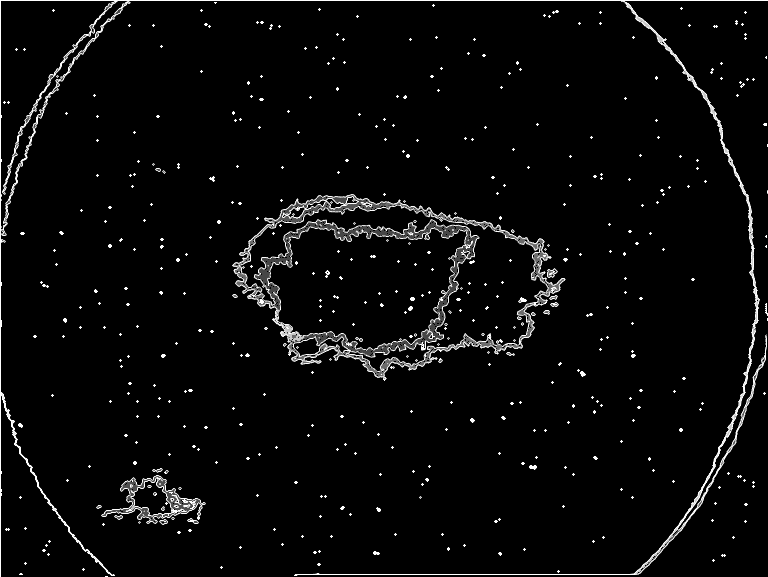} & \includegraphics[width=1.8in]{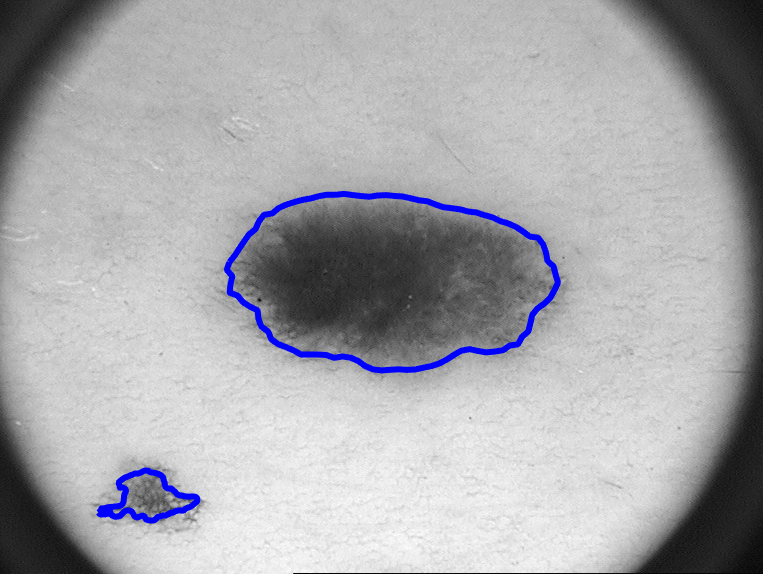} &
\includegraphics[width=1.8in]{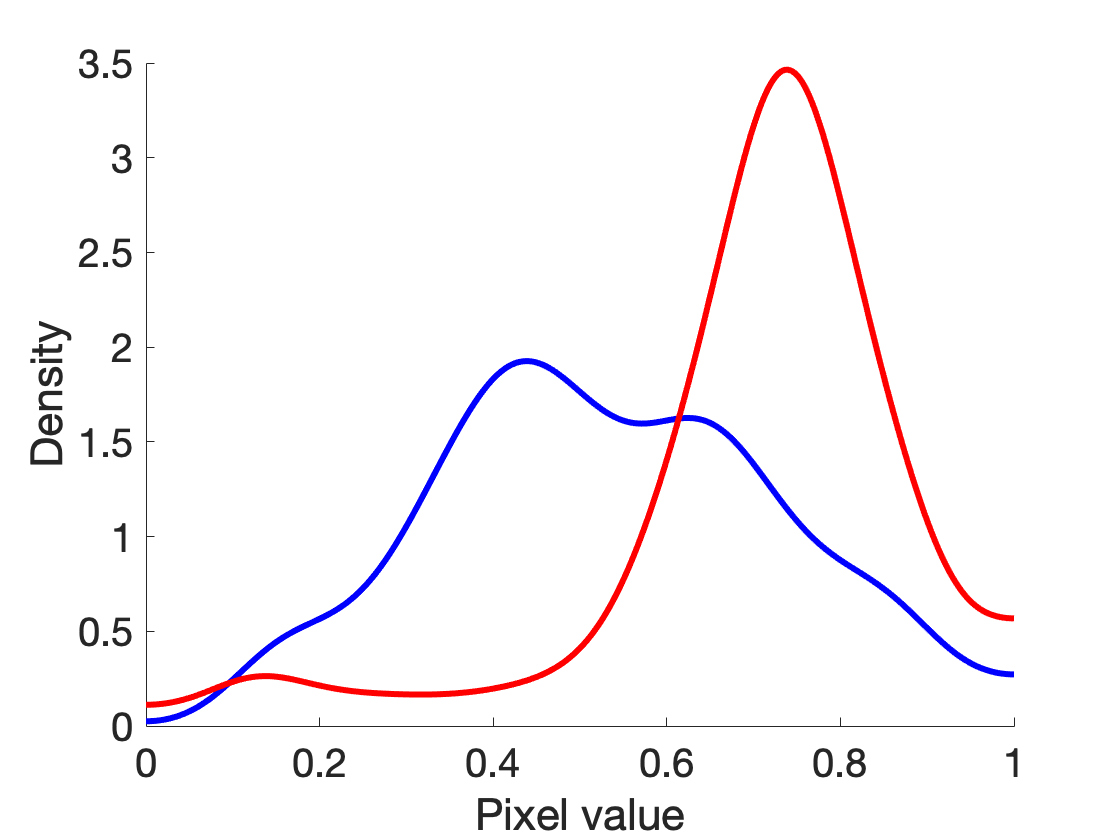}\\
\end{tabular}
\par\end{centering}
\label{fig:SkinLesion2} 
\end{figure}

Figure \ref{fig:SkinLesion2} shows another benign nevus image, again using a set of nine training images to estimate pixel densities. Under the same settings as the previous image, TOP is used to initialize BAC with the two active contours containing the largest area within. In this image, the two regions of skin irritation are very well-separated, and thus, we do not obtain overlapping contour estimates regardless bandwidths.
This is clearly a case in which a single active contour will not represent this lesion appropriately, whereas TOP may identify too many spurious boundaries which do not correspond to the actual target of interest. 
Additional skin lesion examples are presented in Appendix \ref{appendixEx}.

\subsection{Neuron Images}
We can use the proposed TOP+BAC method as a tool to approach difficult real-world boundary estimation problems. 
Electron microscopy produces 3-dimensional image volumes which allow scientists to assess the connectivity of neural structures; thus, topological features arise naturally. Researchers seek methods for segmenting neurons at each 2-dimensional slice (i.e., section) of the 3-dimensional volume, in order to reconstruct 3-dimensional neuron structures. Unfortunately, we know neither the noise generation process nor  ground truth of this image dataset.

At the image slice level, this kind of dataset presents a serious challenge from a segmentation perspective, due to the tight packing of neurons, as well as the presence of extraneous cellular structures not corresponding to neuron membranes. Numerous methods, as discussed in \citet{jurrus2013semi}, have been proposed for extracting these boundaries, many of which require user supervision. 
We demonstrate use of TOP+BAC on this data to illustrate how one can initialize separate active contour algorithms in order to more finely estimate neural boundaries.

We perform TOP to initialize BAC for refined estimates of neuron boundaries. It is clear that BAC alone would encounter great difficulty. Additionally, BAC converges very slowly when we experiment with simple initializations. Unfortunately, boundaries estimated by TOP alone are also not useful -- it is difficult to extract closed contours that represent neuron boundaries, as the neurons are tightly packed with little separation with membranes. In addition, the image features regions with no neurons, as well as additional noise.
Finally, many of the boundaries have thickness associated with them, making it difficult to identify closed contours through the pixel feature space. To address this final issue, we first apply Gaussian blur (with kernel standard deviation $5$) to the image in order to thin boundaries. Then, we obtain boundary estimates via 
TOP. 
This results in a large quantity of candidates for initial boundaries, many of which do not pertain to neurons. Thus, we further refine the boundaries in the following way:
\begin{enumerate}
    \item Remove boundaries which have the smallest enclosed areas. 
    \item Identify boundaries with center of mass closest to the middle of the image. 
    \item Select the most elliptical shape contours by computing their elastic shape distances to a circle.
\end{enumerate}
The first two refinements screen out small bubble-like pixels and miscellaneous features at the edge of the lens, which are not likely to correspond to neural structures, but noise. From examination of the images, neuron cells tend to lie in the middle of the images due to the image processing procedure. For the final refinement, we note that neurons tend to have a generally smooth, elliptical shape. However, some of the boundaries estimated from TOP are quite rough, and may pertain to irregularly-shaped regions between neurons. Thus, we narrow down to the 35 remaining boundaries with the most circular shapes. One way to do this is by computing the elastic shape distance between each contour and a circle shape, via Equation \ref{eq:ESD} in Appendix \ref{appendixPM}, and selecting the 35 which are smallest. This distance is invariant to the contour size, and thus does not favor contours which are smaller in total length \citep{younes-elastic-distance, KurtekJASA}.

We note that many other methods, including the proposed method in \citet{jurrus2013semi}, require user supervision to accept or reject various boundary estimates. In a similar manner, one could take a similar approach with TOP+BAC, by cycling through individual neural membranes estimated from TOP manually and deciding which correctly represent the targeted cellular objects. However, we choose to approach this through automatic filtering of the data as above, as it provides a potentially non-supervised way to extract contours that is relatively easy and intuitive to tune. We also note that using TOP as a initialization for BAC is more appealing than $k$-means, as it is not clear what value of $k$ should be chosen in order to avoid noise and favor elliptical shapes.

\begin{figure}[h]
\caption{(Top left) Boundaries of the top 35 curves under the discussed filtering criterion for segmenting a slice of a neuron image, based on TOP initialization with parameters $\sigma_1=2$, $\sigma_2=4$, and $T=1$, and bandwidth 0.30 for pixel kernel density estimates. (Top right) Estimated contours after performing TOP+BAC for $\lambda_1=0.2/0.1$ (for the first 25 and last 10 curves, respectively), $\lambda_2=0.7$, $\lambda_3=0$. (Bottom) Two zoomed-in examples of neurons, with initial curve (red) and estimated curve (blue).}

\begin{center}
\begin{tabular}{cc}
\toprule
TOP & TOP+BAC \\
\hline
\hline
\includegraphics[width=2.5in]{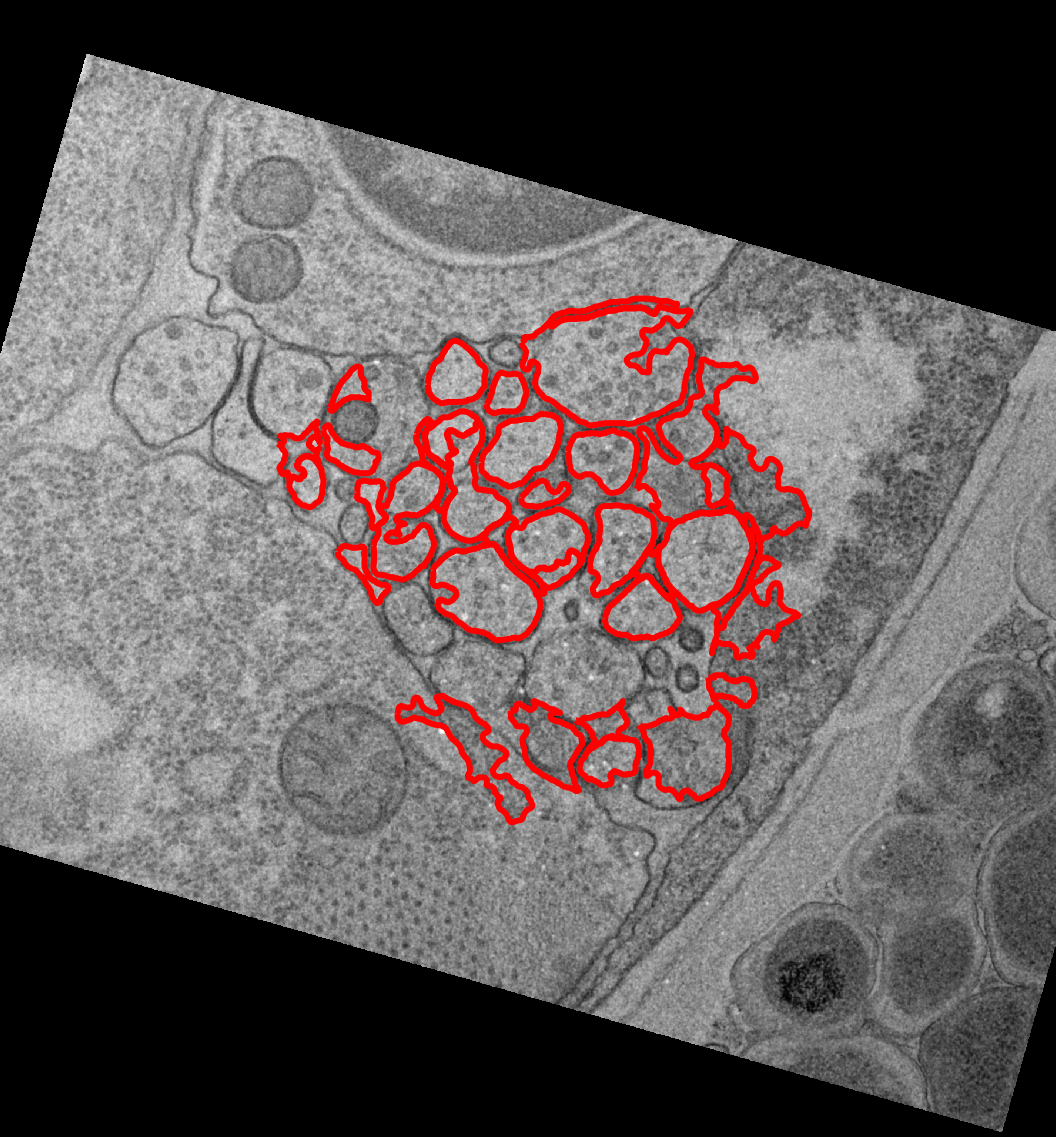} & \includegraphics[width=2.5in]{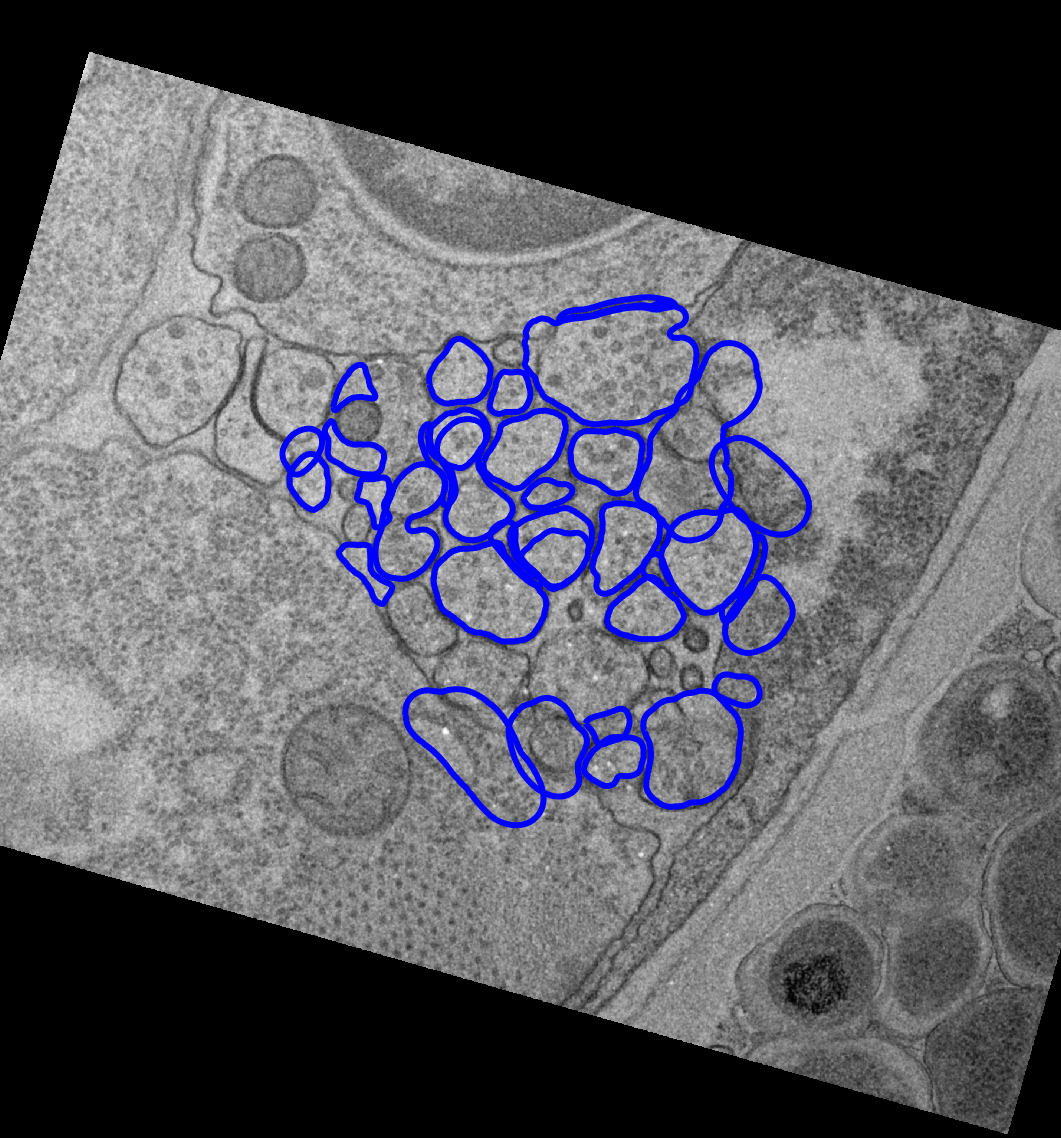} \\

\toprule
Neuron 1 & Neuron 2 \\
\hline
\hline
\includegraphics[width=2.5in]{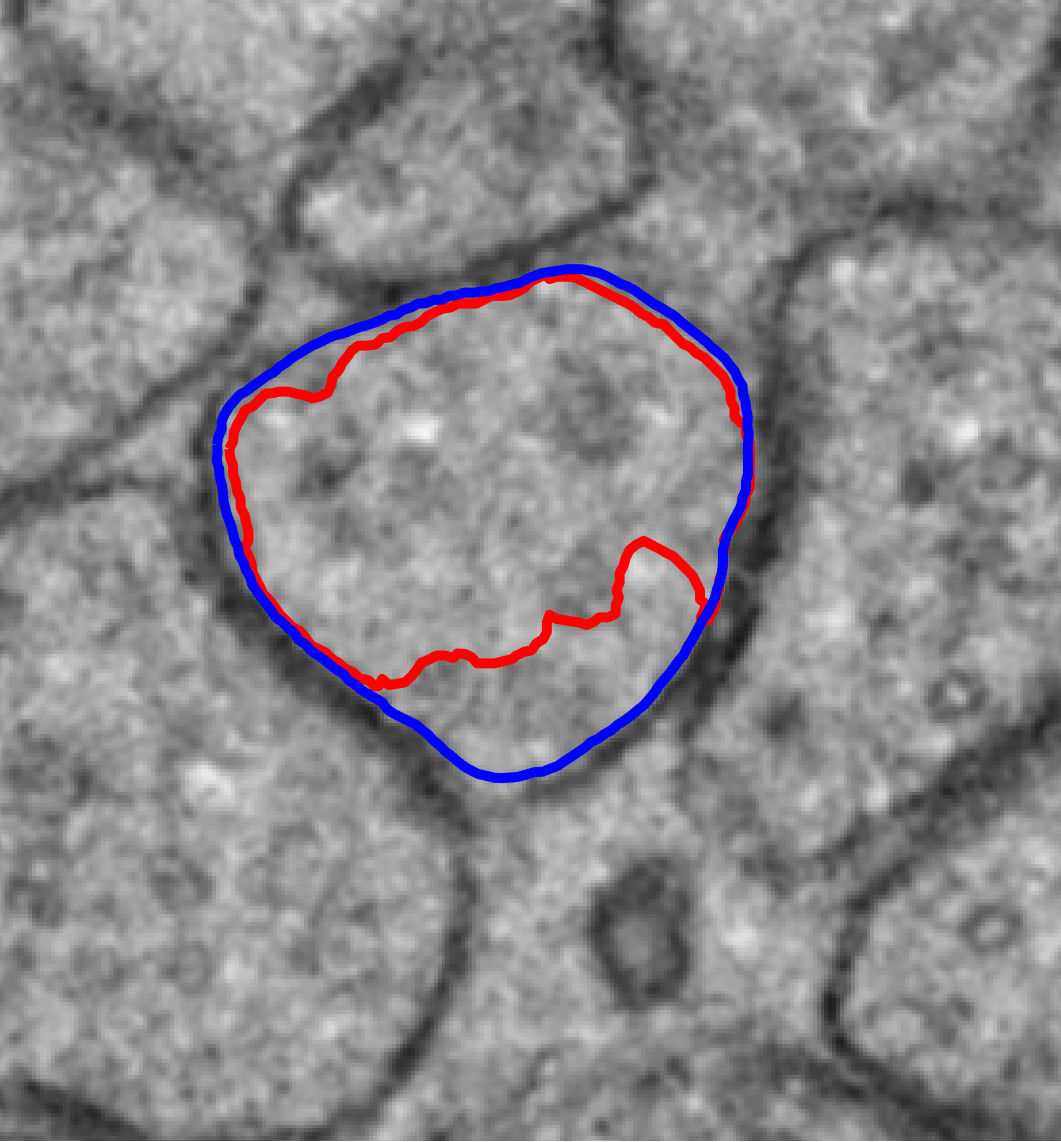} & \includegraphics[width=2.5in]{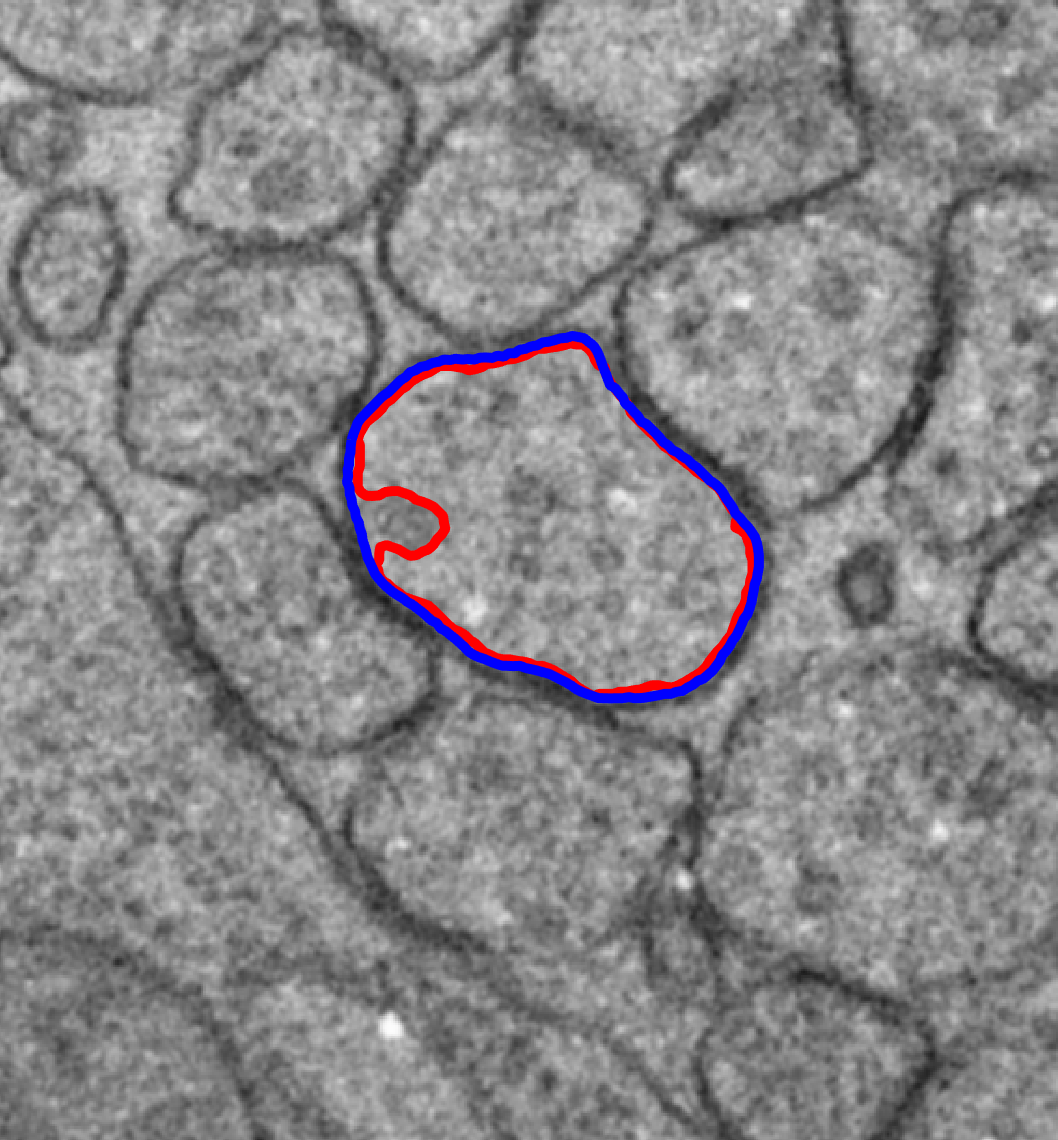}\\
\bottomrule
\end{tabular}
\end{center}
\label{fig:Neuron} 
\end{figure}

Figure \ref{fig:Neuron} summarizes our results after performing TOP+BAC. On the top left are the 35 curves automatically selected by our previously mentioned filtering criterion; while they do not perfectly cover all neurons and may represent extraneous cellular objects, they do cover a majority of them. The top right shows the estimated contours after performing separate BAC algorithms for each. Note that the interior and exterior pixel densities are estimated empirically for each feature separately based on the TOP initializations, due to the lack of availability in training data, with a kernel density estimator bandwidth of $0.30$. While difficult to see when all estimates are superimposed, we are able to refine neuron boundary estimates while imposing some smoothness. In particular, note that some of the TOP boundaries used for initialization are quite rough and may not capture the boundary correctly, as shown by two specific zoomed-in neurons in red on the bottom row of the figure. Applying a separate BAC algorithm smooths these boundaries out, and are more representative of the true neural cell walls than what was obtained by TOP. We use $\lambda_1=0.2$ to update the first 25 contours, which generally have smoother initial contours, and a smaller value ($\lambda_1=0.1$) for the final 10 contours.

\section{Discussion}

\label{sec:Conclusion}

\subsection{Contribution}

The motivation of this work is two-fold: to compare how topological and geometric information are used in the image boundary estimation problem, and attempt to combine these two complementary sources of information, when they are both available. We propose a ``topologically-aware'' initialization for Bayesian active contours through the topoloigcal mean-shift method, which is able to learn the underlying topological structure of objects within the test image in a data-driven manner.

We began by briefly reviewing two different
philosophies of boundary estimation, identifying BAC (treating segmentation as a curve estimation problem) and TOP (treating segmentation as a clustering problem via mean-shift) as representative methods for each. 
BAC \citep{joshisrivBAC,bryner2013elastic}
is a method which combines traditional active contour estimation
approaches with shape analysis, while TOP \citep{Paris&Durand2007} is a topology-based clustering segmentation method.
TOP generally yields topologically-consistent but overly conservative, non-smooth contours and includes many noisy regions. BAC is not necessarily ``topology-aware'' and is highly dependent on gradient-descent algorithm initialization, but can produce smooth contours and is not as sensitive to noise. It also is capable of incorporating any prior information about the shape of the targeted object. We proposed TOP+BAC as a way to combine the two.

The proposed TOP+BAC method uses the boundary estimates from TOP as an initialization of BAC. This two-step method can also be thought
of as using BAC to ``smooth-out'' the result of TOP, with initial curves provided for each topological feature to allow for independent
BAC algorithms to be run. We illustrate that topological information
can be used to guide the choice of a good starting point of the gradient-descent
algorithm. 

We support the use of our TOP+BAC method
by demonstrating its performance using both artificial and real data. For
artificial data, we show that TOP+BAC can identify complex
topological structures within images correctly, enhancing
the interpretability of the boundary estimates. For the skin lesion and neural cellular datasets,
TOP+BAC correctly identifies multiple objects (i.e., connected components)
in the image with interpretable parameters. This also emphasizes the general idea that a well-chosen pre-conditioner can improve performance of the gradient-descent optimization algorithm.

\subsection{Future Work}

One possible direction for future work
is the problem of uncertainty quantification of estimated boundaries.
To our knowledge, in the setting of multiple objects and a single
object with nontrivial topological features, not much work has been
done in this area. One particular difficulty is that uncertainty arises both in how the boundary is estimated, as well as the topological properties of the boundary.
For example, if we draw a confidence band surrounding the segmented
contour of two discs, their confidence bands may intersect and create
a single large connected component. In addition, BAC is extremely sensitive to numerical issues of gradient-descent
when initialized poorly, so there are scenarios in which it will not
converge at all, or the output results in the formation of loops on the boundary (see Appendix \ref{appendixEx}). Thus, it is not straightforward
to characterize this change of topological features when quantifying
uncertainty. Another issue is that the BAC model \citep{bryner2013elastic} is not fully Bayesian, as it produces just a point estimate (i.e., one contour). Therefore, it is not
obvious how to generalize this method to obtain a full posterior distribution
for the boundary estimates in order to yield natural uncertainty quantification.
To address this issue, we believe that a general, fully Bayesian model
should be proposed and studied in the future. This model has the ability to incorporate prior shape information, like BAC, through a formal, informative prior on the shape space specified via training images with known boundaries.

Interesting observations can be made from the boundary estimates of Figure \ref{fig:SkinLesion1}.
Note that TOP+BAC yielded two contours which partially overlapped. This
behavior is a result of estimation of pixel densities from the training images, as many of the ground truth boundaries
for the skin lesion images are somewhat distant from the portion of
the lesion which is darkest. This may be due to human expert bias
during segmentation. One may inherently be conservative when drawing
a boundary surrounding the lesion, in an attempt to capture the entire
lesion and avoid missing a key part of it. Since the active contour
algorithm depends on the training data to estimate pixel densities,
the final segmentation will also reflect this behavior. 
Also, the expansion of the two contours to include lighter pixels is also partly due to the lack of separation between the two features. If the two connected components are well-separated (e.g., Figure \ref{fig:SkinLesion2}), this behavior will not occur in the gradient-descent algorithm. To address this issue, one idea is
to develop a conditional algorithm based on the elastic distance trick we used for data analysis of neuron cellular dataset. BAC can be used on the ``most typical'' initializations sequentially, constrained to the sub-region
which removes the final contour from the already-segmented feature. 

In a broader sense, our idea is inspired by the more general idea of providing a pre-conditioner for algorithm to improve its performance. The TOP+BAC method specifically shows how a topology-based initialization may assist a geometric method in boundary estimation. We could alternatively consider a geometric pre-conditioning for a topological algorithm. Finding topological representative cycles with certain properties \citep{Deay_etal2010,Dey_etal2011} requires intensive algorithm tuning to achieve a lower complexity. If partial geometric information is known, the the algorithm could be improved by choosing a geometry-induced initial design. For instance, when segmenting the neural data in Figure \ref{fig:Neuron}, we narrowed down an exceedingly large collection of boundaries obtained from TOP by only taking those which are most elliptical, due to prior information about the shape of neuron walls.

As already pointed out by \citet{Gao_et_al2013}, there are other
alternatives in TDA to handle topological information present in pixel
images. In this paper, we use the TOP method by \citet{Paris&Durand2007}
which is a representative topological method which focuses on the pixel density estimator for grayscale images, but
can be generalized to color images. To the best of our knowledge, there
has not been any attempt in the literature to compare the performance
and computational costs of different topological methods. Some attempts has been made to
combine topological methods for specific problems at a higher level \citep{Wu2017,clough2020topological,carriere2020perslay}. Yet our approach use the topological information directly from the image data and maintains its interpretability within the optimization framework. Besides, it has not been made clear, in the existing literature, how differently the ``low-level'' and ``high-level'' topology-based methods utilize topological information. 
We believe it is of great interest to carry out such a systematic and comprehensive
study for topological methods in the image segmentation context. 

\subsection{Acknowledgments}
We would like to acknowledge and thank Yu-Min Chung for pointing out
the skin lesion dataset \citet{codella2018} to us. 

We would like to thank Sebastian Kurtek and Abhijoy Saha for their enthusiastic help and suggestions during the process of preparing and revising  the manuscript.
The first author would also like to thank his advisors Steven MacEachern, Mario Peruggia for their support and comments during this research project. 

\subsection{Code and Data}
The code used for this paper could be found at \newline\url{https://github.com/hrluo/TopologicalBayesianActiveContour}. The datasets are all appropriately cited and can be obtained following the original sources.
\bibliography{segmentation}
\newpage

\appendix

\section{Further Details on Elastic Shape Analysis}
\label{appendixESA}
In both BAC and TOP+BAC, the shape prior update aims to move a contour
towards a prior-specified mean shape of a sample of training curves.
The shape prior energy quantifies the discrepancy between a contour
and the mean shape. Explicit specification of these terms requires
knowledge of elastic shape analysis, methods which use a shape representation
that is invariant to contour parameterization. We refer the interested reader to \cite{JoshiSRVF} for details beyond the scope of this section. 

Assume that we have ground truth contours from training images denoted
by $\beta_{1},\ldots,\beta_{M}$, where $\beta_{i}:\mathcal{D}\rightarrow\mathbb{R}^{2}$ and $\mathcal{D}=\mathbb{S}^{1}$
is the parameter space. In the elastic shape analysis literature,
the \emph{square-root velocity function (SRVF)} is a special transformation
$q:\mathcal{D}\rightarrow\mathbb{R}^{2}$ for a curve $\beta$, given
by $q(t)=\dot{\beta}(t)/\sqrt{\vert\dot{\beta}(t)\vert}$, where
$\vert\cdot\vert$ is the Euclidean norm, $\dot{\beta}$ is the time-derivative
of $\beta$, and we define $q(t)=0$ wherever $\dot{\beta}(t)$ vanishes.
Given $\beta(0)$, this transformation is invertible: $\beta(t)=\beta(0)+\int_{0}^{t}q(s)\vert q(s)\vert\ ds$.
The SRVF has many desirable properties in the search for an optimal
re-parameterization function to match curves \citep{SrivESA}. The
shape of a curve $\beta$ with SRVF $q$ is defined by the equivalence
class: $[q]=\{O(q\circ\gamma)\sqrt{\dot{\gamma}}\ |\ O\in SO(2),\gamma\in\Gamma,q\in\mathcal{C}\}$,
where $SO(2)$ is the special orthogonal group, $\Gamma$ is the Lie
group of re-parameterization functions given by $\Gamma=\{\gamma:\mathcal{D}\rightarrow\mathcal{D}\ |\ \gamma \ \text{is a diffeomorphism of } \mathcal{D}\}$,
and $\mathcal{C}$ is the space of all square-integrable SRVFs $q$
with unit $\mathbb{L}^2$ norm (i.e., the unit Hilbert sphere). The term $O(q\circ\gamma)\sqrt{\dot{\gamma}}$
is the result of applying rotation $O$ and re-parameterization $\gamma$
to SRVF $q$. The equivalence class $[q]$ describes the shape of
$\beta$, as it collects all curves with identical shape (i.e., only
differing by translation, scale, rotation, and re-parameterization).
The shape space is $\mathcal{S}=\mathcal{C}/(SO(2)\times\Gamma)$,
the collection of all equivalence classes $[q]$ for any $q\in\mathcal{C}$.

A natural shape metric is the \emph{elastic shape distance}. Given
two SRVFs $q_{1},q_{2}$: 
\begin{align}
d_{\mathcal{S}}([q_{1}],[q_{2}]) & =\underset{O\in SO(2),\gamma\in\Gamma}{\text{min}}\cos^{-1}\Big(\langle\langle q_{1},O(q_{2}\circ\gamma)\sqrt{\dot{\gamma}}\rangle\rangle\Big),\label{eq:ESD}
\end{align}
where $\langle\langle \cdot,\cdot \rangle\rangle$ is the $\mathbb{L}^2$ inner product. The optimal values of $O$ and $\gamma$ constitute a solution to
the \emph{registration problem}; see \citet{SrivESA} for implementation
details. As discussed in Appendix \ref{appendixPM}, this distance can be used to as
a performance evaluation measure for contour estimation, since the output of these algorithms is either a single contour or a collection of contours. The elastic shape distance between true contour $\beta_{\text{true}}$
and estimated contour $\beta_{\text{est}}$ with SRVFs $q_{\text{true}}$
and $q_{\text{est}}$ is given by $d_{\mathcal{S}}([q_{\text{true}}],[q_{\text{est}}])$.

The ``mean shape'' used in the definition of the $E_{\text{prior}}$
term in Section \ref{subsubsec:Energy} is defined as the intrinsic
sample Karcher mean of training shapes, $[\bar{q}]=\arg\min_{[q]\in\mathcal{S}}\sum_{i=1}^{M}d_{\mathcal{S}}([q],[q_{i}])^{2}$,
where $[q_{1}],\ldots,[q_{M}]$ are the training shapes. The prior
shape term in BAC also incorporates shape variation. Assume that training
curves are observed in the form of $N$ points in $\mathbb{R}^{2}$,
stored as a $2\times N$-dimensional matrix. Typically, $N=100$ to
$200$ is sufficient to capture small shape features. The sample Karcher
covariance $\hat{K}$ is computed by first projecting observations
in $\mathcal{S}$ to a linear tangent space centered at the Karcher
mean $[\bar{q}]$, denoted $T_{[\bar{q}]}(\mathcal{S})$. Let $v_{1},\ldots,v_{M}\in T_{[\bar{q}]}(\mathcal{S})$,
where $v_{i}=\exp_{[\bar{q}]}^{-1}([q_{i}])$ for $i=1,\ldots,M$
is the inverse-exponential map which sends shape $[q_{i}]$ to a tangent
vector in $T_{[\bar{q}]}(\mathcal{S})$. Then, $\hat{K}=\frac{1}{M-1}\sum_{i=1}^{M}w_{i}w_{i}^{\top}$,
where $w_{i}=\text{vec}(v_{i})$ denotes a $2N$-dimensional column
vector obtained by concatenating the two rows of the $2\times N$
matrix $v_{i}$ (known as vectorization). Dominant modes of shape variation are found by singular
value decomposition (SVD), i.e., writing $\hat{K}=U\Sigma U^{\top}$,
where $U$ is the matrix of eigenvectors corresponding to the diagonal
matrix $\Sigma$ of eigenvalues.

\textbf{Prior energy term (from Section \ref{subsubsec:Energy}):} Let $U_{J}$ and
$\Sigma_{J}$ be truncations of $U$ and $\Sigma$ corresponding to
the $J$ eigenvectors with the largest eigenvalues, respectively.
The prior energy term is given by: 
\begin{align*}
E_{\text{prior}}(u) & =\frac{1}{2}w^{T}(U_{J}\Sigma_{J}^{-1}U_{J}^{\top})w+\frac{1}{2\delta^{2}}|w-U_{J}U_{J}^{\top}w|^2,
\end{align*}
where $w$ is the vectorization of $v=\exp_{[\bar{q}]}^{-1}([q])$, $[q]$ is the shape of contour
$u$, and $\delta$ is selected to be less than the smallest eigenvalue
in $\Sigma_{m}$. As mentioned in Section \ref{subsubsec:Energy}, this term is equivalent
to the negative log-likelihood of a truncated wrapped Gaussian density
on finite-dimensional approximations of contours on the shape space. To further elaborate, the first term projects the current contour $w$ onto a $J-$dimensional principal subspace basis for $T_{[\bar{q}]}(\mathcal{S})$ (spanned by the columns of $U_J$, obtained from the prior sample covariance matrix), and measures its distance from the origin (adjusted by the covariance), which represents the prior shape mean $[\bar{q}]$. The second term measures the error between the true representation of $w$ and its $J-$dimensional reconstruction $U_J U_J^{\top} w$, i.e., the variation not accounted for by the first $J$ eigenvectors of the prior covariance. It should be noted that this model assumes the contour has already been discretized to $N$ points, and as noted in \citet{FSDA}, the second term only makes sense in practice when contours are assumed to be finite-dimensional (i.e., post-discretization).

\textbf{Prior update term (from Section \ref{subsubsec:GD}):} In order to obtain
$\nabla E_{\text{prior}}(u^{(i)}(t))$, first compute the gradient
vector $a=\Big(U_{J}\Sigma_{J}^{-1}U_{J}^{\top}+\big(I-U_{J}U_{J}^{\top}\big)/\delta^{2}\Big)w$,
where $w$ is the vectorization of $v=\exp_{[\bar{q}]}\big([q^{(i)}]\big)$ for SRVF $q^{(i)}$
corresponding to $u^{(i)}$, and again, $\delta$ is chosen to be
less than the smallest eigenvalue in $\Sigma_{m}$. This update is
computed on $T_{[\bar{q}]}(\mathcal{S})$; however, we wish to update
the curve $u^{(i)}$ (instead of $[\bar{q}]$), and thus, must parallel
transport this gradient vector to $T_{[q^{(i)}]}(\mathcal{S})$. Similar
to \citet{bryner2013elastic}, call this new vector $b=\Pi(a:[\bar{q}]\rightarrow[q^{(i)}])$,
where $\Pi$ denotes the parallel transport operator (an expression
of which is known for spherical spaces). This yields an updated vector
to apply to the SRVF $q^{(i)}$, but ultimately, one wants to apply
this update to the \textit{curve} $u^{(i)}$. The operation of parallel
transport is done as follows: 
\begin{enumerate}
\item Move a small amount along the negative gradient direction $-b$ via
the exponential map: $q_{\text{new}}=\exp_{[q^{(i)}]}(-\epsilon b)$,
for $\epsilon$ small (e.g., $\epsilon=0.3$).
\item Map back to its curve representation $\beta_{\text{new}}$. 
\item Re-scale and translate $\beta_{\text{new}}$ to have the same length
and centroid as $u^{(i)}$. 
\item Define prior update $\nabla E_{\text{prior}}(u^{(i)})(t)=\frac{1}{\epsilon}\big(u^{(i)}(t)-\beta_{\text{new}}(t)\big)$. 
\end{enumerate}
\noindent This conversion back to the original curve representation
is necessary so that the update can be combined with the image and
smoothness update terms, which are defined based on the original contour
(rather than SRVFs).

\section{Further Details on Topological Segmentation (TOP)}
\label{appendixTOP}
\begin{figure}
\caption{Left: Salt-and-pepper-noised donut image to be segmented. Right: Resulting boundary map under TOP with parameters $\sigma_1=3$, $\sigma_2=5$, $T=5$. Note that, due to the large amount of noise, it is impossible to identify boundaries to initialize TOP+BAC.}

\begin{centering}
\begin{tabular}{cc}
\includegraphics[width=2.5in]{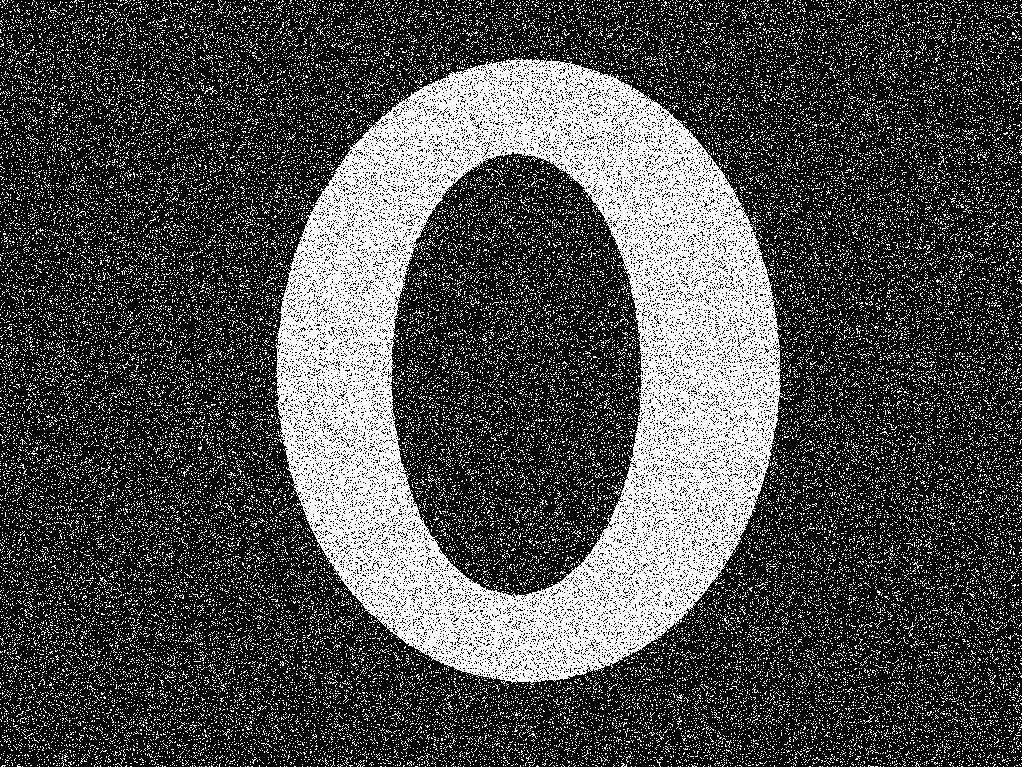}  & \includegraphics[width=2.5in]{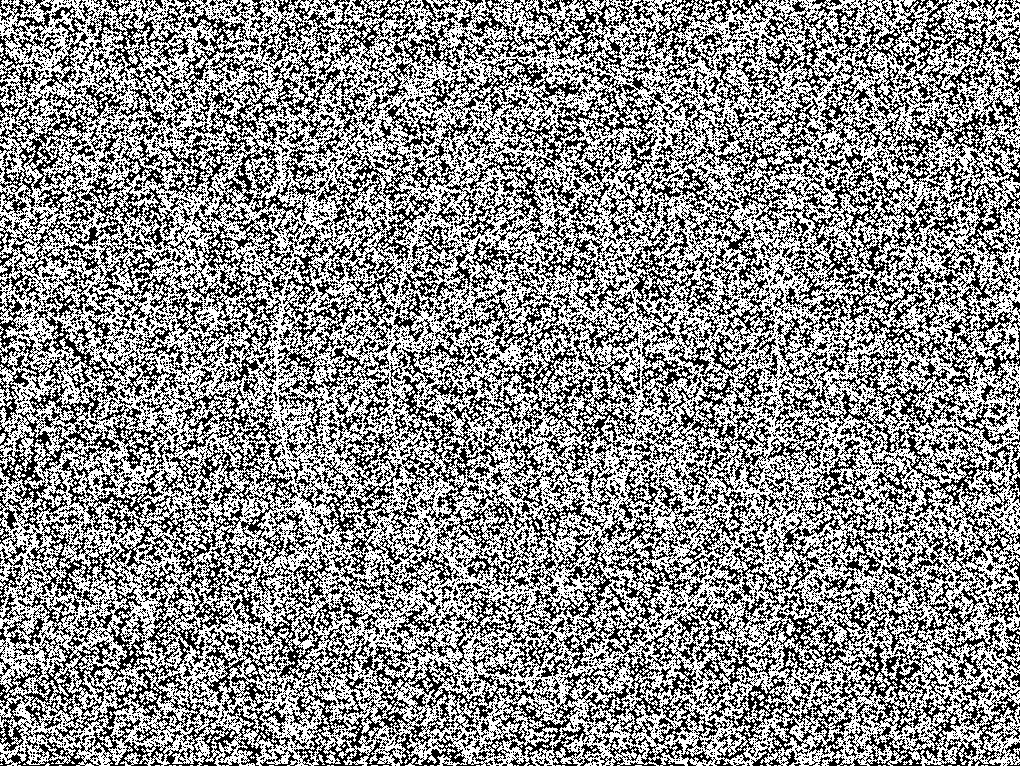}\tabularnewline
\end{tabular}
\par\end{centering}
\label{fig:SPDonut} 
\end{figure}

At the end of Section \ref{subsec:SimImages}, we mention that TOP can be difficult to use as a initialization for BAC when there is a large amount of pixel noise. In the left panel of Figure \ref{fig:SPDonut}, we the same donut object from Figure \ref{fig:BlurDonutTOP}, but perturbed by salt-and-pepper noise, as in Figure \ref{fig:PriorEffect-1}. The right side shows clustering produced by TOP. This segmentation result contains numerous small boundaries due to the image noise, making it difficult to select contours to use as an automatic initialization, as none of the contours will correspond to something similar to the two donut boundaries. Thus, to segment images under similar extreme noise settings, one might resort to a prior dominated BAC model or a user-specified initialization for BAC, avoiding TOP and TOP+BAC.

\section{Performance Evaluation Measures}

\label{appendixPM} 
\subsection{Hausdorff Distance}

The Hausdorff distance can be used to measure the proximity of two
sets are in a metric space. Let $A$ and $B$ be two non-empty subsets
of a metric space $(M,d)$. We define their Hausdorff distance $d_{H}(A,B)$
by, 
\begin{align*}
{\displaystyle d_{H}(A,B)=\max\{\sup_{x\in A}\inf_{y\in B}d(x,y),\sup_{y\in B}\inf_{x\in A}d(x,y)\}} \geq 0& ,
\end{align*}
where $d(\cdot,\cdot)$ is the metric distance. Equivalently, the
Hausdorff distance can be written as ${\displaystyle d_{H}(A,B)=\inf\{\varepsilon\geq0\,;\ A\subseteq B_{\varepsilon}\ \mbox{and}\,B\subseteq A_{\varepsilon}\}}$,
where ${\displaystyle A_{\varepsilon}:=\bigcup_{x\in A}\{z\in M\,;\ d(z,x)\leq\varepsilon\}}$.
We use this as a naive measure between the point sets of the segmentation result and the ground truth. Hausdorff distance reflects both local
and global discrepancy between two point sets, so we expect it to
be an overall measure of discrepancy between two sets of points in
the same space $\mathbb{R}^{2}$. The set $A$ refers to the clusters
of the estimated region and set $B$ consists of the points from the
ground truth clustering. One drawback of this measure is that it does not have a range, but its value depends on the size of sets $A,B$. A smaller value of this measure means regions $A$ and $B$ are more similar.

\subsection{Hamming Distance}

\citet{huang1995quantitative} introduced the idea of the (directional)
Hamming distance between two segmentations consisting of clusters.
Let the directional distance between two segmentations represented by pixel clusters: $S_{1}=\{s_{1}^{1},\cdots,s_{1}^{n}\},S_{2}=\{s_{2}^{1},\cdots,s_{2}^{m}\}$
be given by: 
\begin{align*}
D_{H}(S_{1} & \rightarrow S_{2})\coloneqq\sum_{\{i;s_{2}^{i}\in S_{2}\}}\sum_{\{(k,j);s_{1}^{k}\neq s_{1}^{j},s_{1}^{k}\cap s_{2}^{i}\neq\emptyset\}}|s_{2}^{i}\cap s_{1}^{k}|.
\end{align*}
A region-based performance evaluation measure based on normalized
Hamming distance is defined as, 
\begin{align*}
p_{H}(S_{1},S_{2})\coloneqq & 1-\frac{D_{H}(S_{1}\rightarrow S_{2})+D_{H}(S_{2}\rightarrow S_{1})}{2\cdot|S|},
\end{align*}
where $|S|$ is the image size. Here, we take $S_{1}={A}$ as the estimated
region $A$ and $S_{2}={B}$ as the ground truth region $B$. $A$ is a set of discrete pixel points and $B$ is the ground truth set of pixel points.
 This measure has value between 0 and 1, a value closer to 1 means regions $A$ and $B$ are more similar.
\subsection{Jaccard Distance}

The Jaccard index metric \citep{jaccard1901distribution,spath1981minisum},
also known as the Jaccard similarity coefficient, is a widely used
measure \citet{AMFM2011}. The Jaccard coefficient measures similarity
between finite sample sets; in our case, the clusters from the final
segmentation, and is given by: 
\begin{align*}
{\displaystyle J(A,B)=\frac{|A\cap B|}{|A\cup B|}} & =\frac{|A\cap B|}{|A|+|B|-|A\cap B|},
\end{align*}
where we regard $A$ as the estimated region consisting of discrete pixel points and $B$ as the ground
truth set of pixel points. If $A$ and $B$ are both empty, we define $J(A,B)=1$. One
can use the Jaccard distance to measure dissimilarity between sample
sets, by defining $d_{J}(A,B)=1-J(A,B)$. This measure has value between 0 and 1, a value closer to 1 means regions $A$ and $B$ are more similar.

\subsection{Performance Measure via Discovery Rates}

Since we regard the problem of interest as a curve or region estimation
problem over an image, a natural measure of such an approach is the
false positive rate, treating each pixel as an estimation location.
In this direction, \citet{abdulrahman2017contours} suggests the performance
measure, 
\begin{align*}
PM(A,B) & =1-\frac{TP}{TP+FP+FN},
\end{align*}
where if we regard $A$ as the estimated region consisting of discrete pixel points and $B$ as the ground
truth set of pixel points, the cardinality quantities shown in the formula above can be summarized
as follows: $TP=|A\cap B|$ (true positive); $FP=|A\cap B^{c}|$ (false
positive); $FN=|A^{c}\cap B|$ (false negative).
 This measure has value between 0 and 1, a value closer to 1 means regions $A$ and $B$ are more similar.
\subsection{Elastic Shape Distance}

Finally, since the result of the proposed TOP+BAC algorithm is an
estimated contour, we can compare this to the true contour's shape
by computing their elastic shape distance, which was defined in Section
\ref{subsubsec:Energy}. Given true contour $\beta_{\text{true}}$
and estimated contour $\beta_{\text{est}}$ with SRVFs $q_{\text{true}}$
and $q_{\text{est}}$ respectively, recall that the elastic shape
distance is given by: 
\begin{align*}
d_{\mathcal{S}}([q_{\text{true}}],[q_{\text{est}}])=\underset{O\in SO(2),\gamma\in\Gamma}{\text{min}}\cos^{-1}\Big(\langle\langle q_{\text{true}},O(q_{\text{est}}\circ\gamma)\sqrt{\dot{\gamma}}\rangle\rangle\Big)\geq 0,
\end{align*}
where $\langle\langle\cdot,\cdot\rangle\rangle$ is the $\mathbb{L}^{2}$
inner product and $SO(2)$ is the special orthogonal group. Note that this does not take pixel position into consideration,
and so if the estimated contour is simply translated to a different
portion of the image, the elastic shape distance will not change.
However, this quantifies shape dissimilarity within the image, which
can be useful in certain applications. Also note that for topologically
non-trivial objects with boundaries represented by multiple contours,
we can simply compute the elastic shape distance between corresponding
estimated and ground truth contours separately.
Although this measure is also a distance metric like Hausdorff distance, it is bounded because the space of shapes are compact, therefore this distance is bounded. A smaller value of this measure means regions $A$ and $B$ are more similar.

\FloatBarrier

\section{Additional Data Analysis Examples}
\label{appendixEx}

\subsection{MPEG-7 Images}

\label{subsec:MPEG}

In this section, we present some results for  bone images from the MPEG-7 dataset, referenced by \citet{MPEG}. In this setup, we start with the contours of 20 bone curves (represented by $N=200$ points each), and construct binary images with the interior of each bone in white, and the exterior in black. To examine the performance of contour detection under the models discussed in Section \ref{sec:Model-Specification-and}, we select one image as the test image, leaving the remaining 19 as training images. Since the images have been constructed based on the underlying bone contours, we have ground truth boundaries for both training and test sets, thus allowing us to quantify performance using the previously-mentioned evaluation metrics.

\begin{figure}
\caption{\label{fig:Final-segmented-contour-1-1}Final segmented contour (in blue) for a binary, no-noise bone image using BAC (left), TOP (middle), and TOP+BAC (right). The left panel also shows the BAC initialization in red. BAC methods are performed using $\lambda_{1}=0.3,\ \lambda_{2}=0.3,\ \lambda_{3}=0$; TOP methods are performed using $\sigma_{1}=5,\sigma_{2}=5,T=5$. Using a convergence tolerance of $10^{-7}$, BAC requires 60 steps to converge, while TOP+BAC requires 6 steps. Pixel densities for interior and exterior regions use a histogram estimator with bandwidth of 0.0039 (equivalent to 255 bins).}

\begin{centering}
\begin{tabular}{ccc}
\toprule 
BAC & TOP & TOP+BAC \\
\hline
\hline
\includegraphics[width=1.8in]{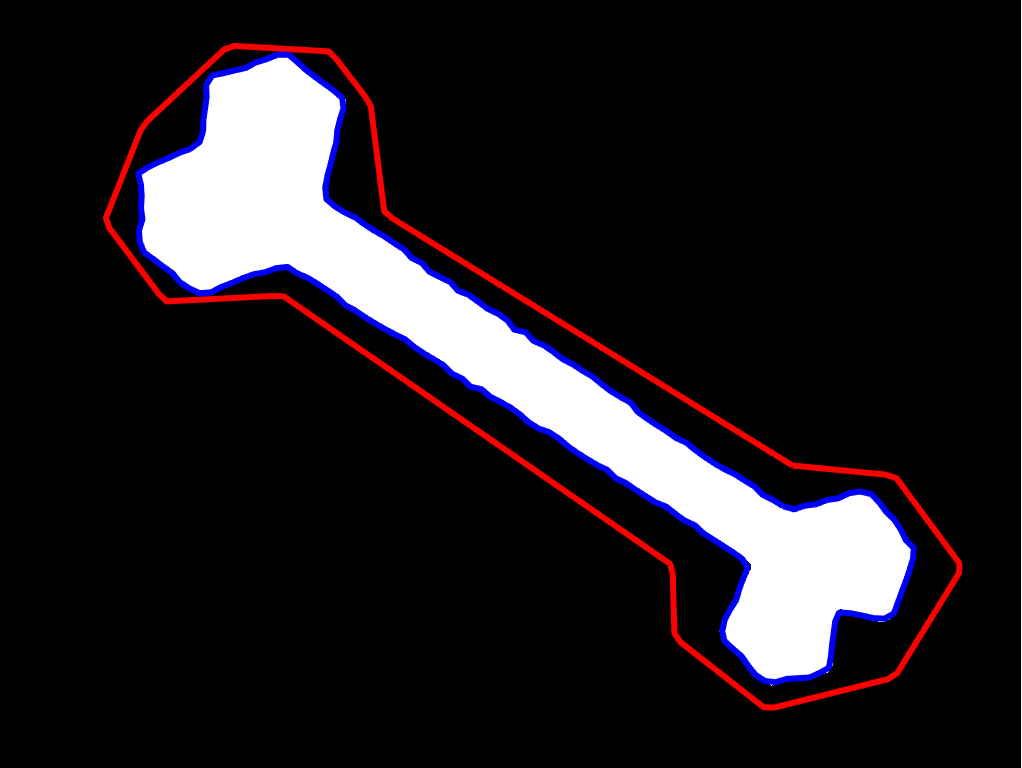} & \includegraphics[width=1.8in]{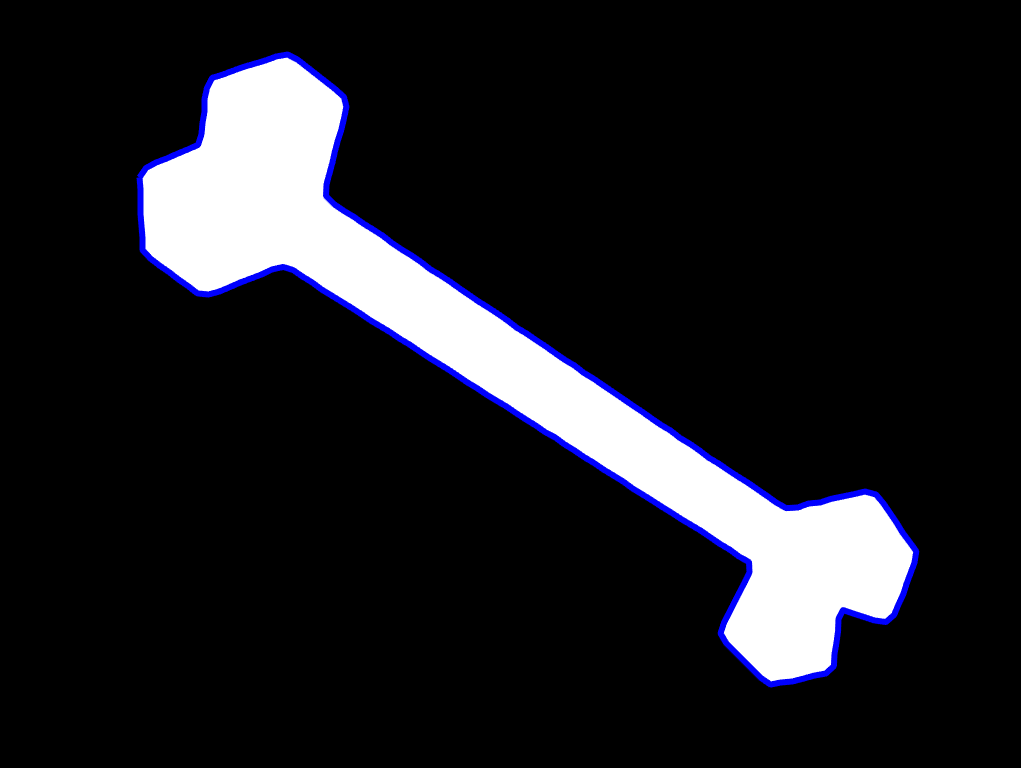} & \includegraphics[width=1.8in]{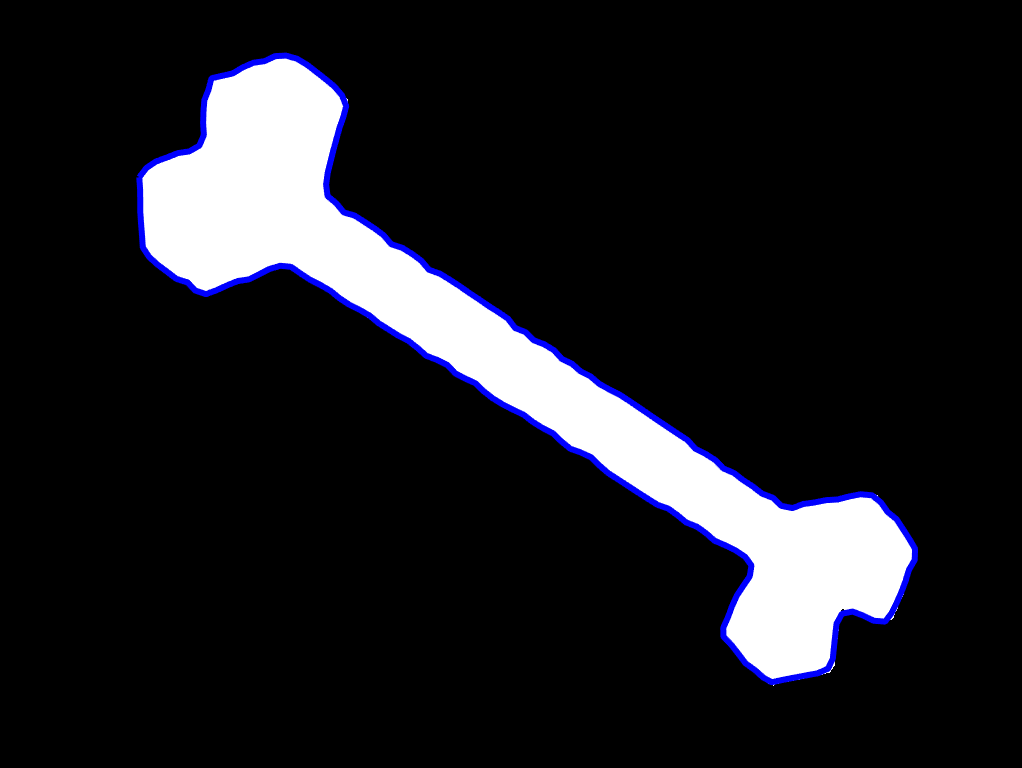}\tabularnewline
\bottomrule
\end{tabular}
\par\end{centering}
\label{fig:NoNoiseBone-1}
\end{figure}

Figure \ref{fig:NoNoiseBone-1} compares the boundary estimate under the BAC (using a pre-specified initialization shown in red on the left panel), TOP and TOP+BAC approaches. For BAC portions of algorithms, we selected $\lambda_{1}=0.3,\ \lambda_{2}=0.3$, and $\lambda_{3}=0$ (i.e., no shape prior update). For TOP, we empirically
selected $\sigma_{1}=5,\sigma_{2}=5,T=5$. As with the donut example in the main text, a shape prior is unnecessary here due to the quality of the test image. Qualitatively, all three approaches obtain similar final estimates of the bone outline -- however, while BAC under the pre-specified initialization required 60 steps to converge, using TOP to initialize only requires 6 steps, since TOP already perfectly identifies the boundary (ignoring convergence and numerical issues). Quantitatively, the similarity in final results is verified by the performance measures found in the left panel of Table \ref{tab:NoNoiseBonePM-1}. Note that pixel densities are estimated using histograms with 255 bins, as these images are binary.

\begin{figure}
\caption{Final segmented contour (in blue) for a binary, no-noise, perturbed bone image using BAC (left), TOP (middle), and TOP+BAC (right). The left panel also shows the BAC initialization in red. BAC methods are performed using $\lambda_{1}=0.3,\ \lambda_{2}=0.3,\ \lambda_{3}=0$; TOP methods are performed using $\sigma_{1}=5,\sigma_{2}=5,T=5$. Using a convergence tolerance of $10^{-7}$, BAC requires 67 steps to converge, while TOP+BAC requires 53 steps. Pixel densities for interior and exterior regions use a histogram estimator with bandwidth of 0.0039 (equivalent to 255 bins).}

\begin{centering}
\begin{tabular}{ccc}
\toprule 
BAC & TOP & TOP+BAC \\
\hline
\hline
\includegraphics[width=1.8in]{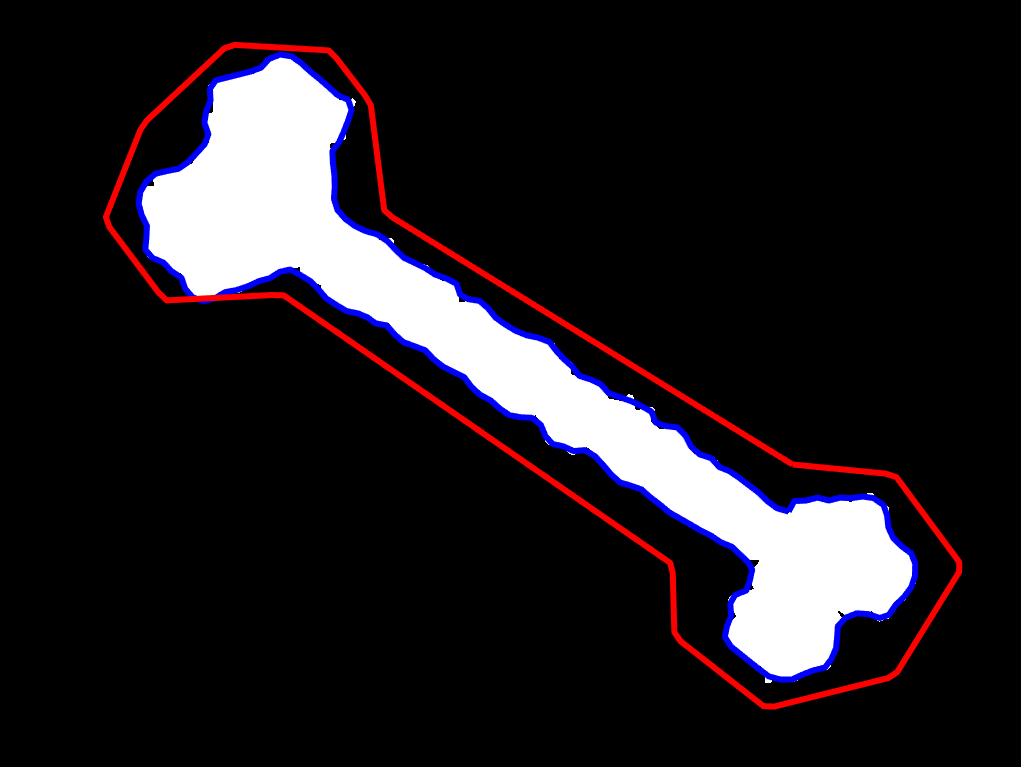} & \includegraphics[width=1.8in]{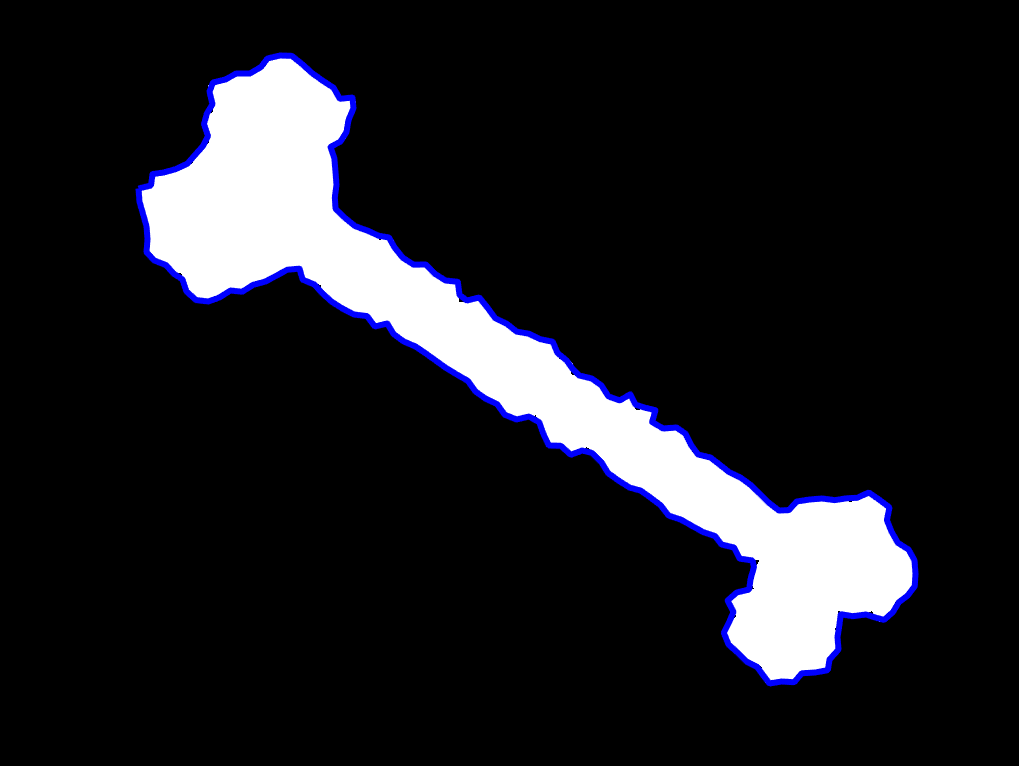} & \includegraphics[width=1.8in]{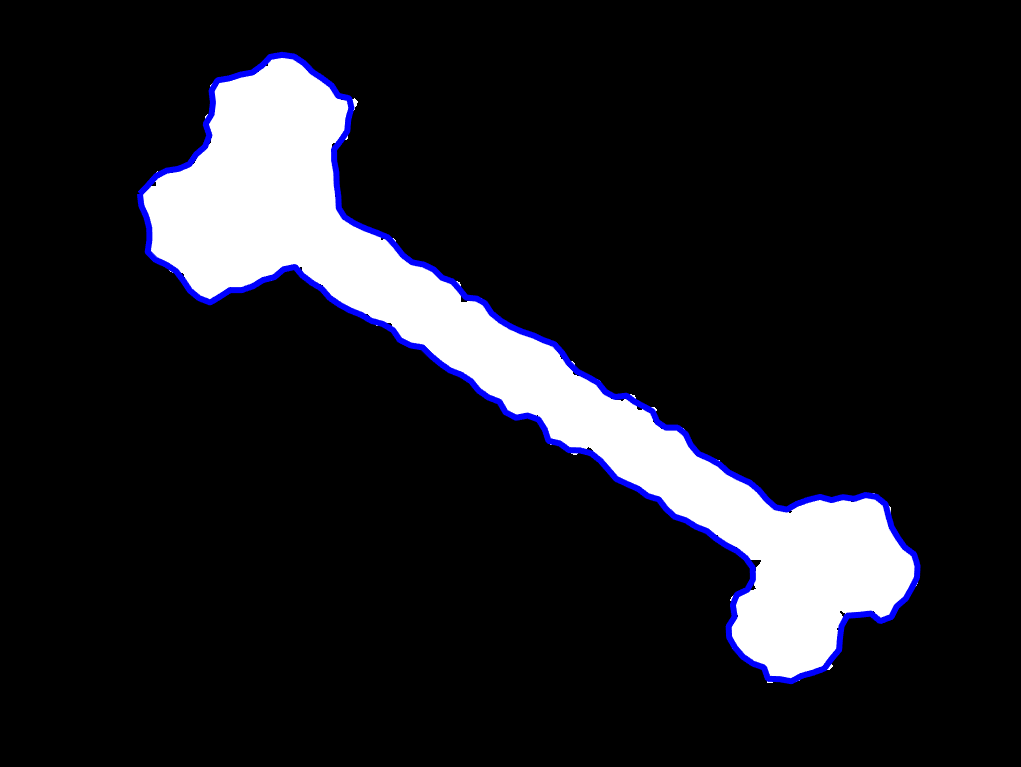}\tabularnewline
\bottomrule
\end{tabular}
\par\end{centering}
\label{fig:PertBone}
\end{figure}

One way to add noise to the underlying image is to perturb the underlying contour by simply adding a bivariate mean-zero Gaussian random number with standard deviation 3 to each column (corresponding to Euclidean coordinates) of the vector representing this contour. This creates a jagged bone from the smooth bone contour (see Figure \ref{fig:PertBone}). However, since the image is still binary and has no noise, we expect BAC, TOP, and TOP+BAC to perform equally well. This is the case, as illustrated in Figure \ref{fig:PertBone}, with identical BAC settings $\lambda_{1}=0.3,\ \lambda_{2}=0.3,\ \lambda_{3}=0$ and TOP settings $\sigma_{1}=5,\sigma_{2}=5,T=5$. Corresponding performance measures are found in the middle portion of Table \ref{tab:NoNoiseBonePM-1}.

\begin{figure}
\caption{Final segmented contour (in blue) for a bone image with artificial Gaussian blur using BAC (left), TOP (middle), and TOP+BAC (right). The left panel also shows the BAC initialization in red. BAC methods are performed using $\lambda_{1}=0.3,\ \lambda_{2}=0.3,\ \lambda_{3}=0$; TOP methods are performed using $\sigma_{1}=5,\sigma_{2}=5,T=5$. Using a convergence tolerance of $10^{-7}$, BAC requires 55 steps, while TOP+BAC requires 43 steps. A bandwidth of 0.03 was used for kernel density estimates of interior and exterior pixel values.}

\begin{centering}
\begin{tabular}{ccc}
\toprule 
BAC & TOP & TOP+BAC \\
\hline
\hline
\includegraphics[width=1.8in]{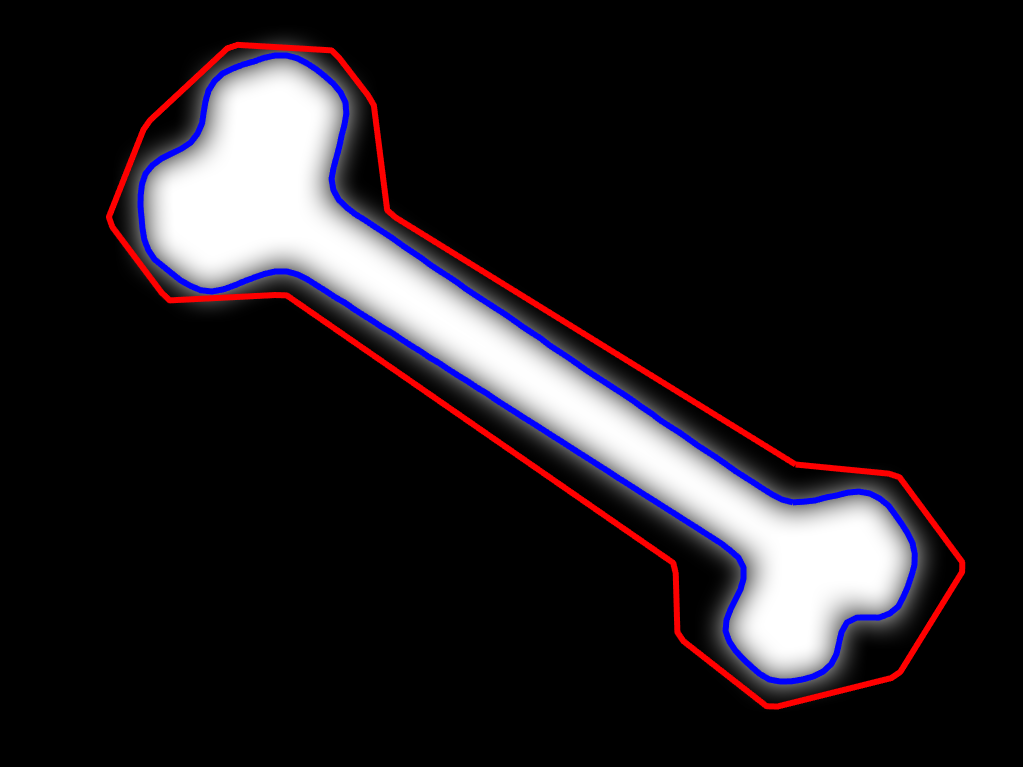} & \includegraphics[width=1.8in]{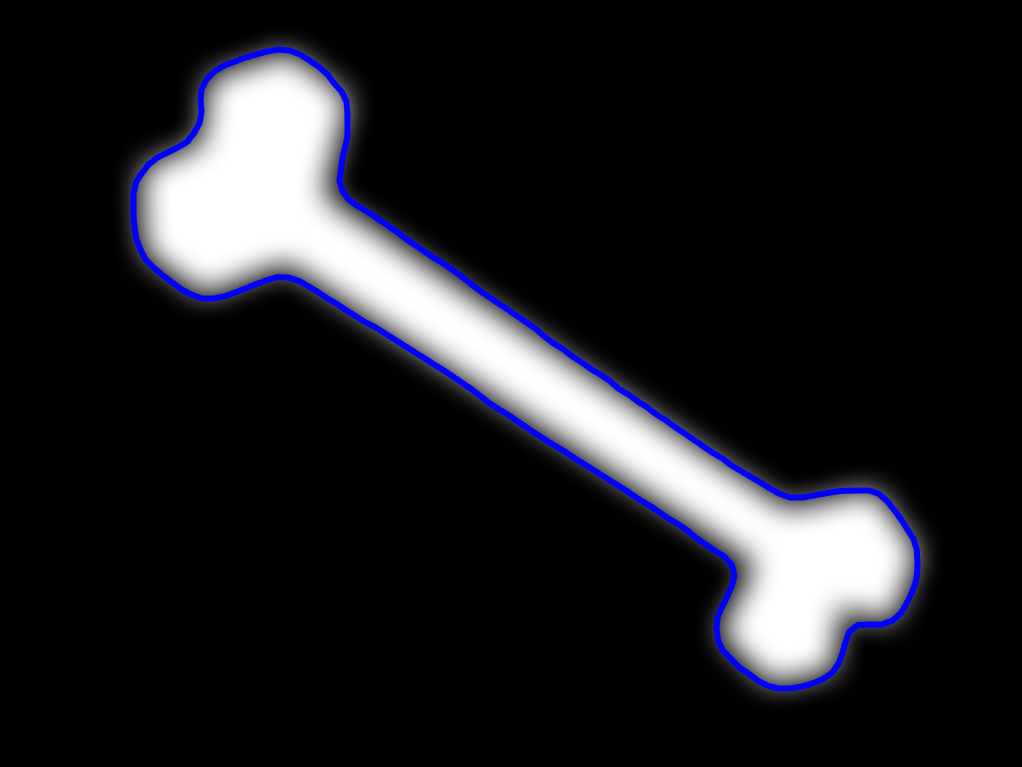} & \includegraphics[width=1.8in]{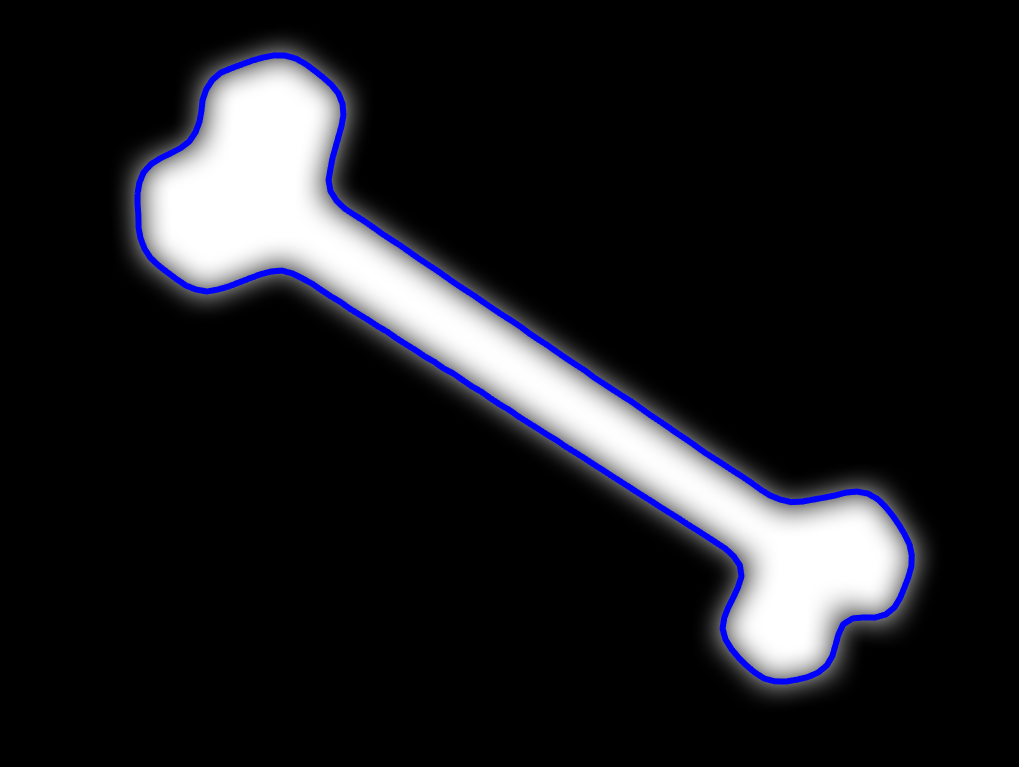}\tabularnewline
\bottomrule
\end{tabular}
\par\end{centering}
\label{fig:BlurBone-1}
\end{figure}

As in the donut image, we also consider adding Gaussian blur with parameter 5 to the binary bone image, yielding a non-binary image. Figure \ref{fig:BlurBone-1} shows results for BAC, TOP and TOP+BAC using the same settings for TOP and BAC as in the previous examples. We note that BAC and TOP perform well on their own, and that TOP+BAC helps to smooth out some of the small bumps that are visible in boundaries estimated from the TOP segmentation. Since TOP treats segmentation as a pixel clustering problem, the resulting contour may not be as smooth as desired, depending on the specified parameters. In addition, as with the donut, varying the TOP parameters can either lead to results which capture the bone perfectly, or estimates a larger bone than the ground truth. In this case, using this as an initialization for TOP+BAC allows one to refine boundary estimates towards the true contour by using the estimated pixel densities from training images. The right panel of Table \ref{tab:NoNoiseBonePM-1} shows performance measures for this example. 

\begin{table}
\caption{Performance measures for the BAC, TOP, and TOP+BAC (abbreviated TAC in the table) methods applied to the bone image (i) without noise (Figure \ref{fig:NoNoiseBone-1}), (ii) with Gaussian contour perturbations (Figure \ref{fig:PertBone}), and (iii) with Gaussian blur (Figure \ref{fig:BlurBone-1}). In bold is the method with the lowest value of the corresponding performance measure.}

\begin{centering}
\begin{tabular}{c|ccccccccc}
\toprule 
 & \multicolumn{3}{c}{(i) No noise} & \multicolumn{3}{c}{(ii) Gaussian perturbation} & \multicolumn{3}{c}{(iii) Gaussian blur}\tabularnewline
\midrule 
Measure & BAC & TOP & TAC & BAC & TOP & TAC & BAC & TOP & TAC\tabularnewline
\midrule
\midrule 
$d_{H}$ & 2.828 & \textbf{2.000} & 2.828 & \textbf{3.464} & 3.873 & 3.873 & \textbf{3.873} & 6.164 & \textbf{3.873}\tabularnewline
$p_{H}$ & 0.003 & \textbf{0.001} & 0.004 & 0.004 & \textbf{0.003} & 0.004 & \textbf{0.006} & 0.020 & \textbf{0.006}\tabularnewline
$d_{J}$ & 0.023 & \textbf{0.009} & 0.027 & 0.032 & \textbf{0.020} & 0.031 & \textbf{0.044} & 0.131 & 0.045\tabularnewline
PM & 0.026 & \textbf{0.011} & 0.031 & 0.036 & \textbf{0.023} & 0.036 & \textbf{0.049} & 0.145 & 0.050\tabularnewline
ESD & 0.084 & \textbf{0.029} & 0.094 & \textbf{0.202} & 0.245 & 0.209 & \textbf{0.111} & 0.121 & 0.113\tabularnewline
\bottomrule
\end{tabular}
\par\end{centering}
\label{tab:NoNoiseBonePM-1}
\end{table}

\begin{figure}
\caption{Final segmented contour (in blue) using BAC for (left) salt-and-pepper-noised bone image and (right) Gaussian contour perturbed bone image, with settings $\lambda_{1}=0.3,\ \lambda_{2}=0.3,\ \lambda_{3}=0.05$. Using a convergence tolerance of $10^{-7}$, the algorithm converged in 37 and 59 steps, respectively. Pixel densities for interior and exterior regions use a histogram estimator with bandwidth of 0.0039 (equivalent to 255 bins).}

\begin{centering}
\begin{tabular}{cc}
\includegraphics[width=2in]{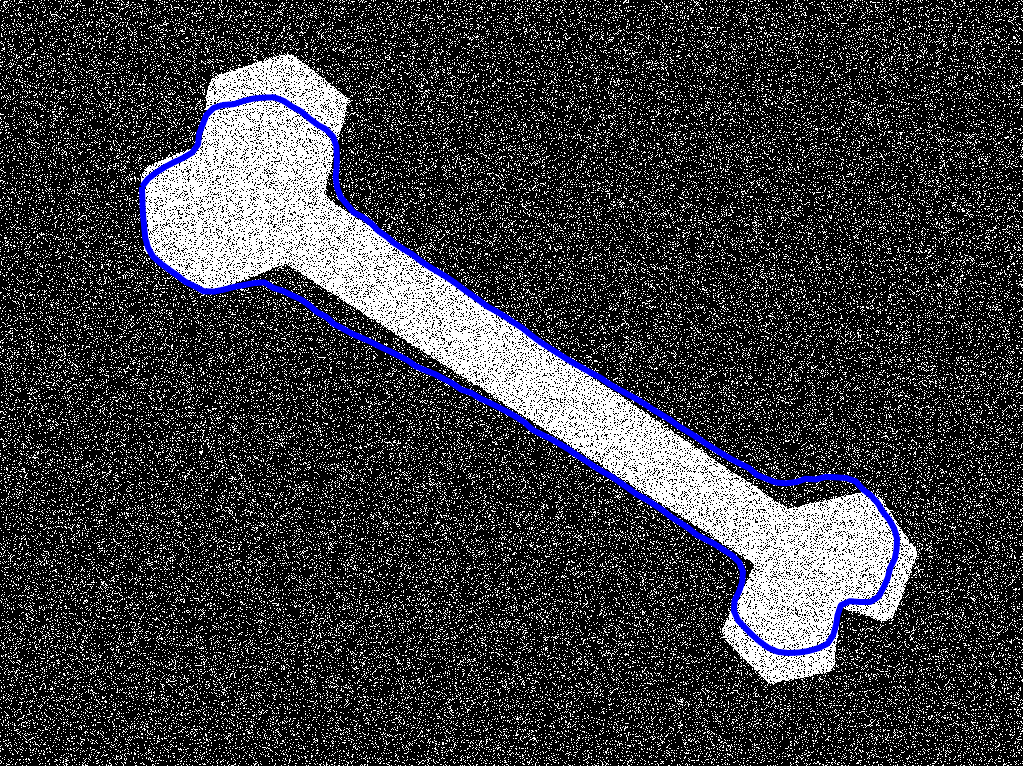} & \includegraphics[width=2in]{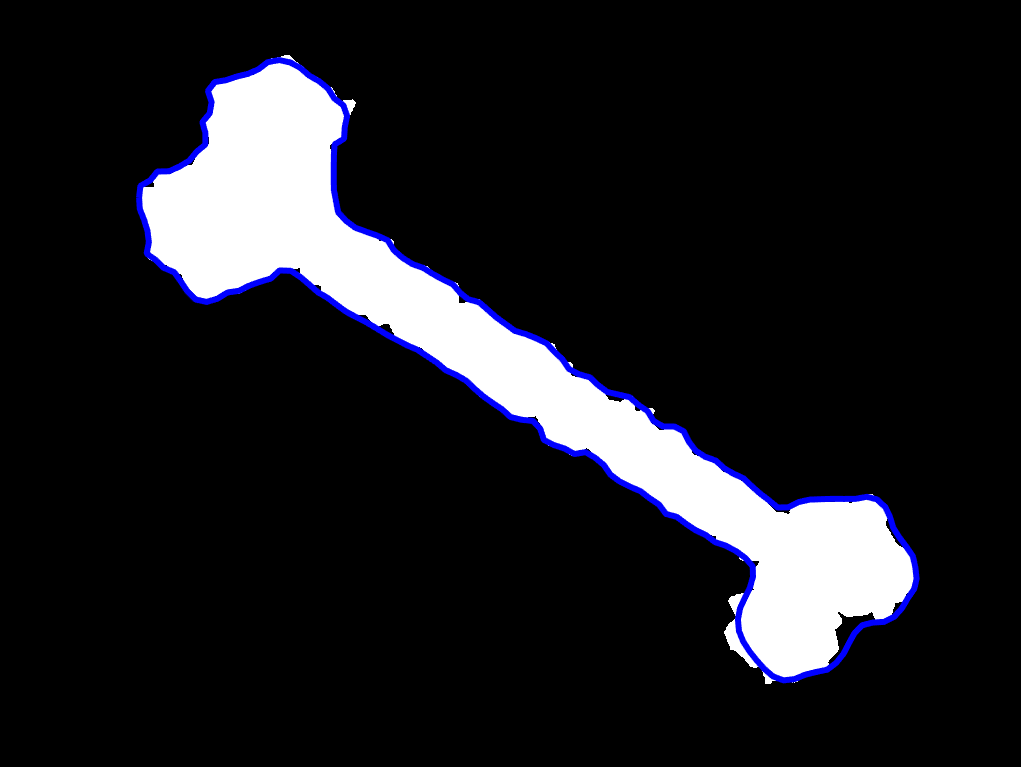}\tabularnewline
\end{tabular}
\par\end{centering}
\label{fig:PriorEffect-1}
\end{figure}

Finally, in Figure \ref{fig:PriorEffect-1}, we demonstrate the crucial effect that the prior term can have on gradient updates in active contour algorithms. In the left is an image with a large amount of salt-and-pepper noise, obtained by applying the $\mathtt{imnoise}$ function (with
a pre-specified noise density $0.3$) in MATLAB to the original binary bone image. Judicious use of prior shape knowledge can help drive the update towards a seomwhat reasonable estimate of the contour in many fewer iterations than without a shape prior. This is because the prior term only involves the contour information from ground truth boundaries of training images, without any regard for the signal-to-noise ratio within the image itself. This is shown in the left panel of Figure \ref{fig:PriorEffect-1}, where we now set $\lambda_{3}=0.05$. While the targeted bone does not perfectly match the shape in the image, we obtain a biased estimate which ignores the noise to some degree. On the right panel, we illustrate the idea that use of the prior can also help average out the random noise along the boundary when segmenting the test image, producing a smooth final contour, while TOP or traditional active contour methods result in a jagged contour. A decision of whether to use the prior term is application-specific. For most images, we suggest setting $\lambda_{3}=0$, particularly when a topological initialization is used. However, if the object is not fully observable in the image, whether due to binary noise along the boundary or occluded features as remarked by \citet{bryner2013elastic}, use of the prior term can be beneficial to boundary estimation. 

\subsection{Simulated Images}
In this section, we show the impact of different choices of the image update parameter $\lambda_{1}$ in BAC on the resulting boundary estimate for no-noise binary donut images similar to Figure \ref{fig:BlurDonutTOP}. Note that under our TOP initializations for the blurred images, setting $\lambda_1=0.15$ and $\lambda_2=0.30$ produced an accurate boundary estimate. However, one must be somewhat cautious with increasing the value of $\lambda_1$ (yielding more aggressive image updates), particularly for initializations which are not very smooth. 

\begin{figure}[h]
\caption{Final segmented contour (in blue) for a no-noise donut image using BAC, with fixed values $\lambda_{2}=0.3, \lambda_{3}=0$. From left to right, $\lambda_{1}=0.15,\ 0.4,\ 0.8$, respectively. Left plot shows the user-specified initialization for both contours in red; the same initialization was used for the other two plots. Using a convergence tolerance of $10^{-7}$, BAC requires 131, 191, and 31 steps to converge, respectively. Pixel densities for interior and exterior regions use a histogram estimator with bandwidth of 0.0039 (equivalent to 255 bins).}

\begin{centering}
\begin{tabular}{ccc}
\toprule
$\lambda_{1}=0.15$  & $\lambda_{1}=0.4$  & $\lambda_{1}=0.8$\\
\hline
\hline
\includegraphics[width=1.8in]{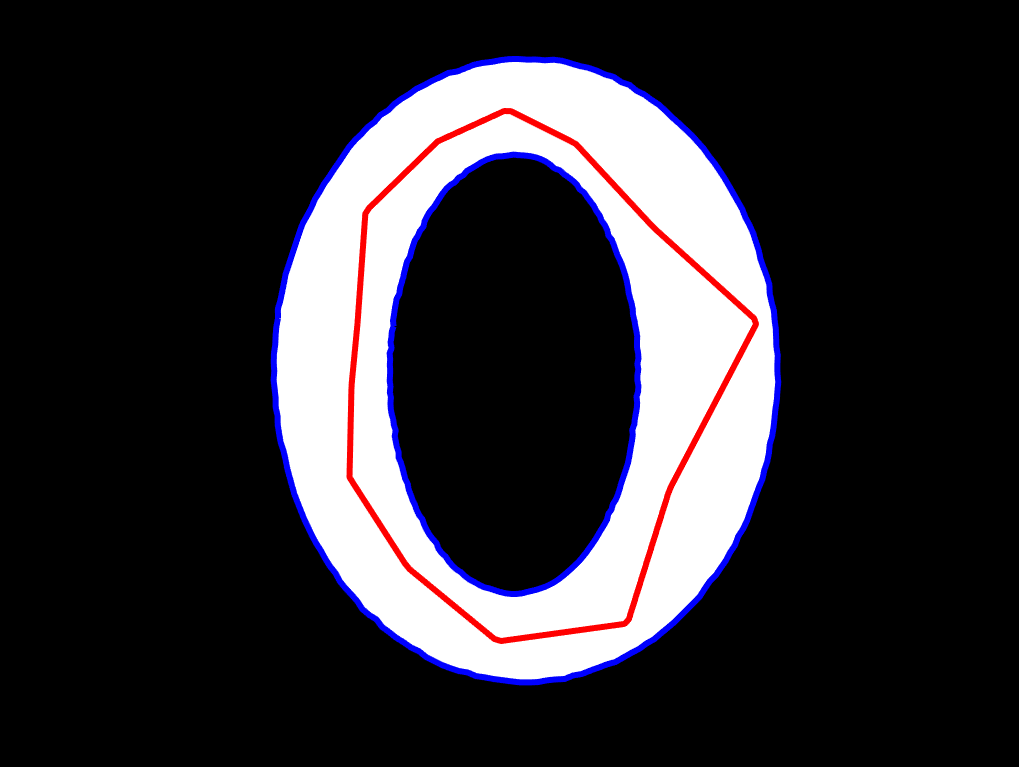}  & \includegraphics[width=1.8in]{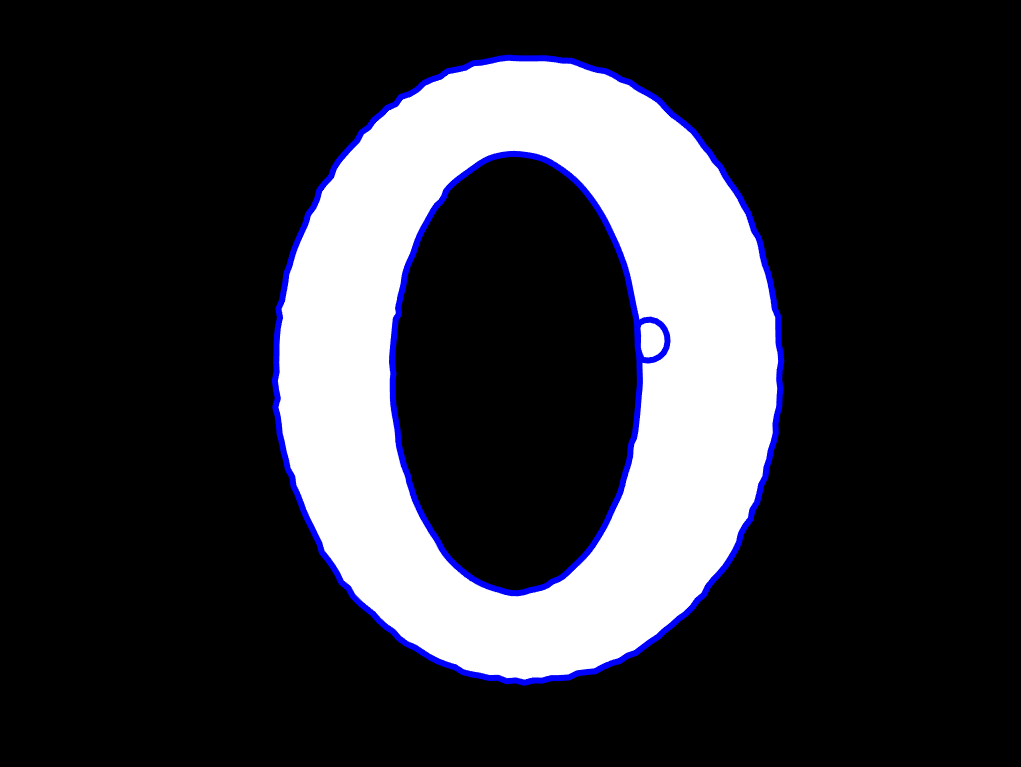}  & \includegraphics[width=1.8in]{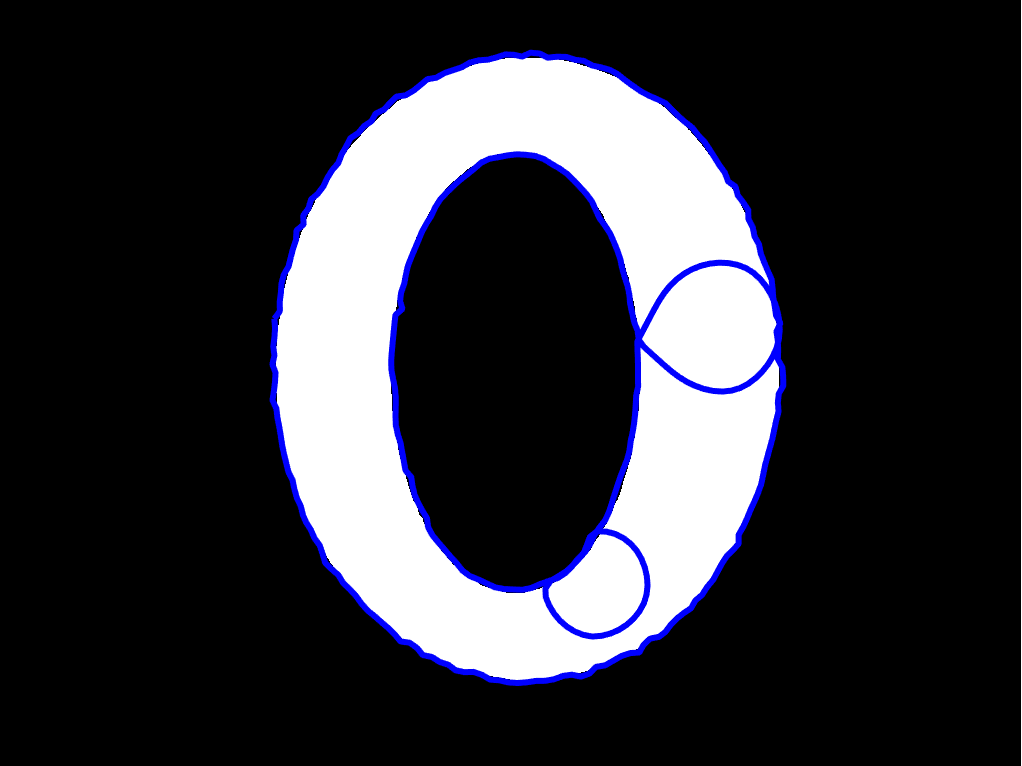}\\ 
\bottomrule
\end{tabular}
\par\end{centering}
\label{fig:DonutLambda1} 
\end{figure}

Figure \ref{fig:DonutLambda1} shows resulting boundary estimates from BAC as we vary $\lambda_1$ under a rough initialization for both boundaries (shown in red on the left plot). This can result in the formation of self-intersections, as seen for the $\lambda_{1}=0.4$ and $0.8$ cases, which is not interpretable from a contour perspective. In order to avoid this behavior, one should choose $\lambda_{1}$ fairly small (in the range of 0.15 to 0.5 seems to be appropriate for many examples
considered by the authors), and if $\lambda_1$ is chosen larger, one may need to increase the value of $\lambda_2$ to mitigate the formation of self-intersections.

\FloatBarrier

\subsection{Skin Lesion Images}
In Section \ref{subsec:Lesion}, we referenced that the choice of bandwidth used for the kernel density estimate of interior and exterior pixel values from training data can be used to control how tightly a contour captures a particular skin lesion site. We illustrate this on two additional skin lesions from the ISIC-1000 dataset by performing the proposed TOP+BAC algorithm. 

\begin{figure}[ht]
\caption{Results of TOP+BAC on a melanoma lesion image from the ISIC-1000 dataset. (Top left) Boundary map obtained from TOP, used to initialize two active contours. (Top middle/right) Estimated final contours based on the topological initializations (TOP+BAC), using parameters $\lambda_{1}=0.4,\ \lambda_{2}=0.3,\ \lambda_{3}=0$ for BAC on the large connected component, and $\lambda_{1}=0.1,\ \lambda_{2}=0.3,\ \lambda_{3}=0$ for BAC on the small connected component; parameters $\sigma_{1}=1,\ \sigma_{2}=5,\ T=5$ for TOP. Bandwidths of the pixel density estimator are 0.05 and 0.30 for the middle and right panels, respectively. BAC is run for a maximum of 1000 and 937 iterations until convergence, respectively. (Middle) Histogram of raw pixel values for interior (left) and exterior (right) regions with a bandwidth of 0.05. (Bottom) Interior (blue) and exterior (red) estimated pixel densities constructed using Gaussian kernel density estimators, with bandwidths 0.05 (left) and 0.30 (right). }

\begin{centering}
\begin{tabular}{ccc}
\includegraphics[width=1.8in]{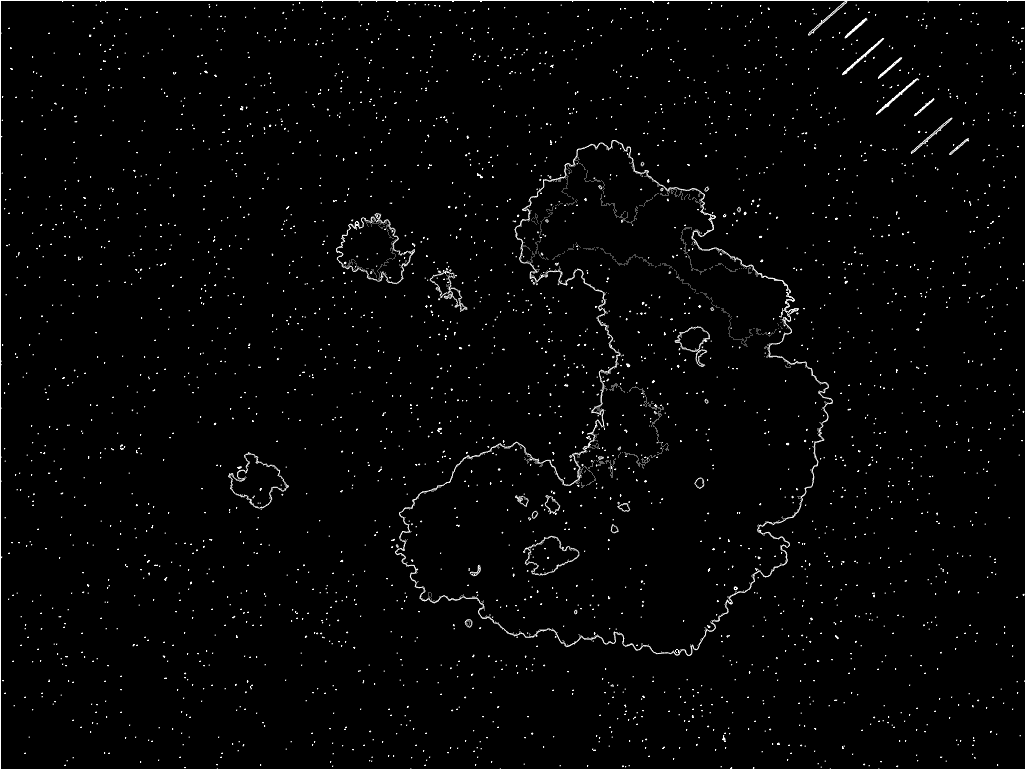}  & \includegraphics[width=1.8in]{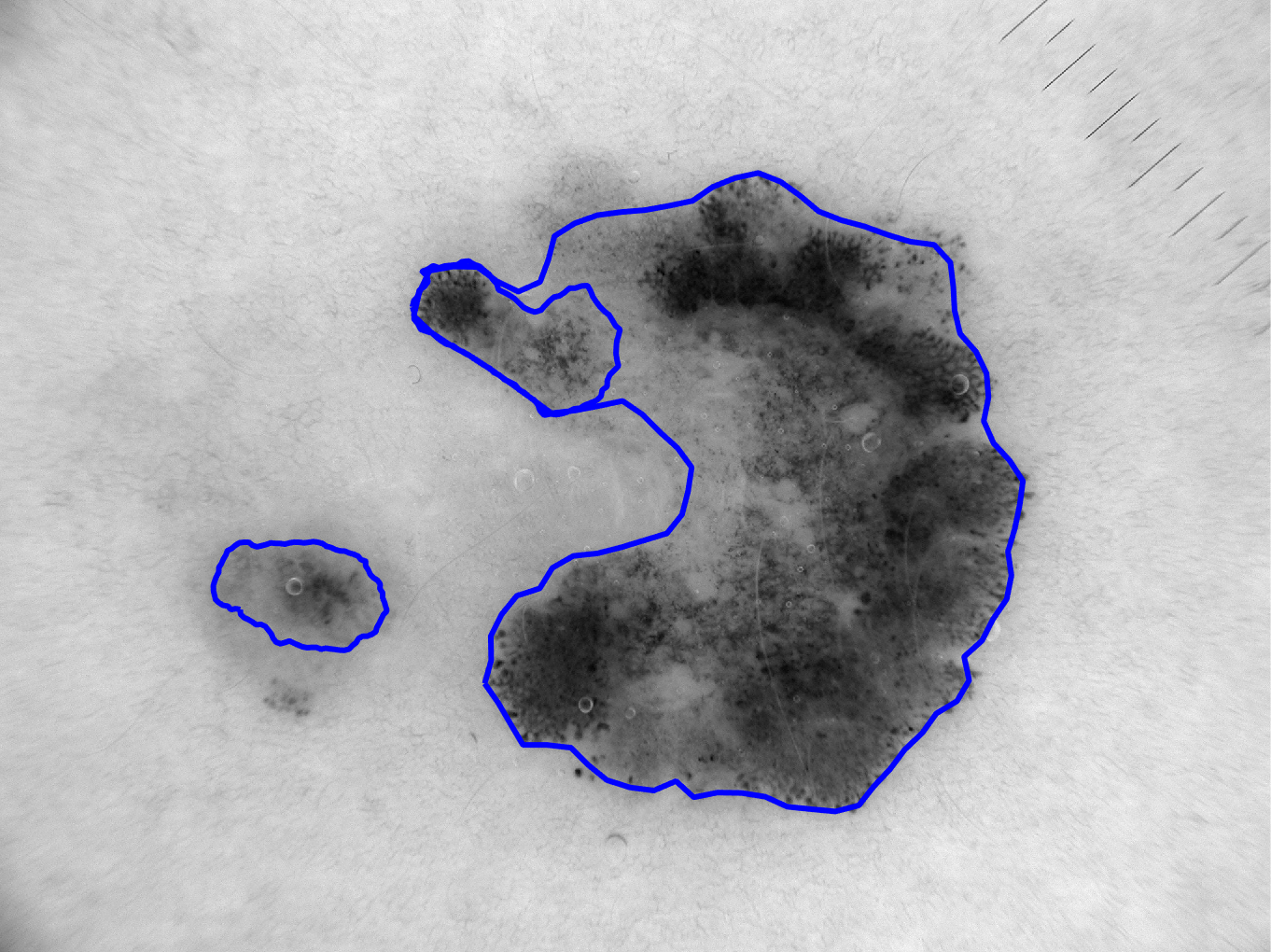}  & \includegraphics[width=1.8in]{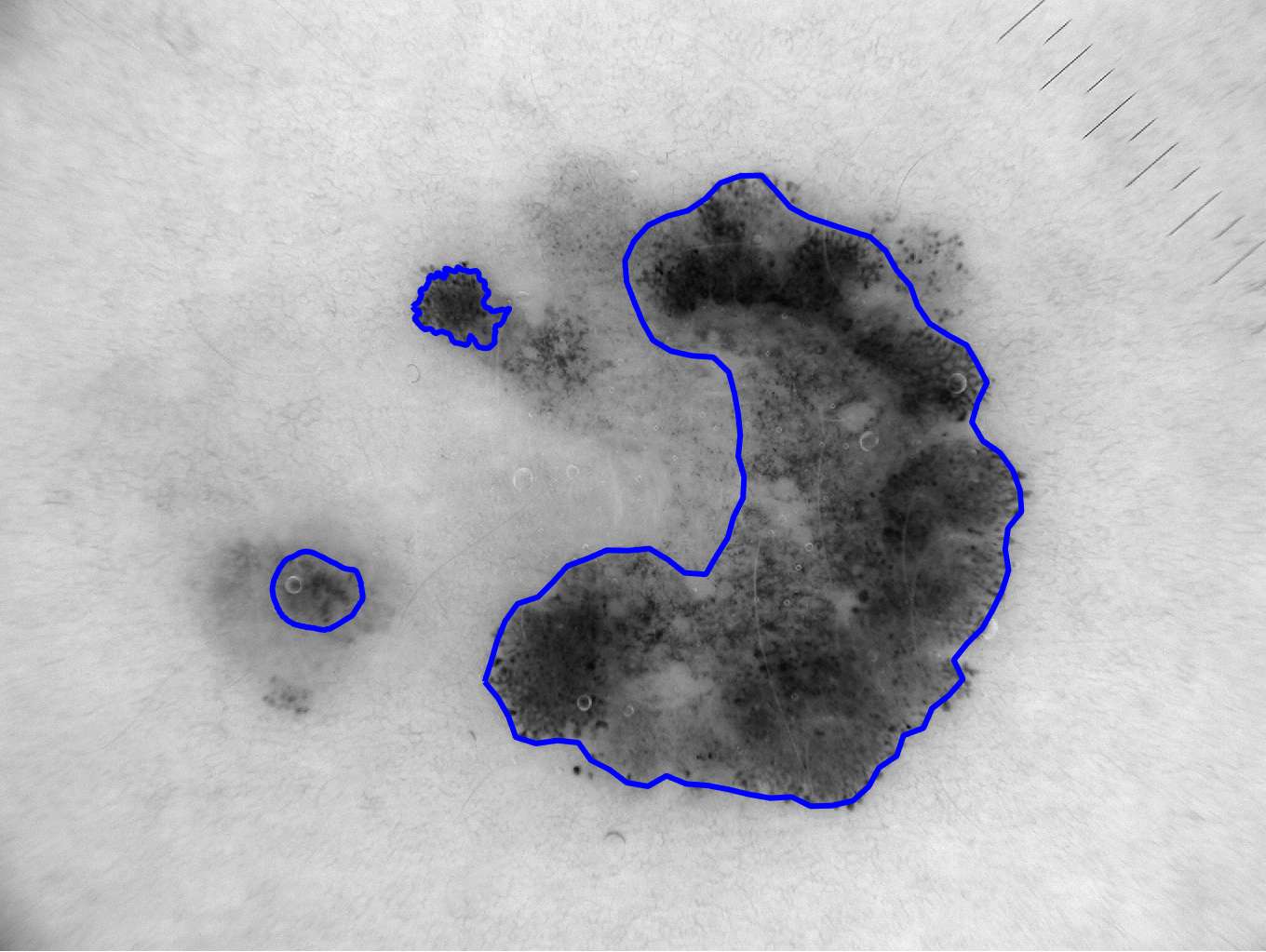}\tabularnewline
\end{tabular}\\
\begin{tabular}{cc}
\includegraphics[width=2.5in]{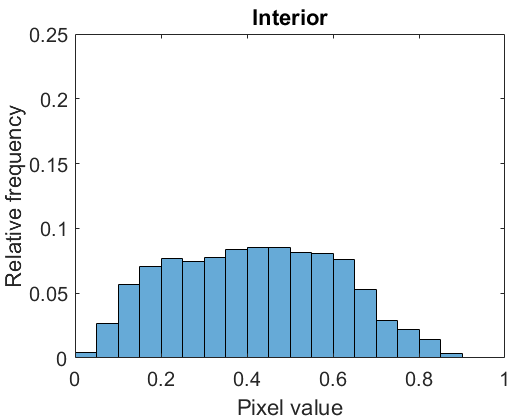}  & \includegraphics[width=2.5in]{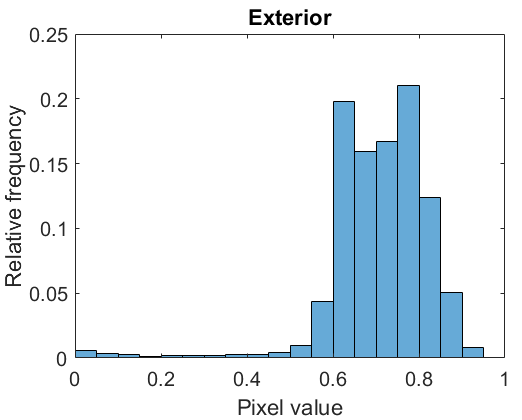}\tabularnewline
\includegraphics[width=2.5in]{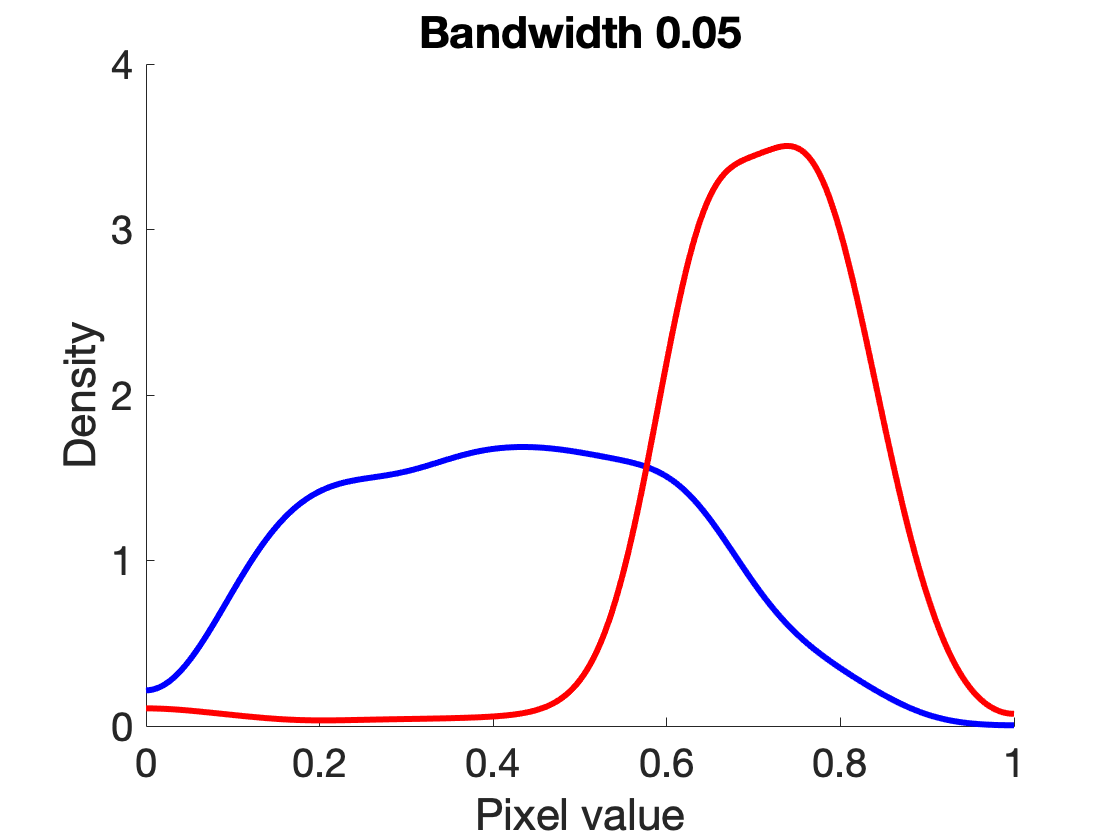}  & \includegraphics[width=2.5in]{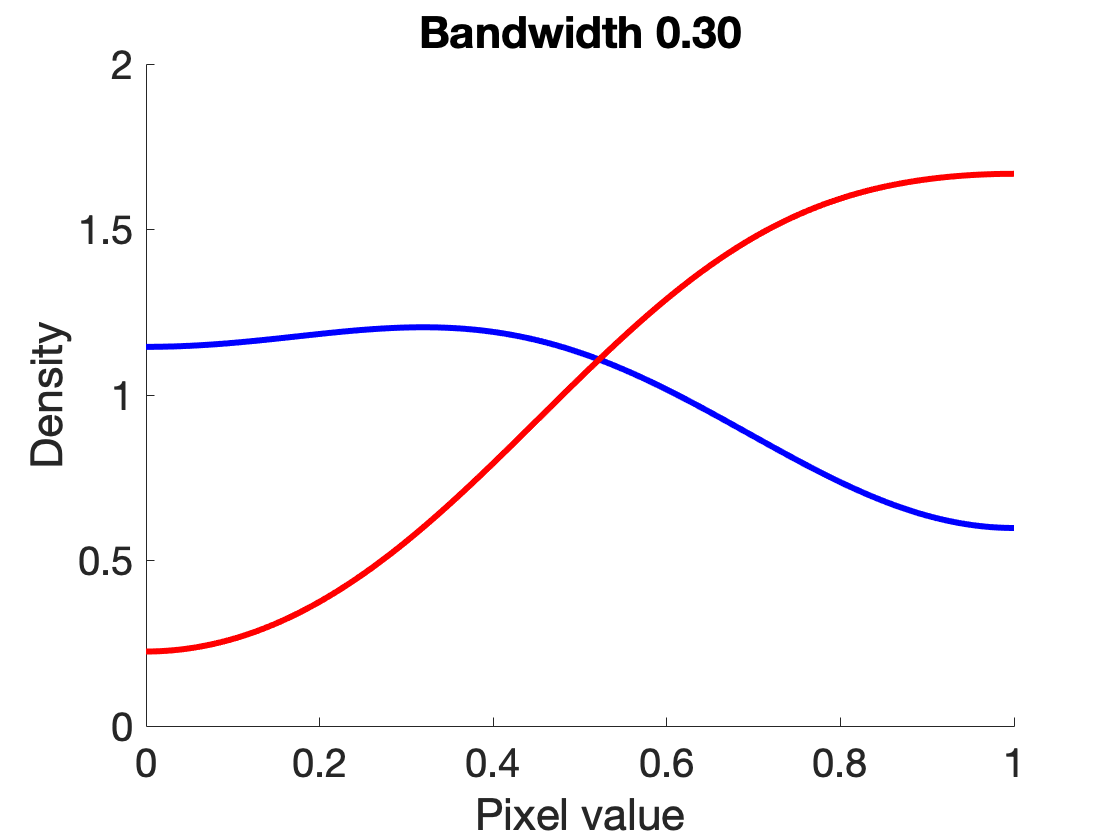}\tabularnewline
\end{tabular}
\par\end{centering}
\label{fig:Lesion12} 
\end{figure}

Figure \ref{fig:Lesion12} shows a patient diagnosed with melanoma. From TOP, there appear to be three well-defined connected components representing the rash -- one large and two small. We initialize the active contour algorithm using these three components, setting $\lambda_{2}=0.3$ and $\lambda_{3}=0$ for all three, but using $\lambda_{1}=0.4$ for the large component, and $\lambda_{1}=0.1$ for the small components in order to avoid self-intersections. We note that in the image, the upper small component is not very well-separated from the large component, as was seen in Figure \ref{fig:SkinLesion1}. One way to resolve this is to change the Gaussian kernel bandwidth used to estimate $p_{\text{int}}$ and $p_{\text{ext}}$, defined in Section \ref{subsubsec:Energy}. Raw pixel values for the interior and exterior from the ground truth boundaries are shown in the middle panel of Figure \ref{fig:Lesion12}. Selecting a Gaussian kernel with bandwidth 0.05, as shown in the bottom left, results in boundary estimates shown
in the top middle plot. In this case, the large component overlaps with the upper small component. However, if we change the kernel bandwidth to 0.30, the resulting boundary estimate more accurately captures the respective
components, and are well-separated. This is in part because it is more difficult to discriminate between the estimated interior and exterior pixel densities, and thus, contours evolve such that only the darkest values are included in their interiors.

\begin{figure}
\caption{Results of TOP+BAC on another benign nevus lesion image from the ISIC-1000 dataset. (Left) Boundary map obtained from TOP, used to initialize two active contours. (Middle/Right) Estimated final contours based on the topological initializations (TOP+BAC), using parameters $\lambda_{1}=0.3,\ \lambda_{2}=0.3,\ \lambda_{3}=0$ for BAC; parameters $\sigma_{1}=1,\ \sigma_{2}=5,\ T=5$ for TOP. Bandwidths of the pixel density estimator are 0.05 and 0.35 for the middle and right panels, respectively. BAC is run for a total of 727 and 743 iterations, respectively. The two contours are shown in different colors to show the overlap in the middle panel, in contrast to more separation with the right panel.}

\begin{centering}
\begin{tabular}{ccc}
\includegraphics[width=1.8in]{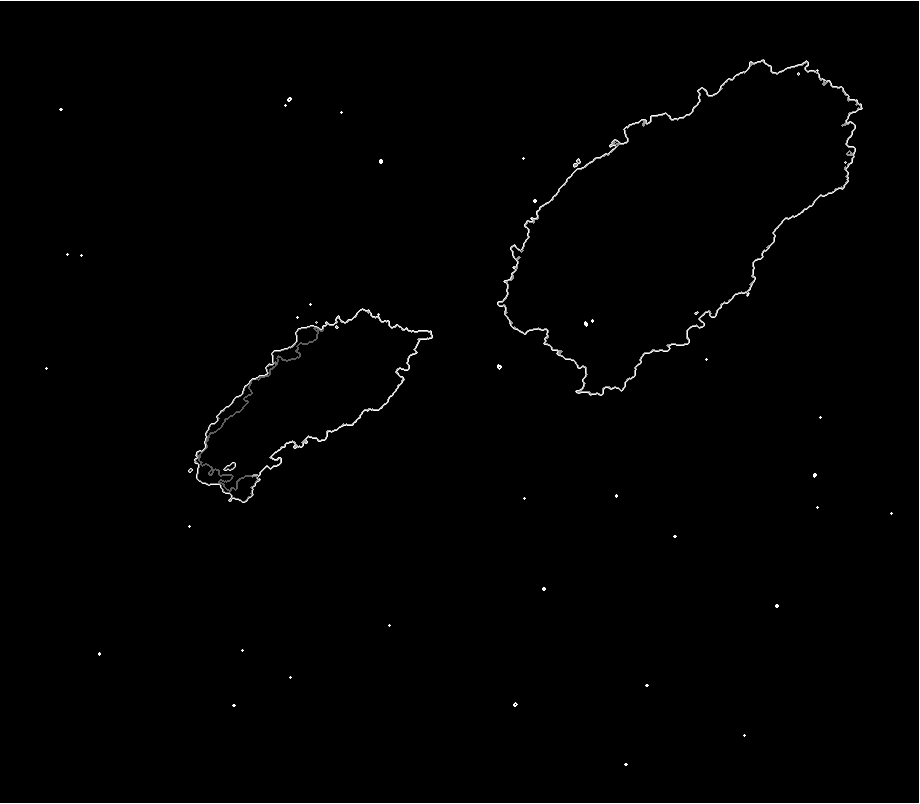}  & \includegraphics[width=1.8in]{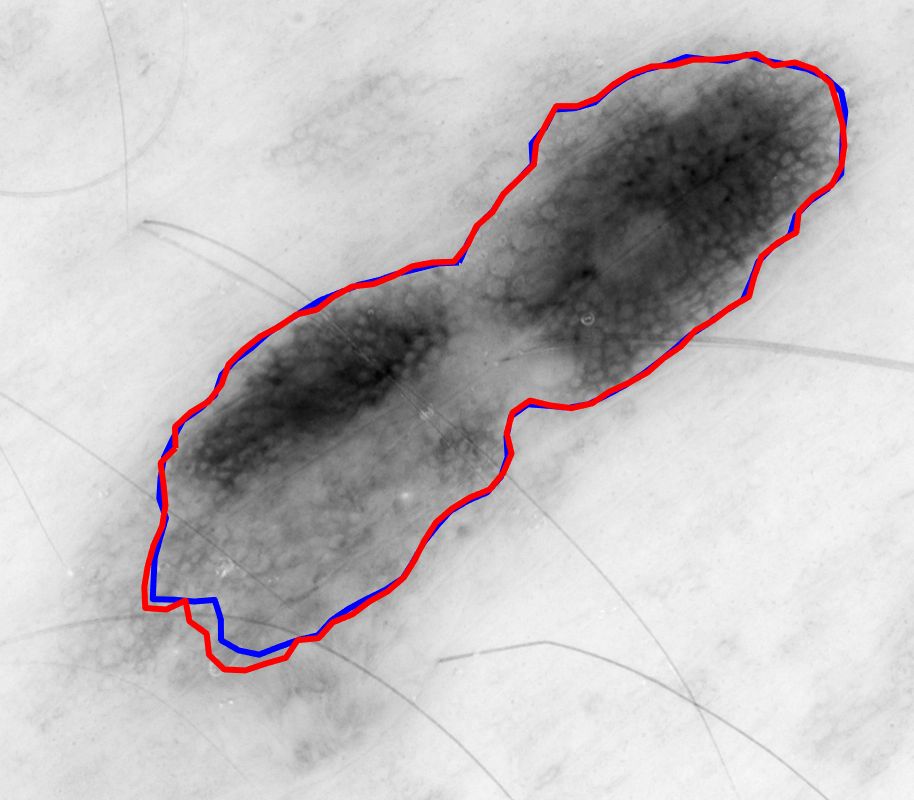}  & \includegraphics[width=1.8in]{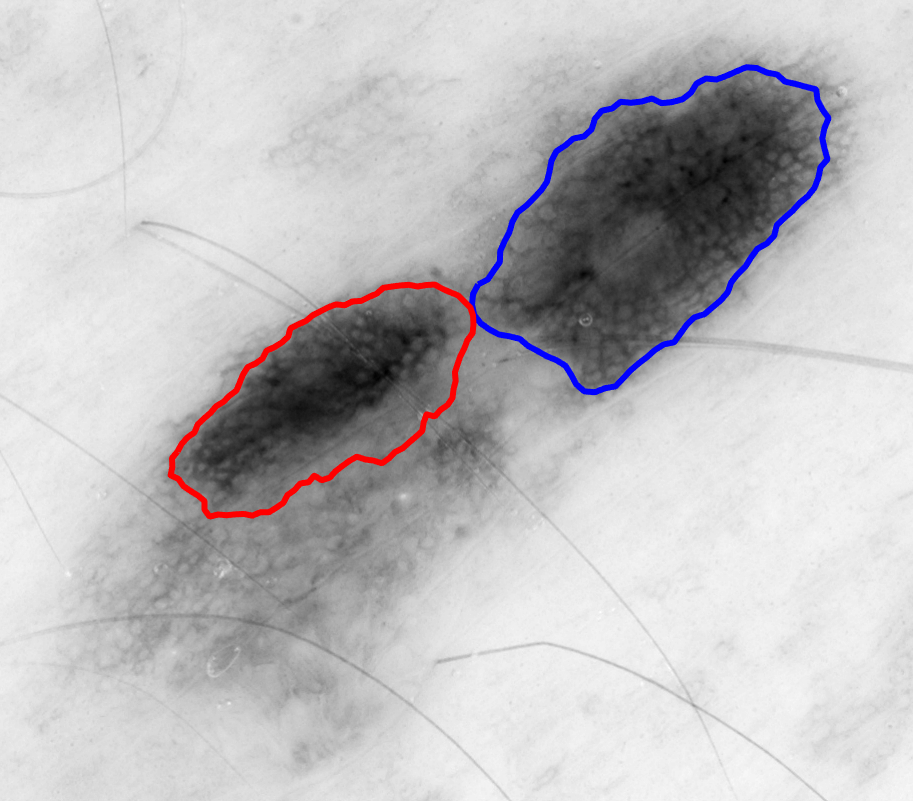}\tabularnewline
\end{tabular}
\par\end{centering}
\label{fig:Lesion2} 
\end{figure}

Similar behavior is displayed for another benign nevus lesion shown in Figure \ref{fig:Lesion2}: with a bandwidth of 0.05, each of the two contours have evolved to coincide with each other. Increasing the bandwidth to 0.35 resolves this issue, resulting in boundary estimate featuring two non-overlapping contours.

\section{Convergence}
\label{appendixConv}
In the simulated examples of Section \ref{sec:Data-Analysis}, TOP generally requires less than 10 seconds to produce a boundary map, meaning the computational cost in estimation here is primarily driven by BAC. In this appendix, we summarize results of the number of steps required for active contour algorithms applied to the MPEG-7 bone images from Appendix \ref{appendixConv}. Table \ref{tab:ConvBone1} shows the number of iterations for the binary bone image with no noise, Gaussian contour perturbation, and Gaussian blur for both BAC and TOP+BAC. Note that in all three noise situations, TOP+BAC requires no more steps than BAC for the bone, as it uses a topological initialization for the BAC algorithm. The setting with no noise requires fewer iterations under TOP+BAC, as the initialization from TOP coincides with the true boundary, as compared to the user-specified initialization which is some distance away from it.

\begin{table}[h]
\caption{Number of steps required for bone examples from Appendix \ref{appendixEx} using BAC and TOP+BAC with convergence tolerance of $10^{-7}$. All examples use BAC settings $\lambda_1=\lambda_2=0.3$ and $\lambda_3=0$. The user-specified initialization for BAC is the red contour shown in the left panel of Figure \ref{fig:NoNoiseBone-1}. The method with the fewest number of iterations for each type of noise is in bold.}

\begin{centering}
\begin{tabular}{cccc}
\toprule 
Noise type & (i) None & (ii) Gaussian perturbation & (iii) Gaussian blur\tabularnewline
\midrule 
\midrule
BAC & 60 & 67 & \textbf{55} \\
TOP+BAC & \textbf{6} & \textbf{53} & \textbf{43} \\
\bottomrule
\end{tabular}
\par\end{centering}
\label{tab:ConvBone1}
\end{table}


\end{document}